\documentclass{IEEEtran}
\usepackage{amsbsy,amsthm,amsmath,amsfonts,amssymb,stmaryrd,pict2e,picture}
\usepackage{graphicx}
\usepackage{textcomp,nicefrac}
\usepackage{outline}
\usepackage{pmgraph}
\usepackage[normalem]{ulem}
\usepackage[utf8]{inputenc}
\usepackage{adjustbox}

\usepackage{hyperref}
\hypersetup{colorlinks,allcolors=black}
\usepackage{enumerate}
\usepackage{times}
\usepackage{xcolor}
\usepackage{xspace}
\usepackage{bm} 
\usepackage[justification=centering]{caption}
\usepackage{subcaption}
\usepackage{epstopdf}
\AppendGraphicsExtensions{.tiff}
\graphicspath{{fig/}} 

\usepackage{epsfig}
\usepackage{tikz}
\usepackage{algpseudocode}
\usepackage[ruled,vlined]{algorithm2e}
\usepackage{mathrsfs}
\usepackage{setspace}
\usepackage[backend=biber,maxbibnames=10,giveninits=true,
doi=true,isbn=false,url=true,eprint=false,
style=numeric-comp,sorting=none
]{biblatex}

\long\def\comment#1{}
\newcommand{\fref}[1] {Fig.~\ref{#1}\xspace}
\newcommand{\tref}[1]{Table~\ref{#1}}

\newcommand{\xtrue}{\xmath{\x_{\text{true}}}}

\DeclareMathOperator*{\argminop}{arg\,min}
\newcommand{\argmin}[1]{\argminop_{#1}} 

\newcommand{\etal}{\textit{et al. }}

\newcommand{\xmath}[1] {\ensuremath{#1}\xspace}
\newcommand{\blmath}[1] {\xmath{\bm{#1}}}
\newcommand{\A}{\blmath{A}}

\newcommand{\x}{\blmath{x}}

\newcommand{\y}{\blmath{y}}

\newcommand{\0}{\blmath{0}}

\newcommand{\ie}{\text{i.e.}}
\newcommand{\eg}{\text{e.g.}}

\newcommand{\bi} {\xmath{b_i}}

\newcommand{\yi} {\xmath{y_i}}

\newcommand{\defequ} {\triangleq}

\newcommand{\reals} {\xmath{\mathbb{R}}}

\newcommand{\xk} {\xmath{\blmath{x}_k}}
\newcommand{\xkk} {\xmath{\blmath{x}_{k+1}}}
\newcommand{\uk} {\xmath{\blmath{u}_{k}}}
\newcommand{\ukk} {\xmath{\blmath{u}_{k+1}}}

\renewcommand{\u} {\blmath{u}}
\renewcommand{\v} {\blmath{v}}

\newcommand{\bd}{\boldsymbol{d}}
\newcommand{\be}{\boldsymbol{e}}

\newcommand{\bp}{\boldsymbol{p}}

\newcommand{\br}{\boldsymbol{r}}

\newcommand{\bmu}{\boldsymbol{\mu}}

\newcommand{\btheta}{\xmath{\bm{\theta}}}

\def\b{{\blmath b}} 

\newcommand{\paren}[1]{\left( #1 \right)}

\definecolor{mich-blue-high}{HTML}{0027CC}
\long\def\blue#1{\bgroup\color{mich-blue-high}#1\egroup}
\long\def\red#1{\bgroup\color{red}#1\egroup}
\newcommand{\ith}{$i$th\xspace}

\makeatletter
\newcommand{\pictslash}[2]{%
  \vcenter{%
    \sbox0{$\m@th#1\varobslash$}\dimen0=.55\wd0
    \hbox to\wd 0{%
      \hfil\pictslash@aux#2\hfil
    }%
  }%
}
\newcommand{\pictslash@aux}[2]{%
    \begin{picture}(\dimen0,\dimen0)
    \roundcap
    \put(0,#1\dimen0){\line(1,#2){\dimen0}}
    \end{picture}%
}
\makeatother

\newcommand{\gtheta}{\xmath{\bm{g}_{\btheta}}}
\newcommand{\bbr}{\bar{\br}}

\newcommand{\xjk}{x_j^{(k)}}

\newcommand{\pluseq}{\mathrel{+}=}
\newcommand{\asteq}{\mathrel{*}=}
\newcommand{\nx}{\xmath{n_{\mathrm{x}}}}
\newcommand{\ny}{\xmath{n_{\mathrm{y}}}}
\newcommand{\nz}{\xmath{n_{\mathrm{z}}}}
\newcommand{\nview}{\xmath{n_{\mathrm{view}}}}
\newcommand{\px}{p_{\mathrm{x}}}
\newcommand{\pz}{p_{\mathrm{z}}}
\newcommand{\lu}{$^{177}\mathrm{Lu}$\xspace}
\newcommand{\ytt}{$^{90}\mathrm{Y}$\xspace}
\providecommand{\DefineBibliographyStrings}[2]{}
\DefineBibliographyStrings{english}{%
  andothers = {\etal}
}
\newcommand{\Deltay}{\xmath{\Delta_{\mathrm{y}}}}
\def\BibTeX{{\rm B\kern-.05em{\sc i\kern-.025em b}\kern-.08em
T\kern-.1667em\lower.7ex\hbox{E}\kern-.125emX}}
\begin{document}
\title{Training End-to-End
Unrolled Iterative Neural Networks
for SPECT Image Reconstruction}
\author{Zongyu Li, 
\IEEEmembership{Student Member, IEEE}, 
Yuni K. Dewaraja, 
\IEEEmembership{Member, IEEE},
\\
and Jeffrey A. Fessler, \IEEEmembership{Fellow, IEEE}
\thanks{This work involved human subjects in its research. 
Approval of all ethical and experimental procedures and protocols 
was granted by University of Michigan Institutional Review Board (IRB).
}
\thanks{This work was supported in part by NIH Grants
R01 EB022075 and R01 CA240706.}
\thanks{Z. Li and J. A. Fessler are with the Department of Electrical Engineering and Computer Science, 
University of Michigan, 
(e-mails: zonyul@umich.edu, fessler@umich.edu).}
\thanks{Y. K. Dewaraja is with the Department of Radiology,
University of Michigan, (e-mail: yuni@med.umich.edu).}
\thanks{Code for reproducing the results will be available at \url{https://github.com/ZongyuLi-umich/}
after the paper is accepted.
}
}

\maketitle

\begin{abstract}
Training end-to-end unrolled iterative neural networks
for SPECT image reconstruction
requires a memory-efficient forward-backward projector
for efficient backpropagation.
This paper describes
an open-source, high performance
Julia implementation of a 
SPECT forward-backward projector
that supports memory-efficient backpropagation
with an exact adjoint.
Our Julia projector 
uses only $\sim$5\% of the memory
of an existing Matlab-based projector.
%
We compare unrolling a CNN-regularized
expectation-maximization (EM) algorithm
with end-to-end training 
using our Julia projector
with other training methods
such as gradient truncation
(ignoring gradients involving the projector)
and sequential training,
using XCAT phantoms and virtual patient (VP) phantoms
generated from SIMIND Monte Carlo (MC) simulations.
Simulation results
with two different radionuclides 
(\ytt and \lu)
show that:
1)
For \lu XCAT phantoms and \ytt VP phantoms,
training unrolled EM algorithm
in end-to-end fashion
with our Julia projector yields the 
best reconstruction quality
compared to other training methods and OSEM,
both qualitatively and quantitatively.
For VP phantoms with \lu radionuclide,
the reconstructed images using end-to-end training are 
in higher quality
than using sequential training and OSEM,
but are comparable with using gradient truncation.
We also find there exists a trade-off 
between computational cost and reconstruction accuracy
for different training methods.
End-to-end training has the highest accuracy 
because the correct gradient is used in backpropagation;
sequential training yields worse 
reconstruction accuracy,
but is significantly faster and uses much less memory.
\end{abstract}


\section{Introduction}
\label{sec:introduction}

\IEEEPARstart{S}{ingle} 
photon emission computerized tomography (SPECT)
is a nuclear medicine technique 
that images spatial distributions of radioisotopes
and plays a pivotal role in
clinical diagnosis,
and in estimating
radiation-absorbed doses in nuclear medicine therapies
\cite{james:21:cso, dewaraja:12:mpn}.
For example, quantitative SPECT imaging 
with Lutetium-177 (\lu) 
in targeted radionuclide therapy
(such as \lu DOTATATE)
is important
in determining dose-response relationships 
in tumors
and holds great potential 
for dosimetry-based individualized treatment.
Additionally, quantitative Yttrium-90 (\ytt) 
bremsstrahlung SPECT imaging
is valuable
for safety assessment 
and absorbed dose verification
after \ytt radioembolization in liver malignancies.
However, SPECT imaging
suffers from noise and limited spatial resolution 
due to the
collimator response;
the resulting reconstruction problem
is hence ill-posed and challenging to solve.

Numerous reconstruction algorithms have been proposed
for SPECT reconstruction,
of which the most popular ones are 
model-based image reconstruction algorithms
such as
maximum likelihood expectation maximization
(MLEM) \cite{shepp:82:mlr}
and
ordered-subset EM (OSEM) \cite{hudson:94:air}.
These methods first construct a mathematical model
for the SPECT imaging system,
then maximize the (log-)likelihood
for a Poisson
noise model.
Although MLEM and OSEM
have achieved great success in clinical use,
they have a trade-off between recovery and noise.
To address that trade-off,
researchers have proposed alternatives
such as regularization-based 
(or maximum a posteriori in Bayesian setting)
reconstruction methods 
\cite{panin:99:tvr, fessler:94:pwl, lalush:93:agb}.
For example, 
Panin \etal \cite{panin:99:tvr}
proposed total variation (TV)
regularization for SPECT reconstruction.
However, TV regularization
may lead to ``blocky" images
and over-smoothing the edges.
One way to overcome blurring edges
is to incorporate 
anatomical boundary side information
from CT images \cite{dewaraja:10:rri},
but that method requires 
accurate organ segmentation.
Chun \etal \cite{chun:12:nlm}
used non-local means (NLM) filters 
that exploit the self-similarity of patches
in images for regularization,
yet that method is computationally expensive
and hence less practical.
In general,
choosing an appropriate regularizer
can be challenging;
moreover, 
these traditional regularized algorithms
may lack generalizability to images 
that do not follow assumptions made by the prior.

With the recent success of 
deep learning (DL) 
and especially 
convolutional neural networks (CNN),
DL methods have been reported to 
outperform conventional algorithms
in many medical imaging applications
such as in MRI \cite{zeng:21:aro, yang:18:ddd, quan:18:csm},
CT \cite{minnema:22:aro, chen:18:lle} 
and PET reconstruction \cite{reader:21:dlf, kim:18:ppr, mehranian:21:mbd}.
\comment{ 
For example, 
Ronneberger \etal 
proposed U-Net \cite{ronneberger:15:unc} 
that achieved state-of-the-art accuracy 
on the 
2012 International Symposium on Biomedical Imaging (ISBI) challenge for 
segmentation of neuronal structures 
in electron microscopic stacks.
Yang \etal \cite{yang:18:ddd}
applied
conditional 
Generative Adversarial Networks (CGAN)
\cite{mirza:14:cga}
for anti-aliasing in fast compressed 
sensing MRI and showed better results
than conventional regularizer like TV and BM3D  \cite{danielyan:12:bfa}.
}However,
fewer DL approaches to SPECT reconstruction
appear in the literature.
Reference~\cite{shao:21:adl}
proposed ``SPECTnet" 
with a two-step training strategy 
that learns the transformation
from projection space to image space
as an alternative to the traditional OSEM algorithm.
Reference~\cite{shao:21:alr}
also proposed a DL method that 
can directly reconstruct the activity image
from the SPECT projection data,
even 
with reduced view angles.
Reference~\cite{mostafapour:22:dlg}
trained a neural network
that maps non-attenuation-corrected SPECT images
to those corrected by CT images
as a post-processing procedure
to enhance the reconstructed image quality.

Though promising results were reported with these methods,
most of them worked in 2D
whereas 3D is used in practice \cite{shao:21:adl, shao:21:alr}.
Furthermore,
there has yet to be an investigation
of end-to-end training
of CNN regularizers
that are
embedded in unrolled SPECT iterative statistical algorithms
such as CNN-regularized EM.
End-to-end training is popular
in machine learning
and other medical imaging fields
such as MRI image reconstruction \cite{shen:19:ete},
and is reported to meet data-driven regularization for inverse problems \cite{mukherjee:21:ete}.
But for SPECT image reconstruction, 
end-to-end training 
is nontrivial to implement
due to its complicated system matrix.
Alternative training methods have been proposed,
such as sequential training 
\cite{lim:20:ilc, sahiner:21:crb, ozturkler:21:sai, corda:22:met}
and
gradient truncation \cite{mehranian:19:mbd};
these methods were shown to be effective,
though they could yield
sub-optimal reconstruction results
due to approximations
to the training loss gradient.
Another approach is to construct a neural network 
that also models the SPECT system matrix, 
like in ``SPECTnet" \cite{shao:21:adl},
but this approach lacks interpretability
compared to algorithms like
unrolled CNN-regularized EM,
\ie, 
if one sets the regularization parameter to zero,
then the latter 
becomes identical to the traditional EM.

As an end-to-end training approach 
has not yet been investigated 
for SPECT image reconstruction,
this paper first describes
a SPECT forward-backward projector
written
in the open-source and high performance Julia language
that enables efficient auto-differentiation.
Then we compare the end-to-end training approach 
with other non-end-to-end training methods.

The structure of this article is as follows.
Section~\ref{sec:methods} 
describes the implementation of our Julia projector
and discusses end-to-end training 
and other training methods 
for the unrolled EM algorithm.
Section~\ref{sec:results} 
compares the accuracy, speed
and memory use of our Julia projector with
Monte Carlo (MC) 
and a Matlab-based projector,
and then compares reconstructed images with
end-to-end training
versus sequential training 
and gradient truncation on different datasets
(XCAT and VP phantoms),
using qualitative and quantitative
evaluation metrics.
Section~\ref{sec:discussion} and 
\ref{sec:conclusion}
conclude this paper
and discuss future works.

\textit{Notation:}
Bold upper/lower case letters (\eg, \A, \x, \y, \b) 
denote matrices and column vectors, respectively.
Italics (\eg, $\mu, y, b$) denote scalars.
\yi and \bi denote the \ith  
element in vector \y and \b, respectively.
$\mathbb{R}^N$ and $\mathbb{C}^N$ denote
$N$-dimensional real/complex normed vector space, respectively.
$(\cdot)^{*}$ denotes the complex  conjugate
and $(\cdot)'$ denotes Hermitian transpose. 

\section{Methods}
\label{sec:methods}

This section
summarizes the Julia SPECT projector,
a DL-based image reconstruction method
as well as the dataset used in experiments
and other experiment setups.

\subsection{Implementation of Julia SPECT projector}

\begin{figure}[hbt!]
    \centering
    \includegraphics[width=0.6\linewidth]{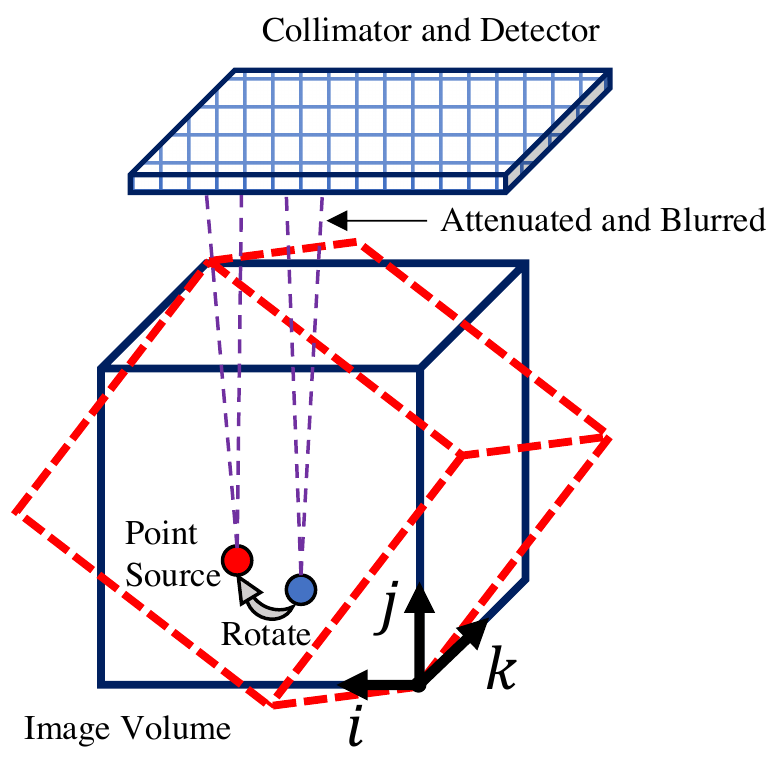}
    \caption{SPECT imaging model for 
    parallel-beam collimators,
    with attenuation 
    and depth-dependent collimator 
    point spread response.}
    \label{fig:intro}
\end{figure}

Our Julia implementation of SPECT projector 
is based on \cite{zeng:91:fdi},
modeling
parallel-beam collimator geometries.
Our projector also accounts for attenuation
and depth-dependent collimator response.
We did not model the scattering events
like Compton scatter and coherent scatter 
of high energy gamma rays within the object.
\fref{fig:intro} illustrates 
the SPECT imaging system modeled in this paper.

For the forward projector,
at each rotation angle,
we first rotate the 3D image matrix
$\x \in \reals^{\nx \times \ny \times \nz}$ 
according to the third dimension
by its projection angle
$\theta_l$
(typically $2\pi (l-1)/\nview$);
$l$ denotes the view index,
which ranges from 1 to \nview
and \nview
denotes the total number of projection views.
We implemented and compared 
(results shown in Section~\ref{sec:results})
both bilinear interpolation
and 3-pass 1D linear interpolation
\cite{dibella:96:aco}
with zero padding boundary condition
for image rotation.
For attenuation correction,
we first rotated 
the 3D attenuation map
$\bmu \in \reals^{\nx \times \ny \times \nz}$ 
(obtained from transmission tomography)
also by $\theta_l$,
yielding a rotated 3D array
$\tilde{\mu}(i,j,k;l)$,
where $i,j,k$ denotes the 3D voxel coordinate.
Assuming \ny is the index 
corresponding to the closest plane of \x
to the detector,
then we model the accumulated attenuation factor $\bar{\mu}$
for each view angle as 
\begin{equation}\label{e,atten}
\bar{\mu}(i,j,k;l) = \mathrm{e}^{-\Deltay \paren{\frac{1}{2}
\tilde{\mu}(i,j,k;l) +
\sum_{s=j+1}^{\ny} \tilde{\mu}(i,s,k;l)}}   
,
\end{equation}
where
\Deltay denotes the voxel size 
for the (first and) second coordinate.
Next,
for each $y$ slice
(an $(x,z)$ plane
for a given $j$ index)
of the rotated and attenuated image,
we convolve
with the appropriate slice
of the depth-dependent point spread function 
$\bp \in \reals^{\px \times \pz \times \ny \times \nview}$
using a 2D fast Fourier transform (FFT).
Here we use replicate padding 
for both the $i$ and $k$ coordinates.
The view-dependent PSF
accommodates non-circular orbits.
Finally,
the forward projection operation
simply sums the rotated,
blurred and attenuated activity image \x
along the second coordinate $j$.
Algorithm~\ref{alg:spect,forw} 
summarizes the forward projector,
where $\circledast$ denotes
a 2D convolution operation.

\begin{algorithm}[ht!]
\SetAlgoLined
\SetInd{0.5em}{0.5em}
 \textbf{Input:} 
 3D image $\x \in \reals^{\nx \times \ny \times \nz}$, \\
 3D attenuation map $\bmu \in \reals^{\nx \times \ny \times \nz}$, \\
 4D point spread function $\bp
 \in \reals^{\px \times \pz \times \ny \times \nview}$, \\
 voxel size $\Deltay$. \\
 \textbf{Initialize: } $\v \in \reals^{\nx \times \nz \times \nview}$ 
 as all zeros. \\
 \For{$l=1,...,\nview$}{
 $\tilde{\x} \leftarrow \text{rotate } \x$ by $\theta_l$
 \\
 $\tilde{\bmu} \leftarrow \text{rotate } \bmu$ by $\theta_l$
 \\ 
 \For{$j=1,...,\ny$}{
 $\bar{\bmu}
 \leftarrow $
 calculate by \eqref{e,atten}
 using $\tilde{\bmu}$
 \\
 $
 \tilde{x}(i,j,k) \asteq
 \bar{\mu}(i,j,k)$ 
 \\
 $v(i,k,l) \pluseq \tilde{x}(i,j,k) \circledast p(i,k;j,l)$
 }
 }
 \textbf{Output:} projection views $\v \in \reals^{\nx \times \nz \times \nview}$
 \caption{SPECT forward projector}
 \label{alg:spect,forw}
\end{algorithm}

All of these steps are linear,
so hereafter, we use \A
to denote the forward projector,
though it is not stored explicitly as a matrix.
As each step is linear,
each step has an adjoint operation.
So the backward projector $\A'$ is 
the adjoint of \A
that satisfies
\begin{equation}\label{e,adj}
\langle \A \x, \y \rangle 
= 
\langle \x, \A' \y \rangle, 
\quad 
\forall \x, \y.
\end{equation}
The exact adjoint of (discrete) image rotation
is not simply a discrete rotation 
of the image by $-\theta_l$.
Instead,
one should also consider the adjoint 
of linear interpolation.
For the adjoint of convolution,
we assume the point spread function
is symmetric along coordinates $i$ and $k$
so that the adjoint convolution operator 
is just the forward convoluation operator
along with
the adjoint of replicate padding.
Algorithm~\ref{alg:spect,back} 
summarizes the SPECT backward projector.

\begin{algorithm}[ht!]
\SetAlgoLined
\SetInd{0.5em}{0.5em}
 \textbf{Input:} 
 Array of 2D projection views $\v \in \reals^{\nx \times \nz \times \nview}$, \\
 3D attenuation map $\bmu \in \reals^{\nx \times \ny \times \nz}$, \\
 4D point spread function $\bp
 \in \reals^{\px \times \pz \times \ny \times \nview}$, \\
 voxel size $\Deltay$. \\
 \textbf{Initialize: } $\x \in \reals^{\nx \times \ny \times \nz}$ 
 as all zeros. \\
 \For{$l=1,...,\nview$}{
 $\tilde{\bmu}$ $\leftarrow$ rotate 
 $\bmu$ by
 $\theta_l$
 \\
 \For{$j=1,...,\ny$}{
 $\bar{\bmu}\leftarrow$ 
 calculate by \eqref{e,atten} using $\tilde{\bmu}$
 \\
 $\tilde{v}(i,k,l)$ $\leftarrow$
 adjoint of $v(i,k,l) \circledast p(i,k,j,l)$
 \\
 $\tilde{x}(i,j,k) \leftarrow \tilde{v}(i,k,l) \cdot \bar{\mu}(i,j,k;l)$
 }
 $\x \pluseq$ adjoint rotate $\tilde{\x}$ by
$\theta_l$
 \\
 }
 \textbf{Output:} $\x \in \reals^{\nx \times \ny \times \nz}$
 \caption{SPECT backward projector}
 \label{alg:spect,back}
\end{algorithm}

To accelerate the for-loop process, 
we used multi-threading 
to enable projecting or backprojecting
multiple angles at the same time.
To reduce memory use,
we pre-allocated necessary arrays 
and used fully in-place operations 
inside the for-loop in forward and backward projection.
To further accelerate auto-differentiation,
we customized the chain rule 
to use the linear operator \A or $\A'$ as the Jacobian 
when calling $\A\x$ or $\A'\y$
during backpropagation.
We implemented and tested our projector 
in Julia v1.6 \comment{\red{let's test with 1.8 when we move it to juliaImageRecon so we can claim the most recent version...  oh, maybe that would change your timing results so maybe we should leave this.  hmm.}};
we also implemented a GPU version 
in Julia (using CUDA.jl)
that runs efficiently on a GPU 
by eliminating explicit scalar indexing.
For completeness,
we also provide a PyTorch version
but without multi-threading support,
in-place operations
nor the exact adjoint of image rotation.

\subsection{Unrolled CNN-regularized EM algorithm}

Model-based image reconstruction algorithms
seek to estimate image $\x \in \reals^N$
from noisy measurements $\y \in \reals^M$ 
with imaging model $\A \in \reals^{M \times N}$.
In SPECT reconstruction,
the measurements $\y$ are often modeled by 
\begin{equation}\label{e,y,poisson}
\y \sim \text{Poisson}(\A \x + \bbr),    
\end{equation}
where $\bbr \in \reals^M$ denotes the vector of means
of background events such as scatters.
Combining regularization with the
Poisson negative log-likelihood
yields the following optimization problem:
\begin{align}\label{e,cost,poisson}
\hat{\x} &= \argmin{\x \ge \0}{f(\x) + R(\x)},
\nonumber \\
f(\x) &\defequ \blmath{1}'(\A\x + \bbr) 
- \y' \log(\A \x + \bbr)
,
\end{align}
where $f(\x)$ is the data fidelity term
and $R(\x)$ denotes the regularizer.
For deep learning regularizers, 
we follow
\cite{lim:20:ilc}
and formulate
$R(\x)$ as
\begin{equation}\label{e,reg}
R(\x) \defequ \frac{\beta}{2} 
\|\x - \gtheta(\x)\|_2^2,
\end{equation}
where $\beta$ denotes the regularization parameter;
\gtheta
denotes a neural network
with parameter \btheta
that is trained
to learn to enhance the image quality.

Based on $\eqref{e,cost,poisson}$,
a natural reconstruction approach 
is to
apply variable splitting with
$\u = \gtheta(\x)$
and then
alternatively update 
the images \x and \u
as follows
\begin{align}\label{e,alternate}
&\ukk = \gtheta(\xk),
\nonumber \\
&\xkk = \argmin{\x \ge 0}{f(\x) + \frac{\beta}{2} 
\| \x - \ukk \|_2^2}
,
\end{align}
where subscript $k$ denotes the iteration number.
To minimize \eqref{e,alternate},
we used the EM-surrogate from
\cite{depierro:95:ame}
as summarized in
\cite{lim:20:ilc},
leading to the following%
\comment{ 
(can be derived from Jensen's inequality):
\begin{align}\label{e,em}
\phi(t;y) &\defequ t - y \log(t),
\quad 
\bar{\yi} \defequ [\A \xk + \bbr]_i,
\\
f(\x) &=
\sum_{i=1}^M 
\phi \paren{
\sum_{j=1}^N
\paren{\frac{a_{ij}\xjk}{\bar{\yi}}}
\frac{x_j}{\xjk}
\bar{\yi}
+\frac{\bbr_i}{\bar{\yi}}\bar{\yi}}
\nonumber
\\
&\le \sum_{j=1}^N F_j(x_j; \xk)
\nonumber,
\\
F_j(x_j; \xk) &\defequ 
\sum_{i=1}^M 
\paren{\frac{a_{ij} \xjk}{\bar{\yi}}}
\paren{\frac{x_j}{\xjk}\bar{\yi} 
- \yi \log\paren{\frac{x_j}{\xjk}\bar{\yi}}}
\nonumber 
,
\end{align}
where $x_j$ denotes the $j$th voxel in $\x$.
Considering the regularization term,
we majorize \eqref{e,alternate} by 
\begin{align}\label{e,Q}
&f(\x) + \frac{\beta}{2} \|\x - \ukk\|_2^2
\le \sum_{j=1}^N Q_j(x_j; \xk),
\nonumber \\
&Q_j(x_j; \xk) \defequ 
F_j(x_j; \xk) 
+ \frac{\beta}{2} (x_j - [\ukk]_j)^2
.
\end{align}
Differentiating $Q_j(x_j; \xk)$
w.r.t. $x_j$
and setting the derivative to zero
leads to
\begin{align}\label{e,deri}
&\beta x_j^2 + 
d_j(\beta) x_j
- \xjk e_j(\xk)
=0,
\nonumber \\
&d_j(\beta) \defequ 
\sum_{i=1}^M a_{ij} - \beta [\uk]_j,
\nonumber \\
&e_j(\xk) \defequ
\sum_{i=1}^M
a_{ij} \frac{\yi}{[\A \xk + \bbr]_i}
.
\end{align}
Solving for $x_j$ yields
\begin{equation}\label{e,xj}
\hat{x}_j = \frac{-d_j(\beta)
+ \sqrt{d_j(\beta)^2 + 4\beta x_j e_j(\xk)}
}{2\beta}
,
\end{equation}
and the resulting 
} 
vector update:
\begin{align}
&\hat{\x}_k = 
\frac{1}{2\beta}
\paren{-\bd(\beta)
+ \sqrt{\bd(\beta)^2 + 
4\beta \xk \odot \be(\xk)}}
,
\label{e,x,update}
\\
&\bd(\beta) \defequ 
\A' \blmath{1} - \beta \uk,
\quad 
\be(\xk) \defequ \A'\paren{\y \varoslash 
\paren{\A \xk + \bbr}}
,
\label{e,e,update}
\end{align}
where $\odot$ and $\varoslash$
denote element-wise multiplication
and division, respectively.
To compute \xkk,
one must substitue
$\hat{\x}_k$
back into
$\be(\cdot)$
in
\eqref{e,e,update},
and repeat.
Hereafter, we refer to 
\eqref{e,alternate}
as one outer iteration
and
\eqref{e,x,update}
as one inner EM iteration.
Algorithm~\ref{alg:spect,regem} 
summarizes the  
CNN-regularized EM algorithm.

\begin{algorithm}[ht!]
\SetAlgoLined
\SetInd{0.5em}{0.5em}
 \textbf{Input:} 
 3D projection measurements $\y$, \\
 3D background measurements $\bbr$, \\
 system model \A, 
 initial guess $\x_0$, \\
 deep neural network \gtheta,
 \\
 outer iterations $K$.
 \\
 \For{$k=0,...,K-1$}{
 $\ukk = \gtheta(\xk)$ \\
 $\xkk$ $\leftarrow$ repeat \eqref{e,x,update}
 until convergence tolerance
 or maximum \# of inner iterations
 is reached
 }
 \textbf{Output:} $\x_K$
 \caption{SPECT CNN-regularized EM algorithm}
 \label{alg:spect,regem}
\end{algorithm}

To train \gtheta,
the most direct way
is to unroll Algorithm~\ref{alg:spect,regem}
and train end-to-end 
with an appropriate target;
this supervised approach
requires backpropagating through 
the SPECT system model,
which is not trivial to implement
with previous SPECT projection tools
due to the memory issues.
Non-end-to-end training methods, \eg, 
sequential training \cite{lim:20:ilc},
first train \uk
by the target
and then plug into \eqref{e,x,update}
at each iteration.
This method must use non-shared weights
for the neural network per each iteration.
Another method is gradient truncation
\cite{mehranian:19:mbd}
that ignores the gradient involving
the system matrix \A and its adjoint $\A'$
during backpropagation.
Both of these training methods,
though reported to be effective,
may be sub-optimal
because they approximate
the overall training loss gradients.

\begin{figure*}[hbt!]
\centering
\subfloat[MC (\lu)]{\includegraphics[width=0.25\linewidth]{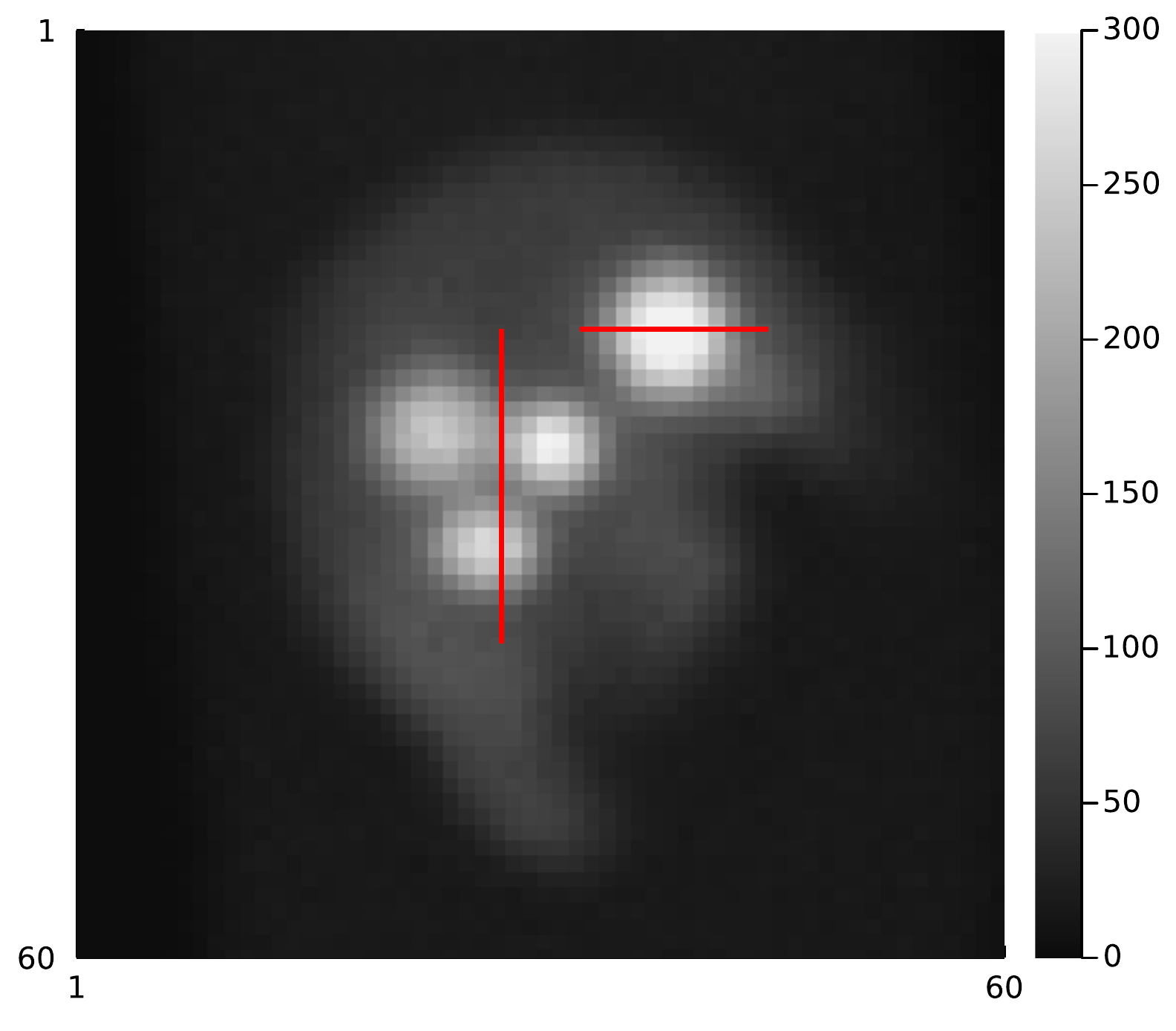}}
\subfloat[Julia 1D (\lu)]{\includegraphics[width=0.25\linewidth]{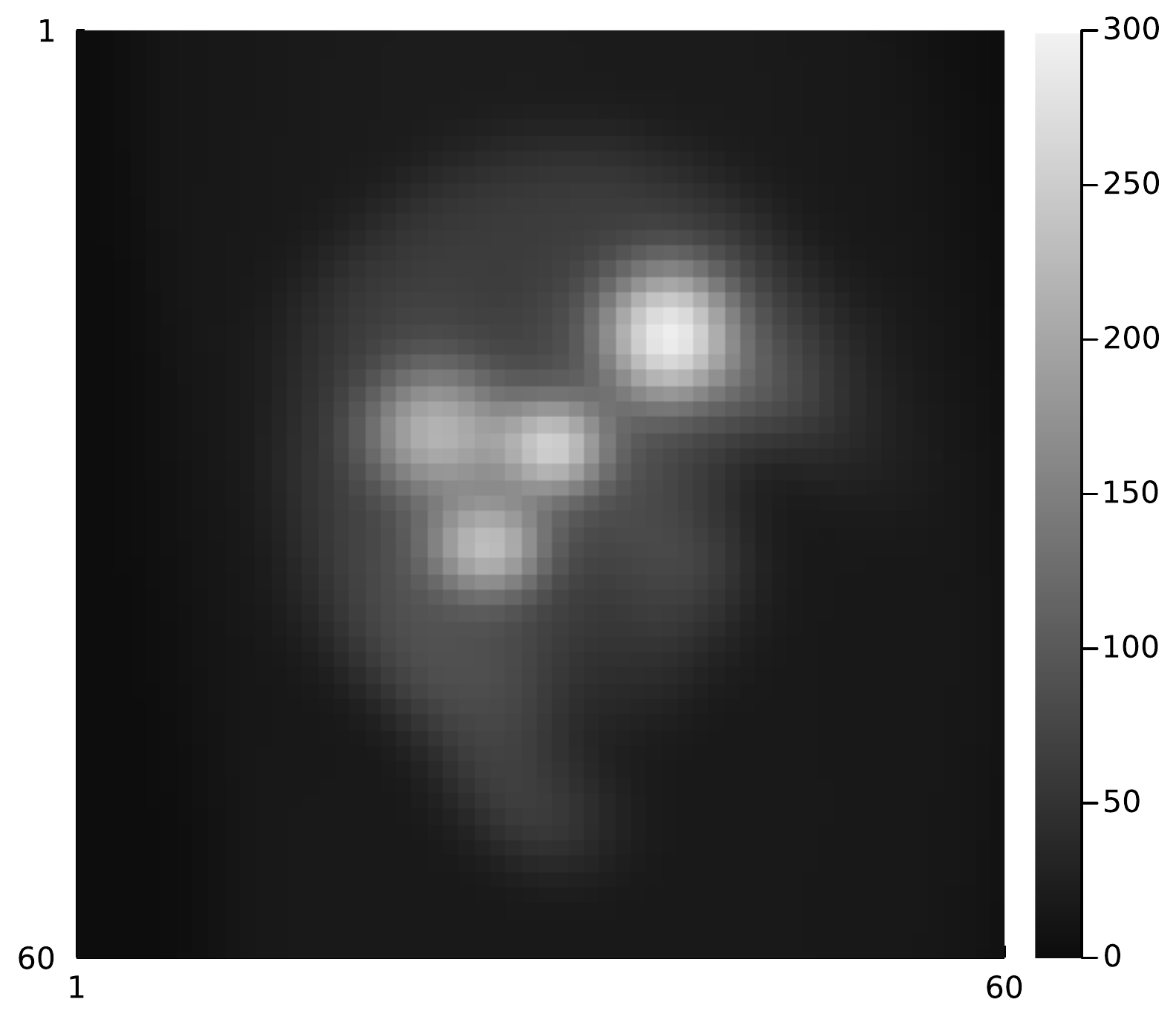}}
\subfloat[Julia 2D (\lu)]{\includegraphics[width=0.25\linewidth]{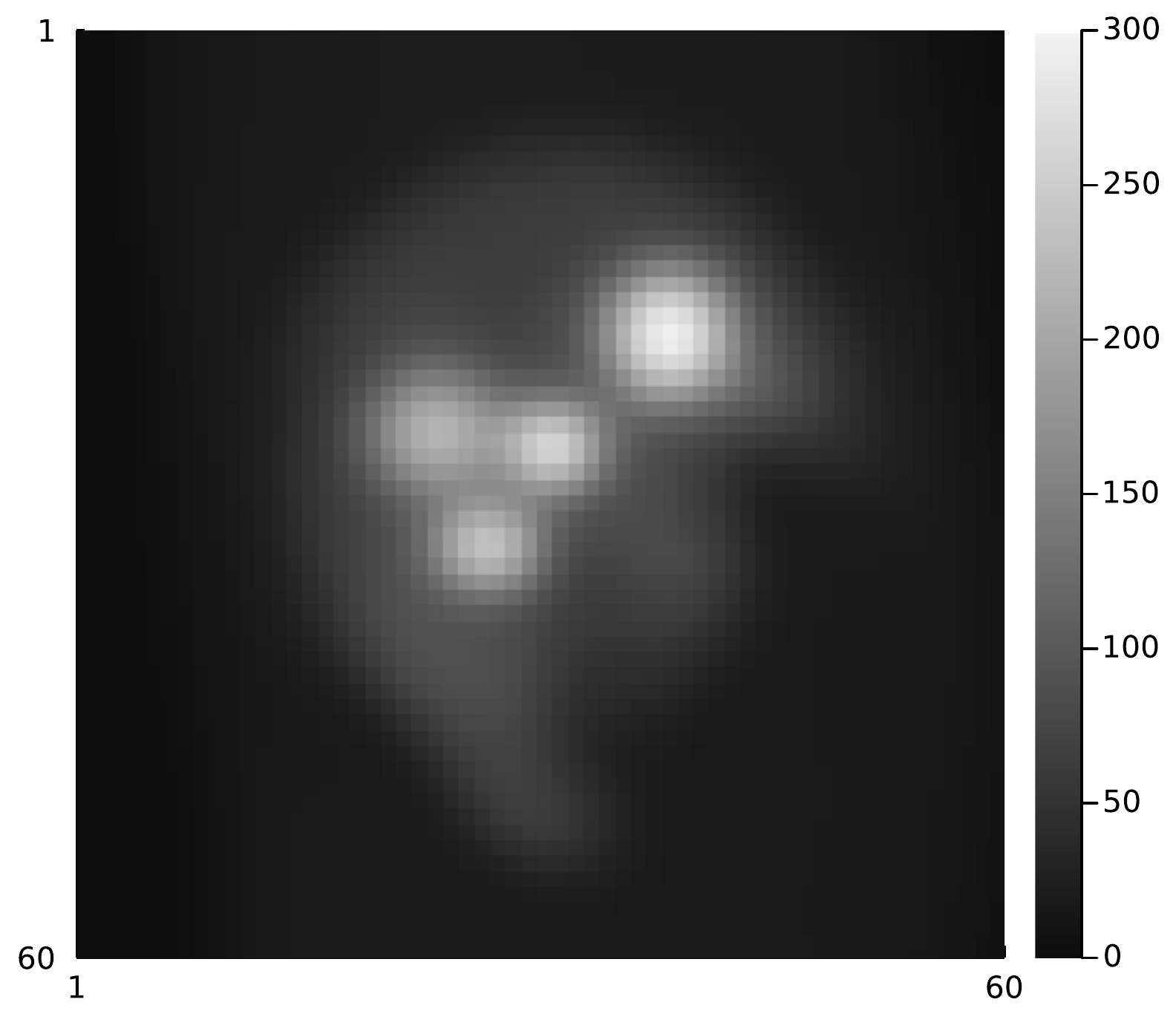}}
\subfloat[Matlab (\lu)]{\includegraphics[width=0.25\linewidth]{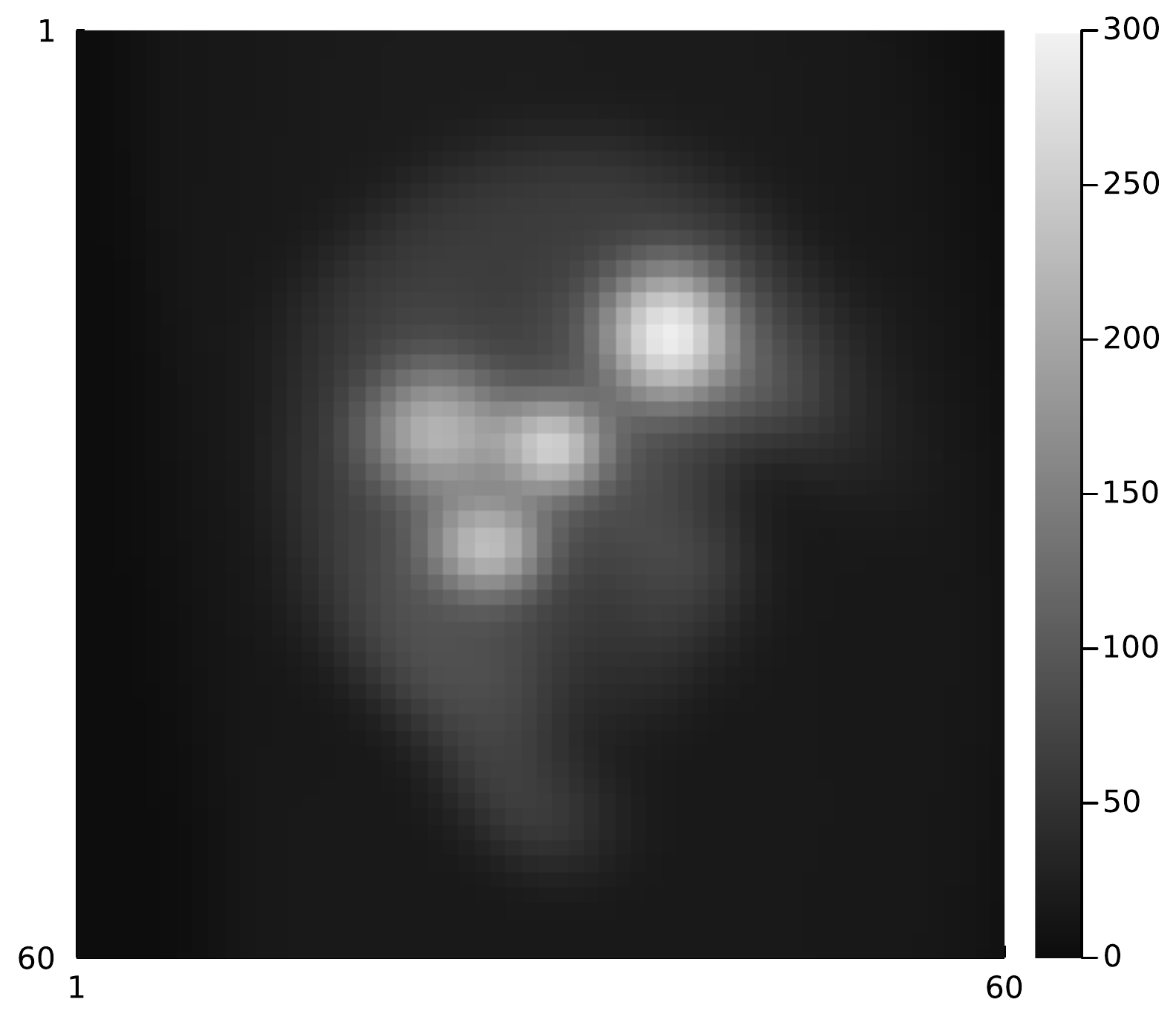}}
\\
\subfloat[MC (\ytt)]{\includegraphics[width=0.25\linewidth]{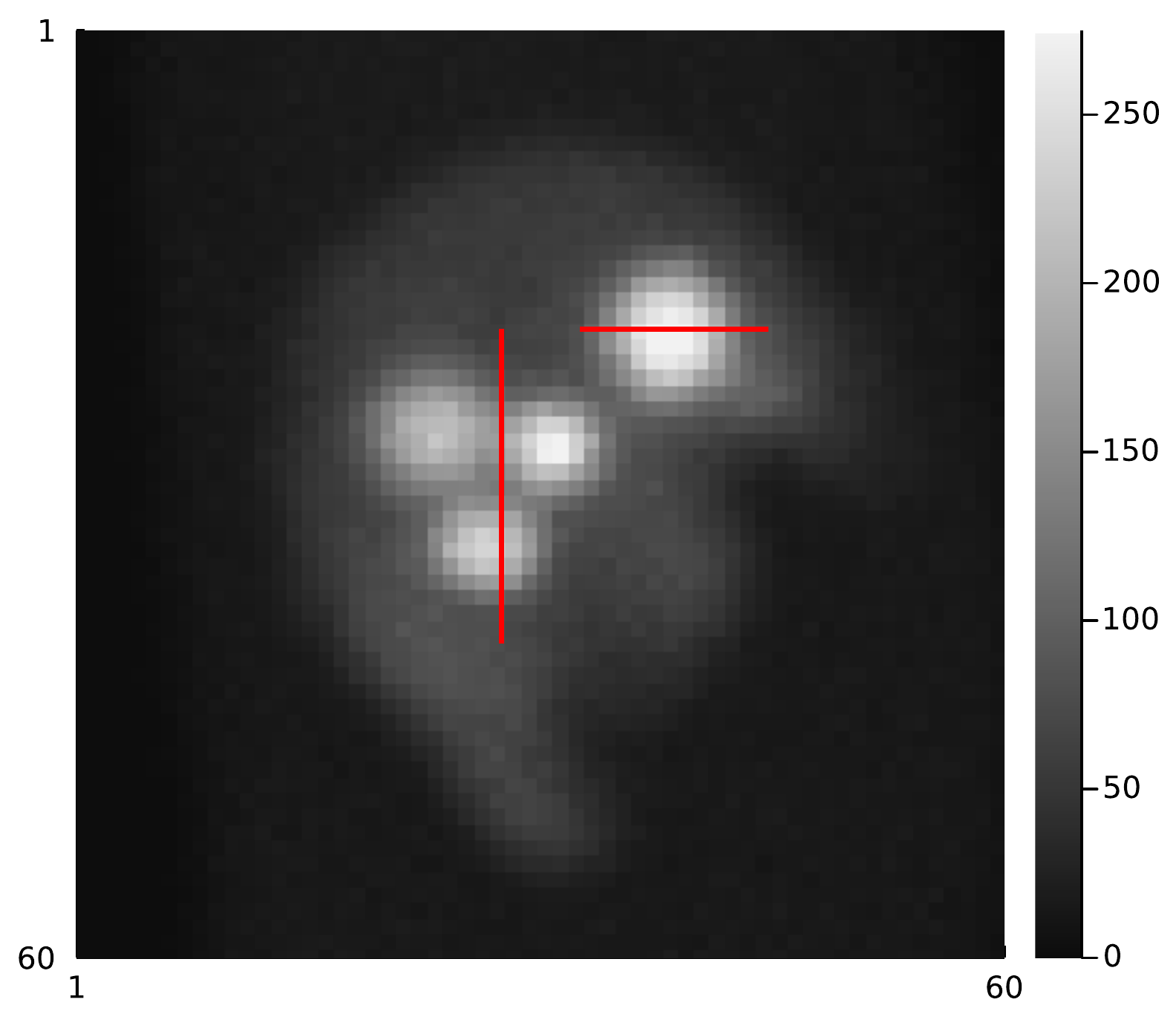}}
\subfloat[Julia 1D (\ytt)]{\includegraphics[width=0.25\linewidth]{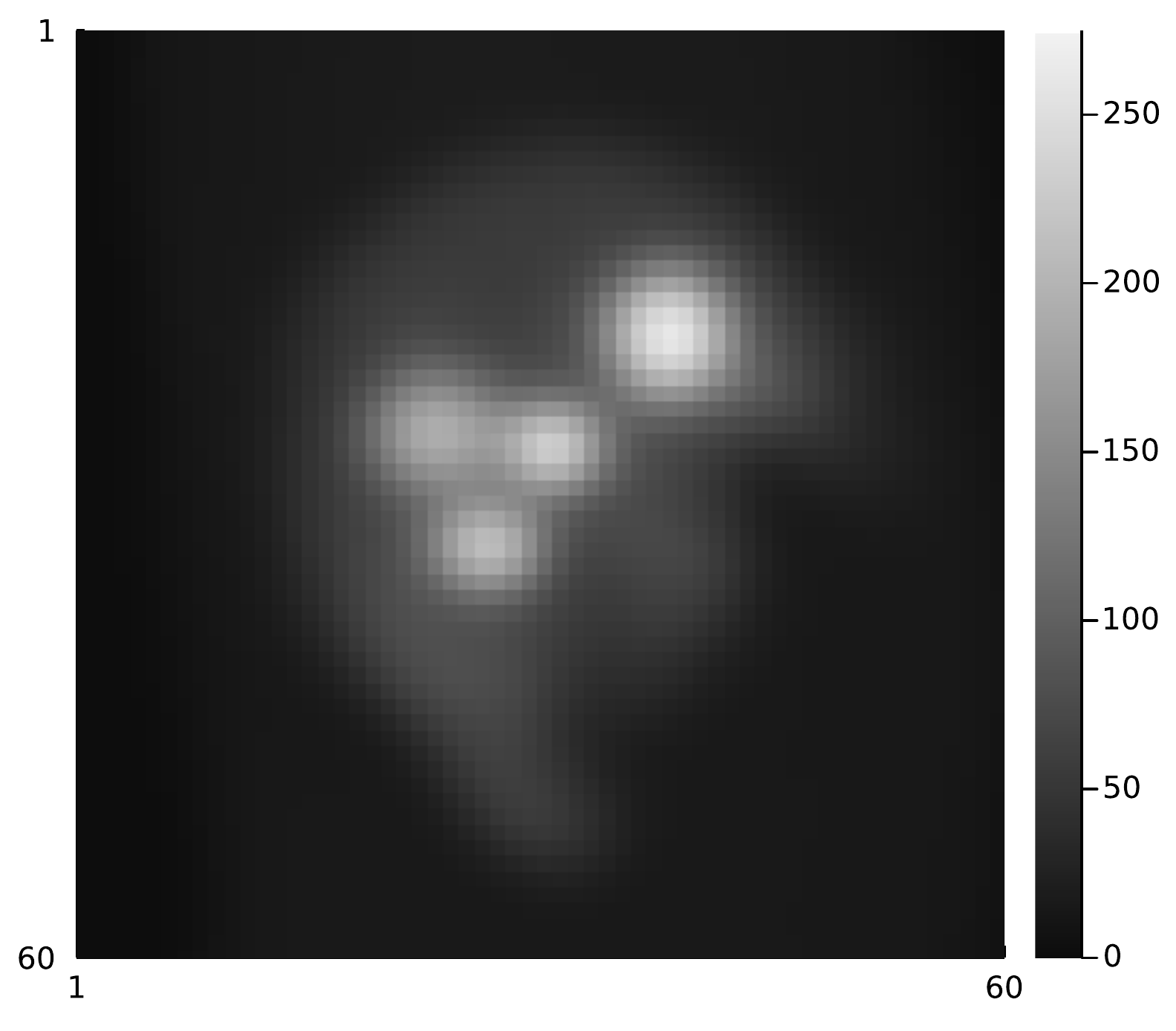}}
\subfloat[Julia 2D (\ytt)]{\includegraphics[width=0.25\linewidth]{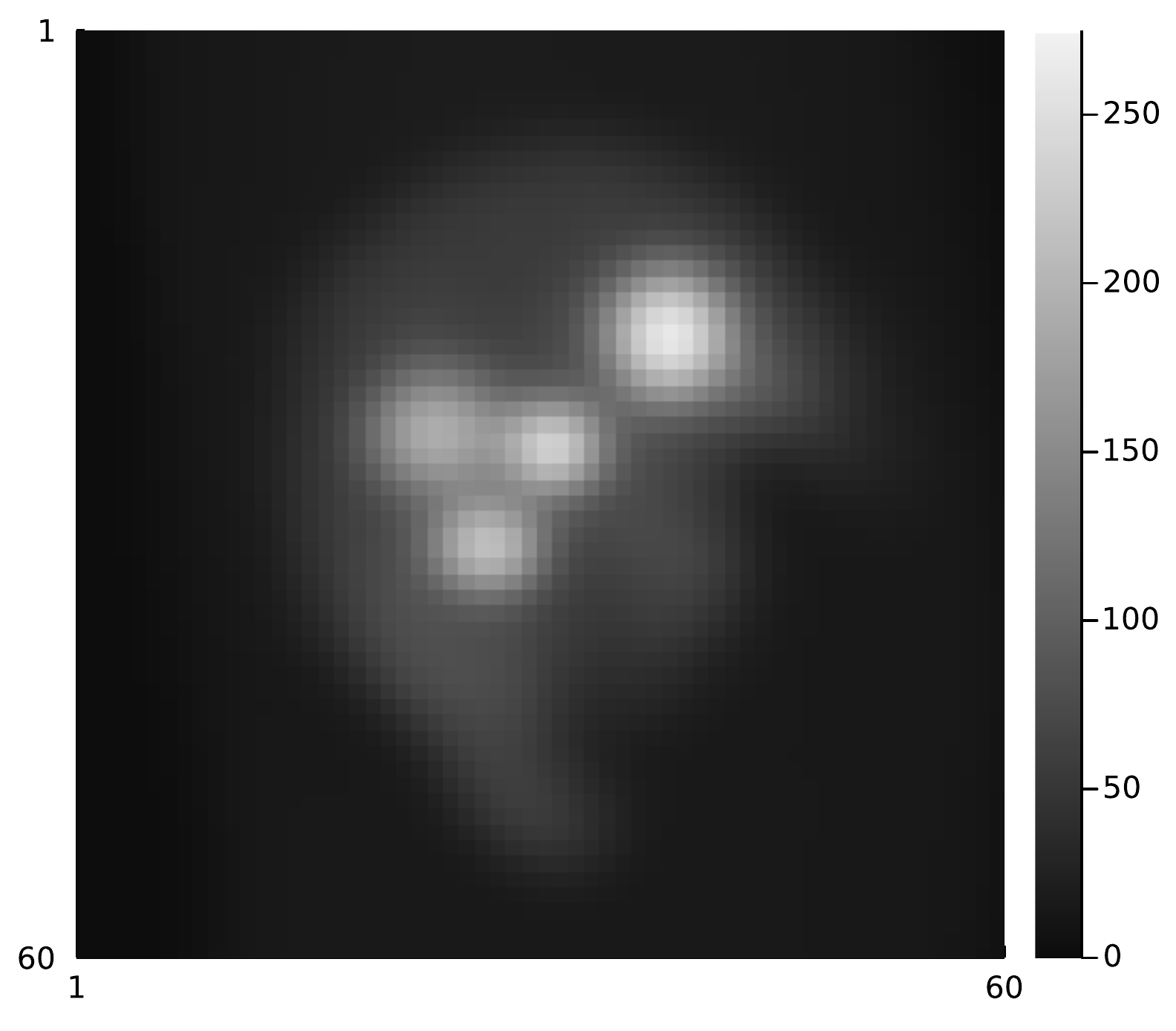}}
\subfloat[Matlab (\ytt)]{\includegraphics[width=0.25\linewidth]{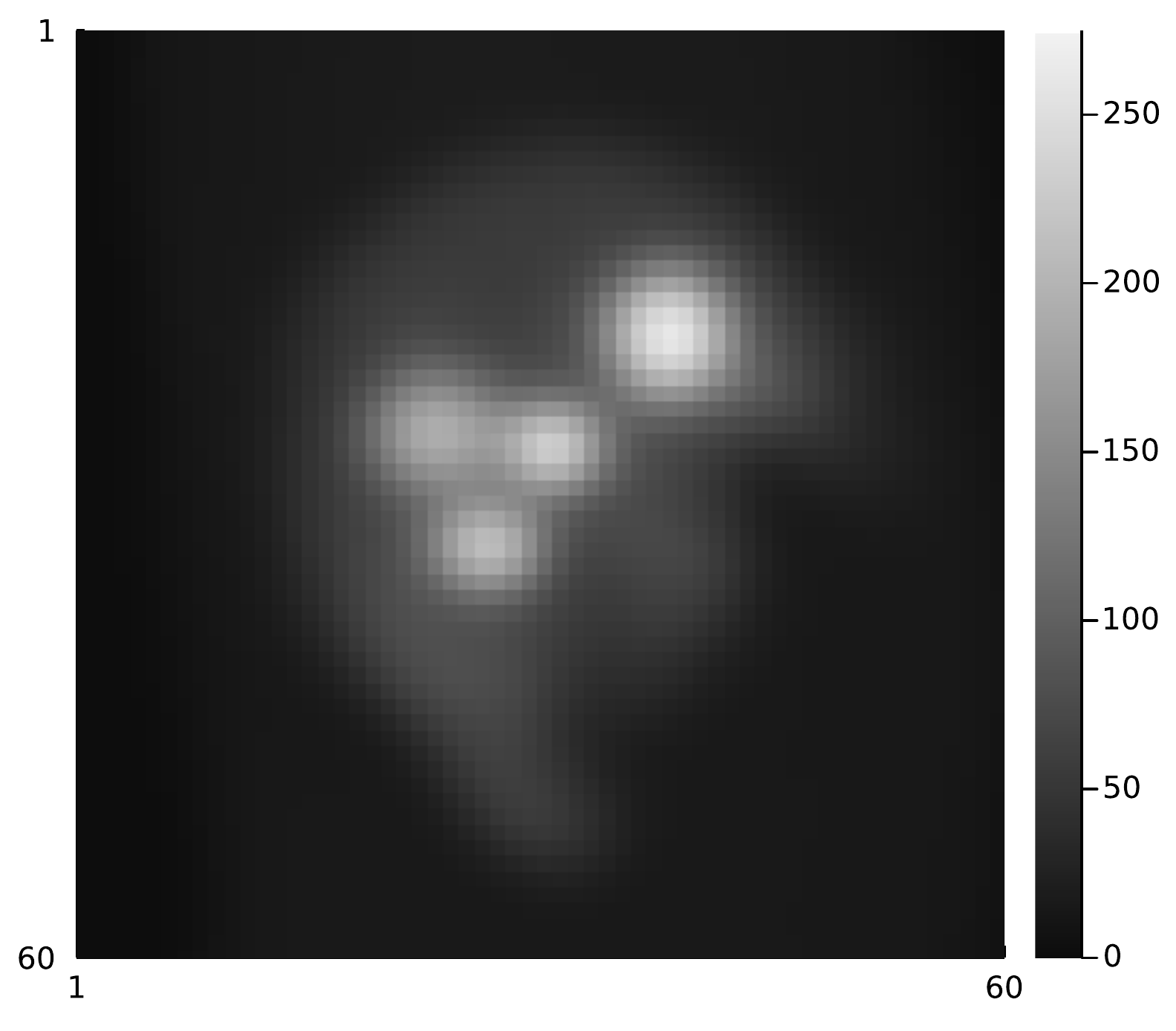}}
\\
\subfloat[Horizontal profile (\lu)]{\includegraphics[width=0.25\linewidth]{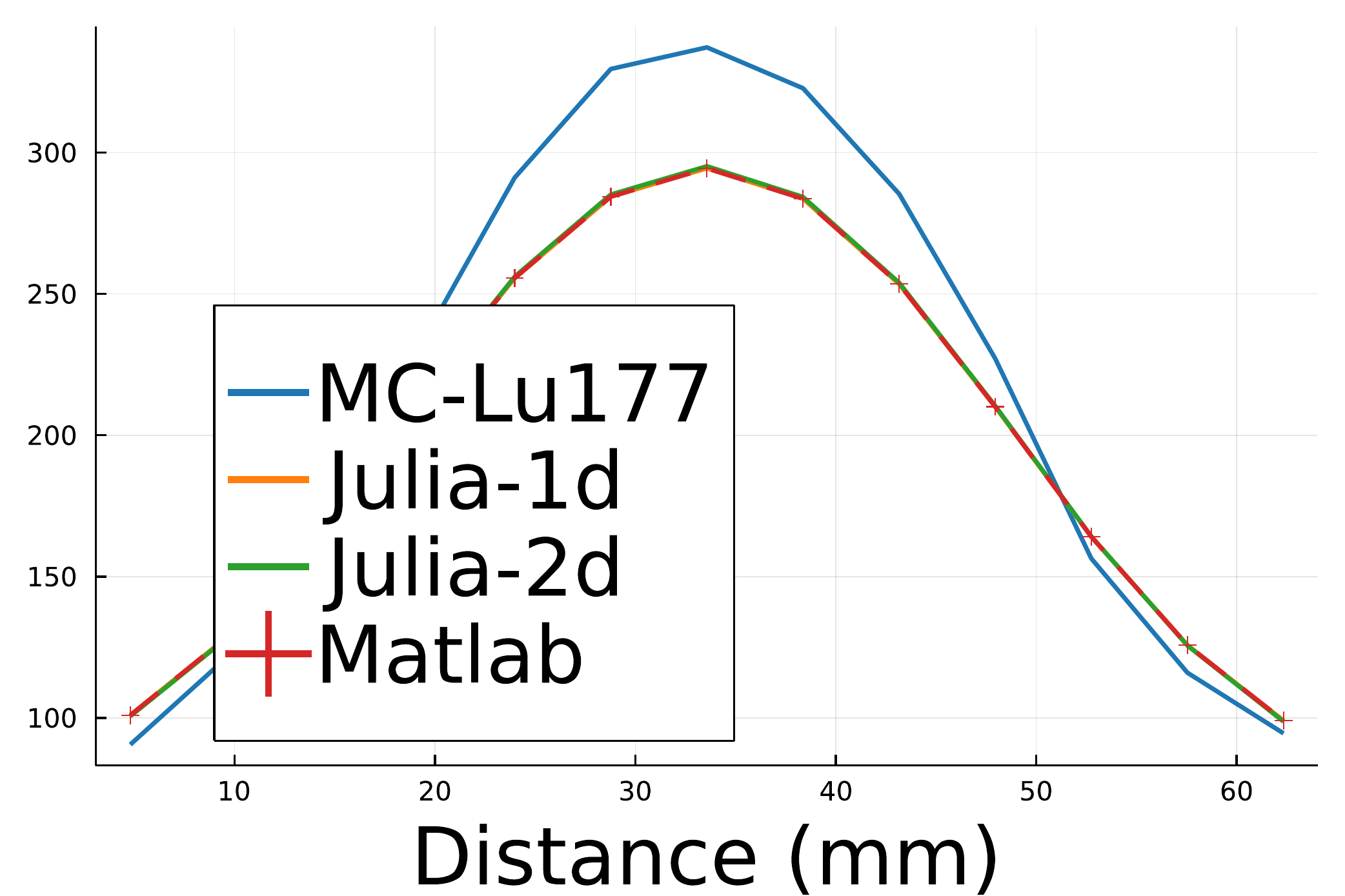}}
\subfloat[Vertical profile (\lu)]{\includegraphics[width=0.25\linewidth]{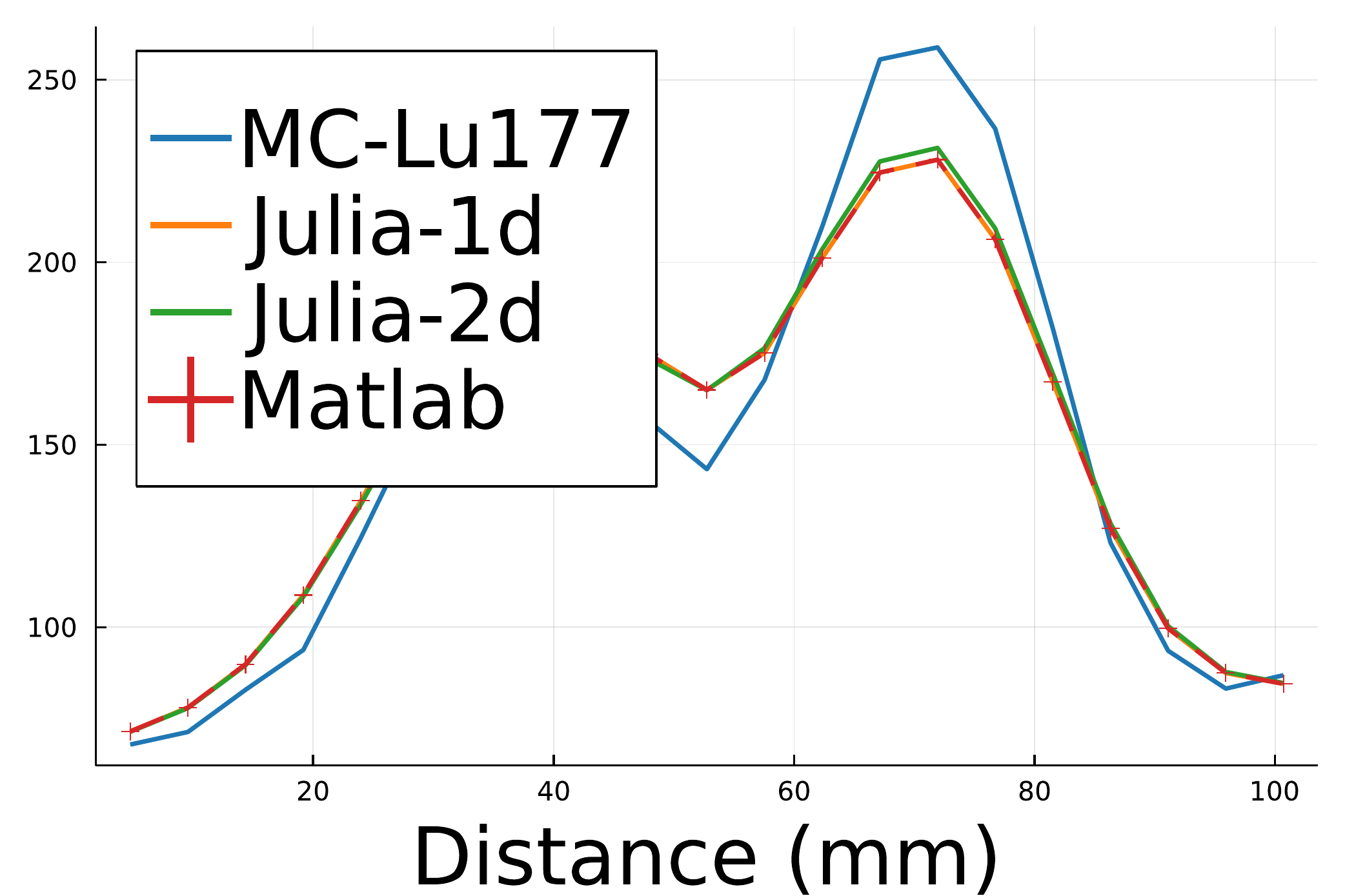}}
\subfloat[Horizontal profile (\ytt)]{\includegraphics[width=0.25\linewidth]{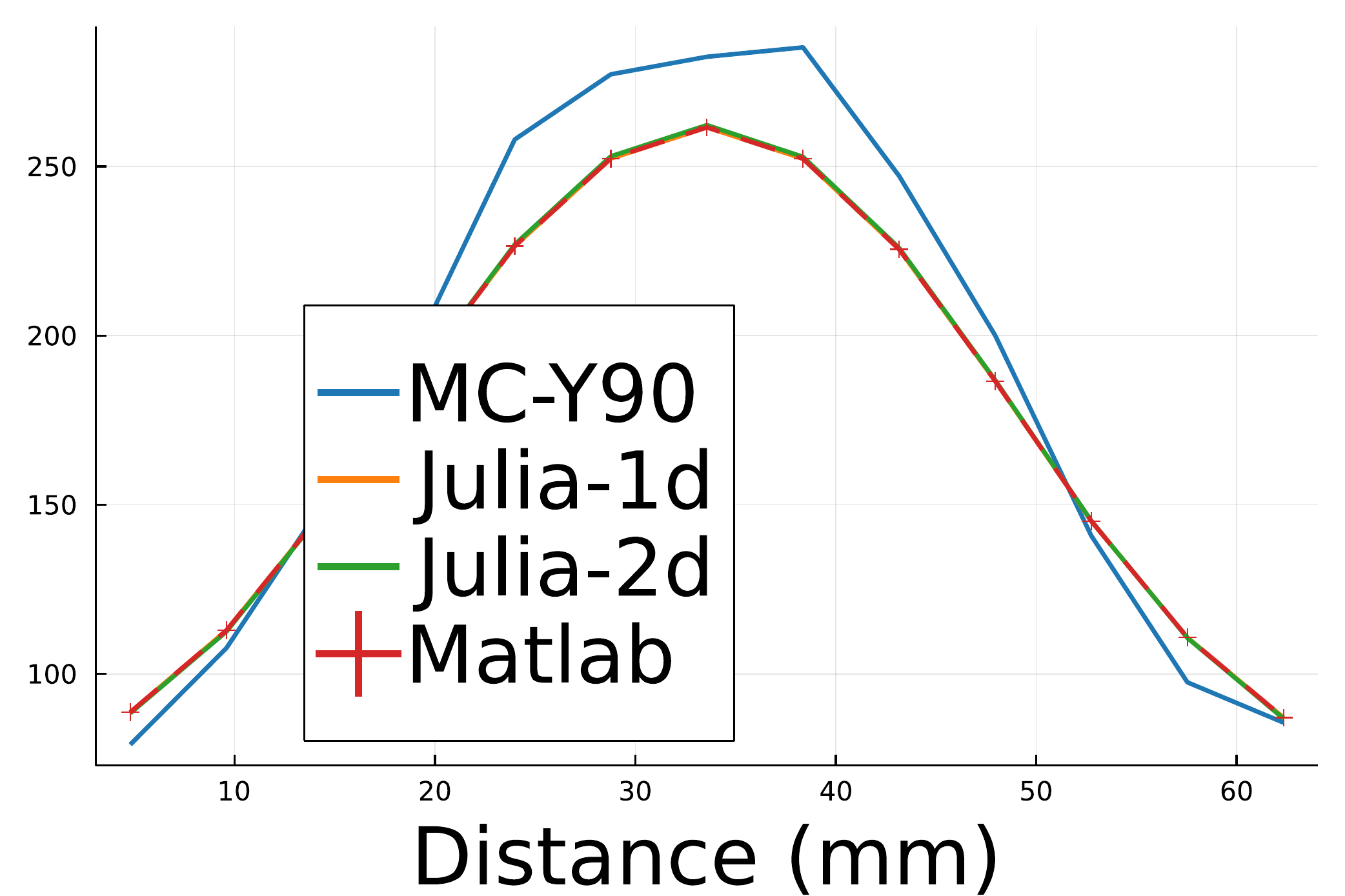}}
\subfloat[Vertical profile (\ytt)]{\includegraphics[width=0.25\linewidth]{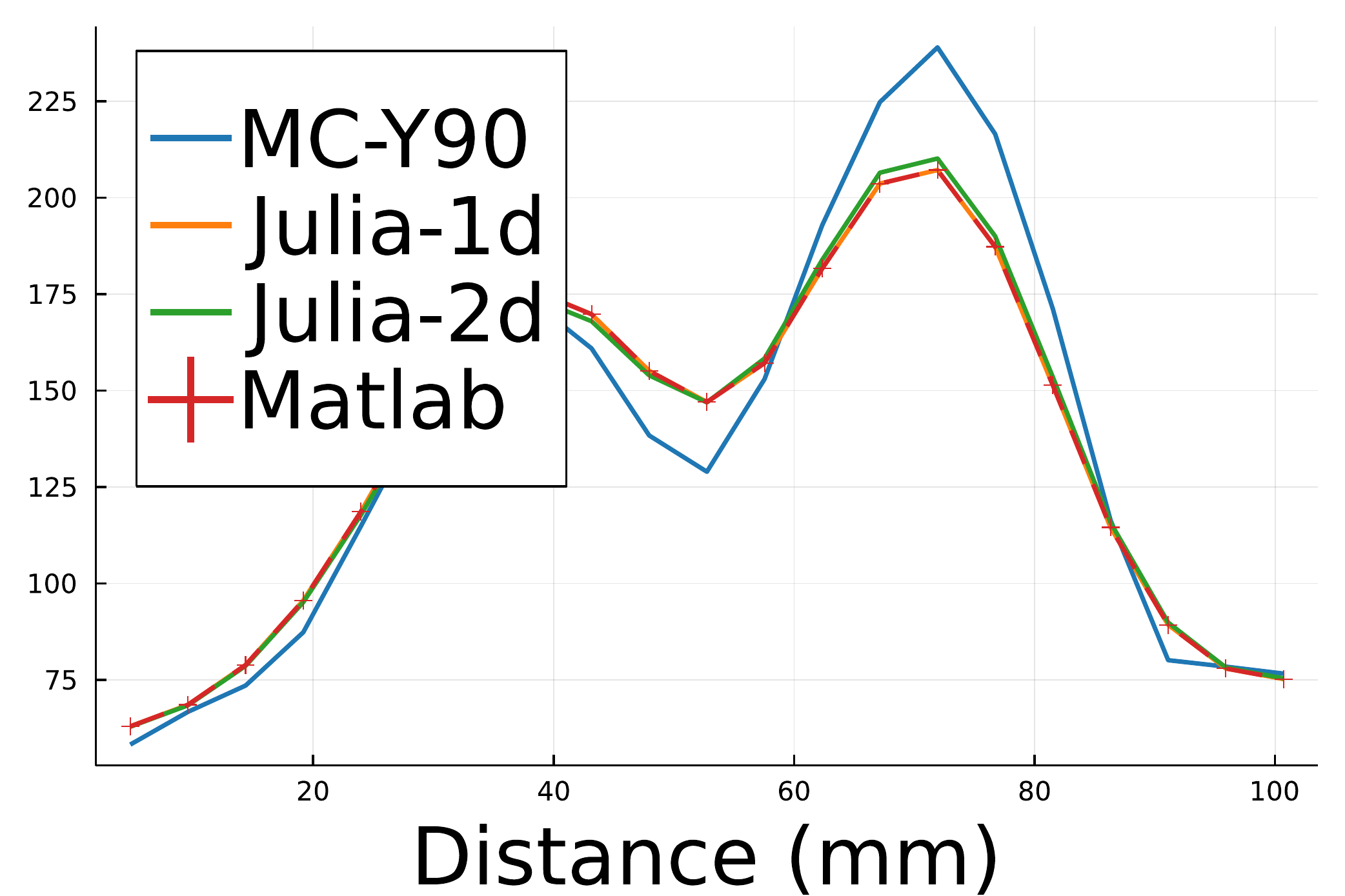}}

    \caption{Primary (scatter-free) 
    projections generated by MC simulation, 
    Matlab projector and our Julia projector 
    with 3-pass 1D linear interpolation and 
    2D bilinear interpolation for image rotation, 
    using $^{177}$Lu and $^{90}$Y radionuclides.
    Subfigure (i)-(l) show line profiles 
    across tumors as shown in subfigure (a) and (e), respectively. 
    MC projections were scaled to have 
    the same total activities as the Matlab projector per field-of-view.}
    \label{fig:proj}
\end{figure*}

\subsection{Phantom Dataset and Simulation Setup}

We used simulated XCAT phantoms \cite{segars:10:4xp} 
and virtual patient phantoms 
for experiment results presented in Section~\ref{sec:results}.
Each XCAT phantom was simulated 
to approximately follow the activity distributions
observed when imaging patients 
after \lu DOTATATE therapy.
We set the image size to $128\times128\times80$
with voxel size $4.8\times4.8\times4.8\mathrm{mm}^3$.
Tumors of various shapes and sizes (5-100mL)
were located in the liver as is typical for 
patients undergoing this therapy.

For virtual patient phantoms, 
we consider two radionuclides: \lu and \ytt.
For \lu phantoms,
the true images were from PET/CT scans of 
patients who underwent diagnostic 
$^{68}$Ga DOTATATE PET/CT imaging 
(Siemens Biograph mCT) 
to determine eligibility 
for \lu DOTATATE therapy.
The $^{68}$Ga DOTATATE distribution in patients
is expected to be similar to \lu 
and hence can provide a reasonable approximation
to the activity distribution of \lu in patients
for DL training purposes but at higher resolution.
The PET images 
had size $200 \times 200  \times 577$
and voxel size
$4.073 \times 4.073 \times 2$ mm$^3$
and were obtained from our Siemens mCT 
(resolution is 5–6 mm FWHM \cite{soderlund:15:b1c}) 
and reconstructed using the standard clinic protocol:
3D OSEM with three iterations, 21 subsets,
including resolution recovery, time-of-flight,
and a 5mm (FWHM) Gaussian post-reconstruction filter. 
The density maps were also generated 
using the experimentally derived 
CT-to-density calibration curve.

For \ytt phantoms, 
the true activity images were reconstructed 
(using a previously implemented 
3D OSEM reconstruction with CNN-based scatter estimation \cite{xiang:20:adn})
from $^{90}$Y SPECT/CT scans of patients
who underwent $^{90}$Y 
microsphere radioembolization
in our clinic.

In total, we simulated 4 XCAT phantoms,
8 \lu and 8 \ytt virtual patient phantoms.
We repeated all of our experiments 3 times 
with different noise realizations.
All image data have 
University of Michigan Institutional Review Board (IRB) approval for retrospective analysis.
For all simulated phantoms,
we selected the center slices 
covering the lung, liver and kidney
corresponding to SPECT axial FOV (39cm).

Then we ran SIMIND Monte Carlo (MC) program
\cite{ljungberg:12:tsm}
to generate the radial position of SPECT camera
for 128 view angles.
The SIMIND model parameters for \lu were based on 
\lu DOTATATE patient imaging in our clinic 
(Siemens Intevo with medium energy collimators, 
a 5/8" crystal, 
a 20\% photopeak window at 208 keV, 
and two adjacent 10\% scatter windows) \cite{dewaraja:22:apf}. 
For \ytt,
a high-energy collimator, 
5/8" crystal, 
and a 105 to 195 keV acquisition energy window 
was modeled as in our clinical protocol for \ytt bremsstrahlung imaging.
Next we approximated the point spread function
for \lu and \ytt by simulating point source at 
6 different distances (20, 50, 100, 150, 200, 250mm)
and then 
fitting a 2D Gaussian distribution
at each distance.
The camera orbit was assumed to be non-circular 
(auto-contouring mode in clinical systems) 
with the minimum distance between 
the phantom surface
and detector set at 1 cm.

\section{Experiment Results}
\label{sec:results}

\subsection{Comparison of projectors}

We used an XCAT phantom 
to evaluate the accuracy and memory-efficiency
of our Julia projector.
\comment{For that XCAT phantom, 
we set
the activity ratio 
between lesion, liver and kidney to be $10:1:1.5$
to make the activity distribution realistic.}

\comment{
\fref{fig:xtrue,atten,psf} 
shows the XCAT activity map,
attenuation map and 
point spread function at different depths.
\begin{figure}[hbt!]
    \centering
    \subfloat[Activity map]{\includegraphics[width=0.5\linewidth]{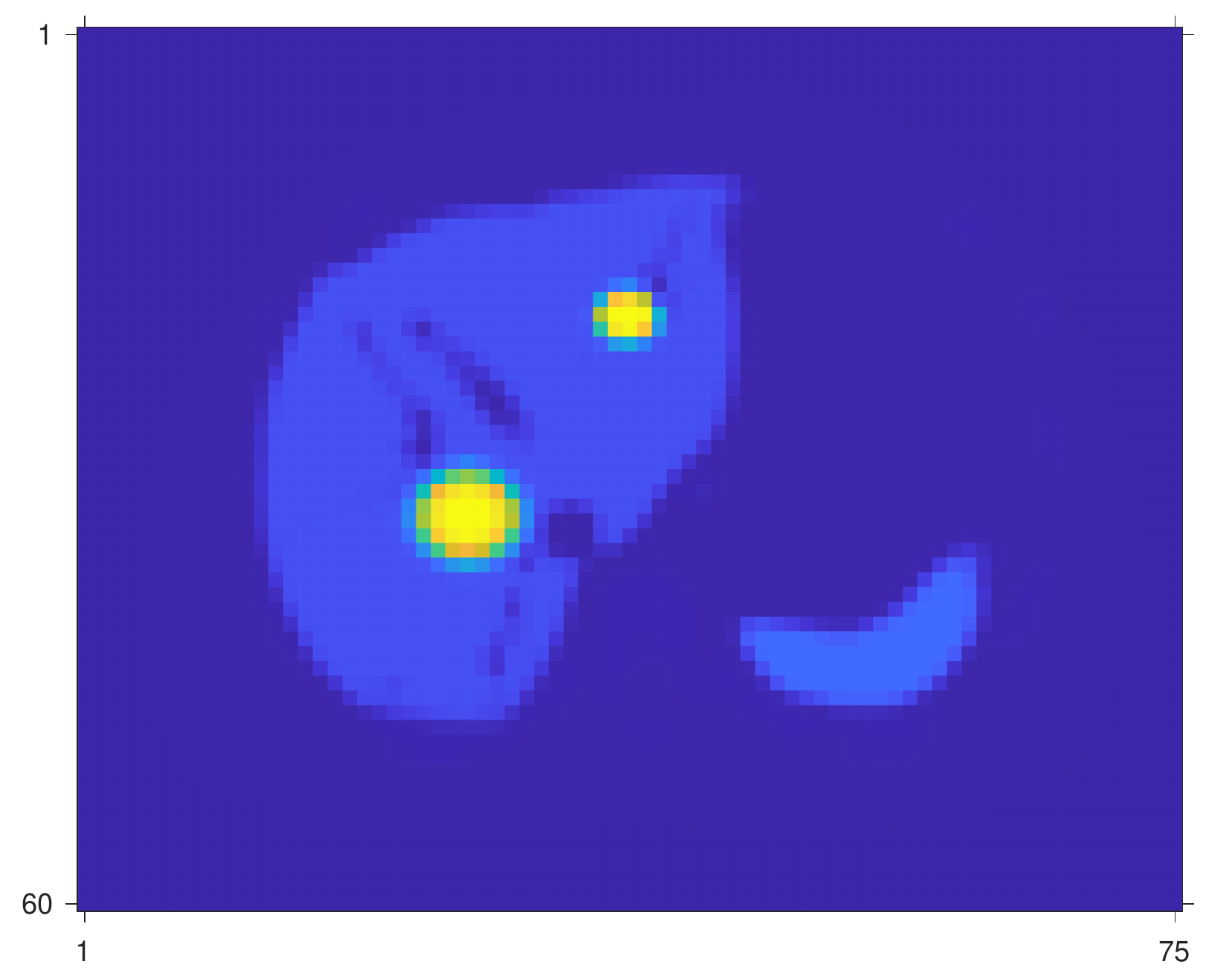}}
    \subfloat[Attenuation map]{\includegraphics[width=0.5\linewidth]{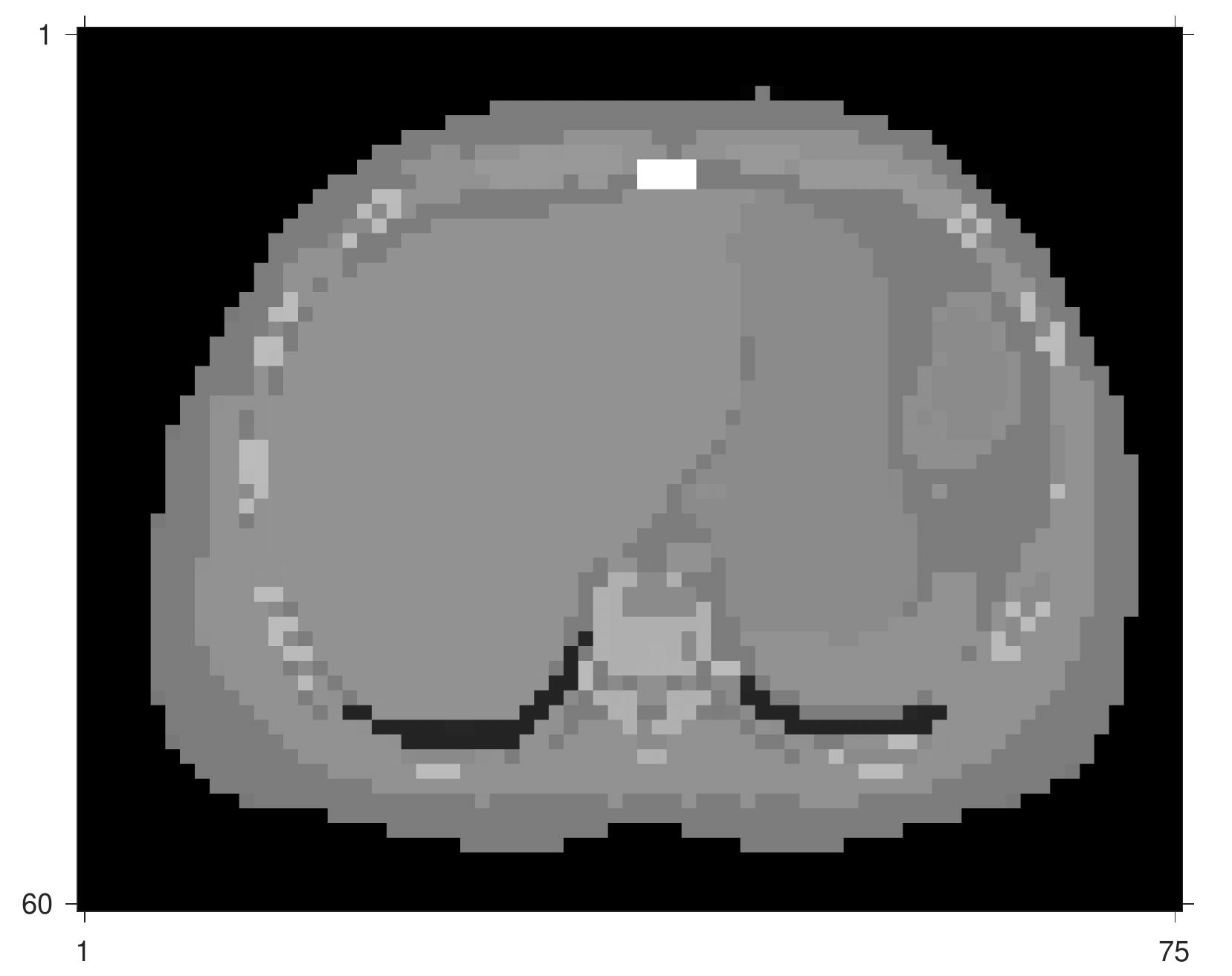}}
    \\
    \subfloat[12cm]{\includegraphics[width=0.25\linewidth]{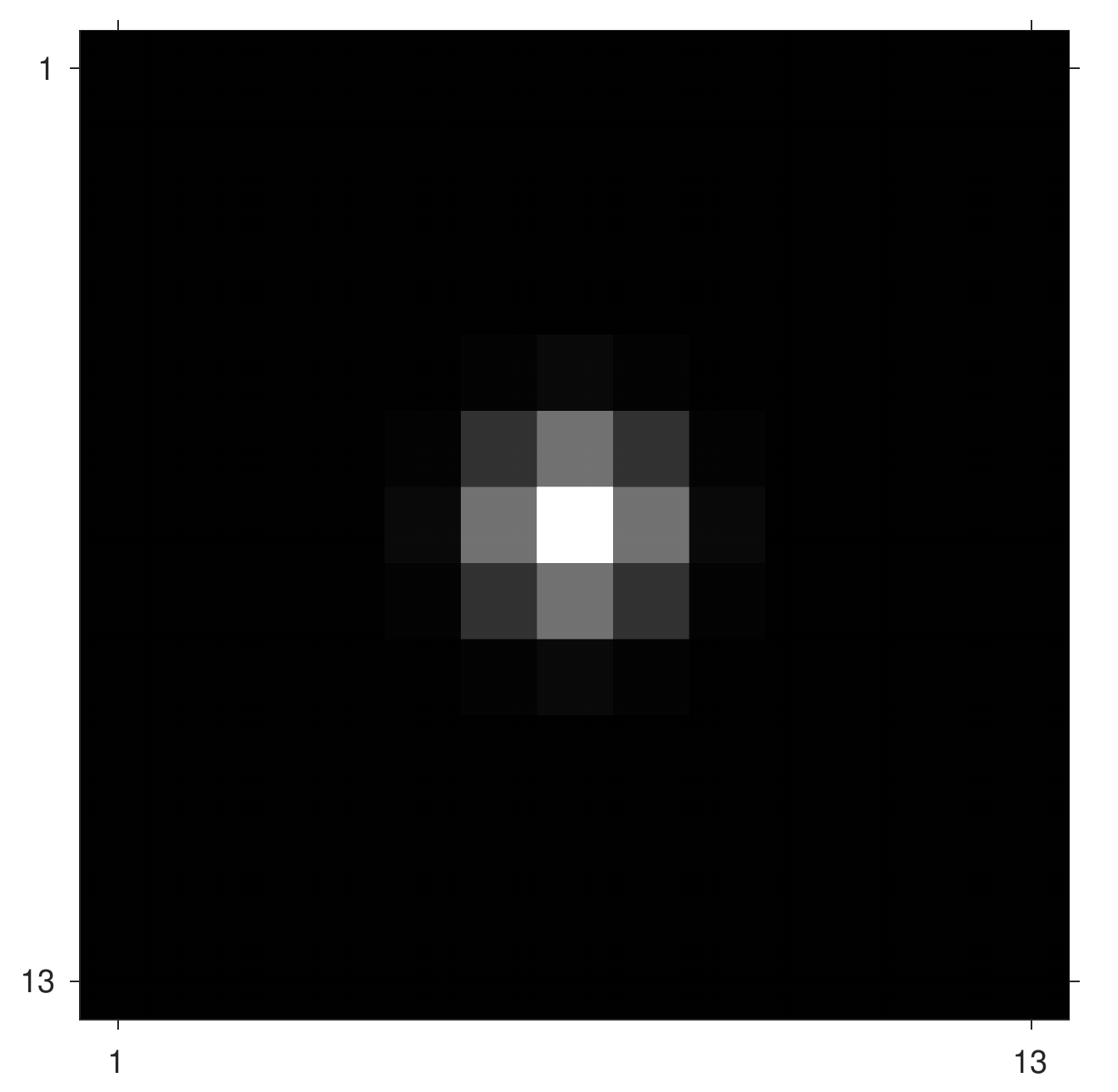}}
    \subfloat[19.2cm]{\includegraphics[width=0.25\linewidth]{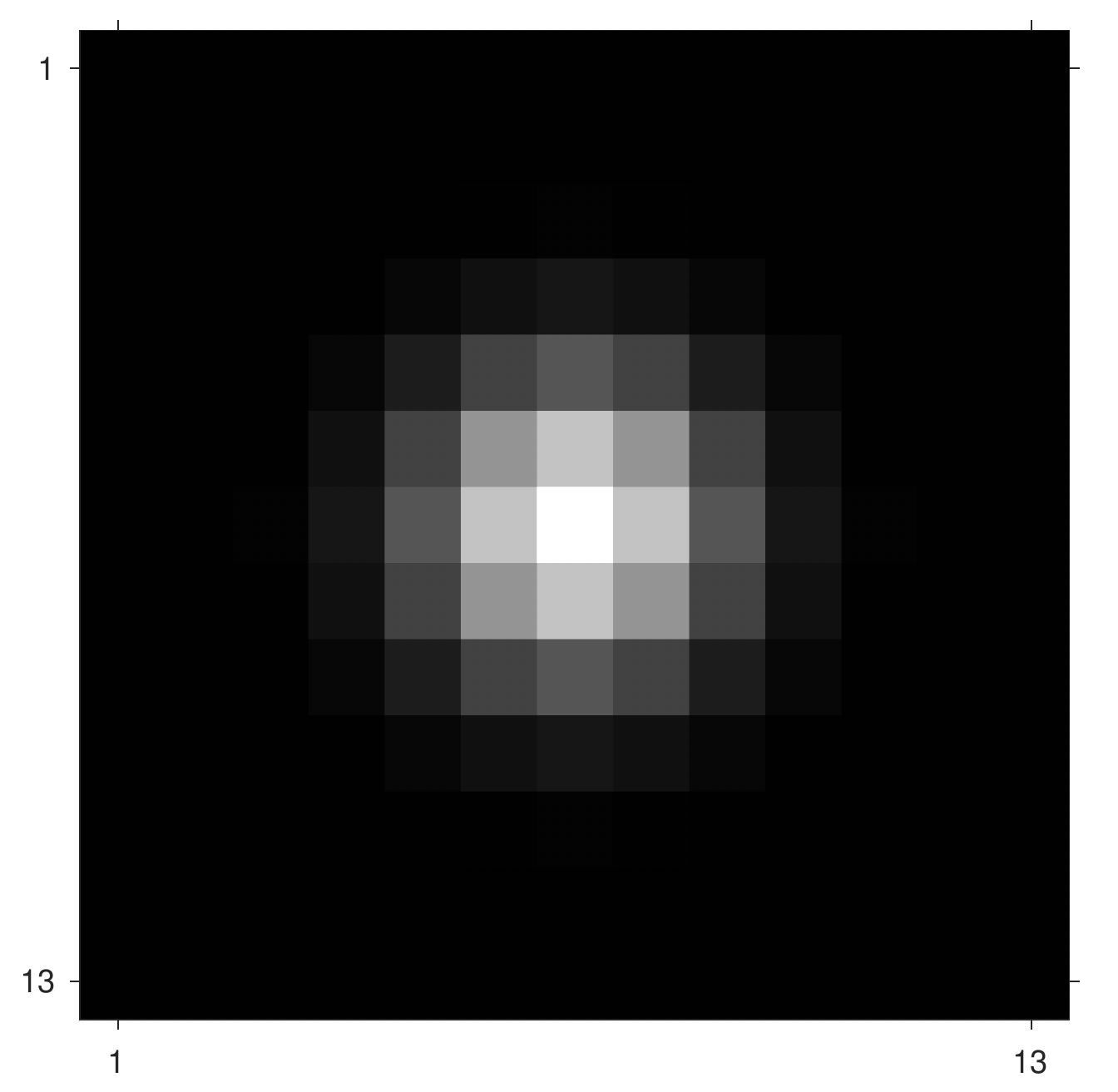}}
    \subfloat[28.8cm]{\includegraphics[width=0.25\linewidth]{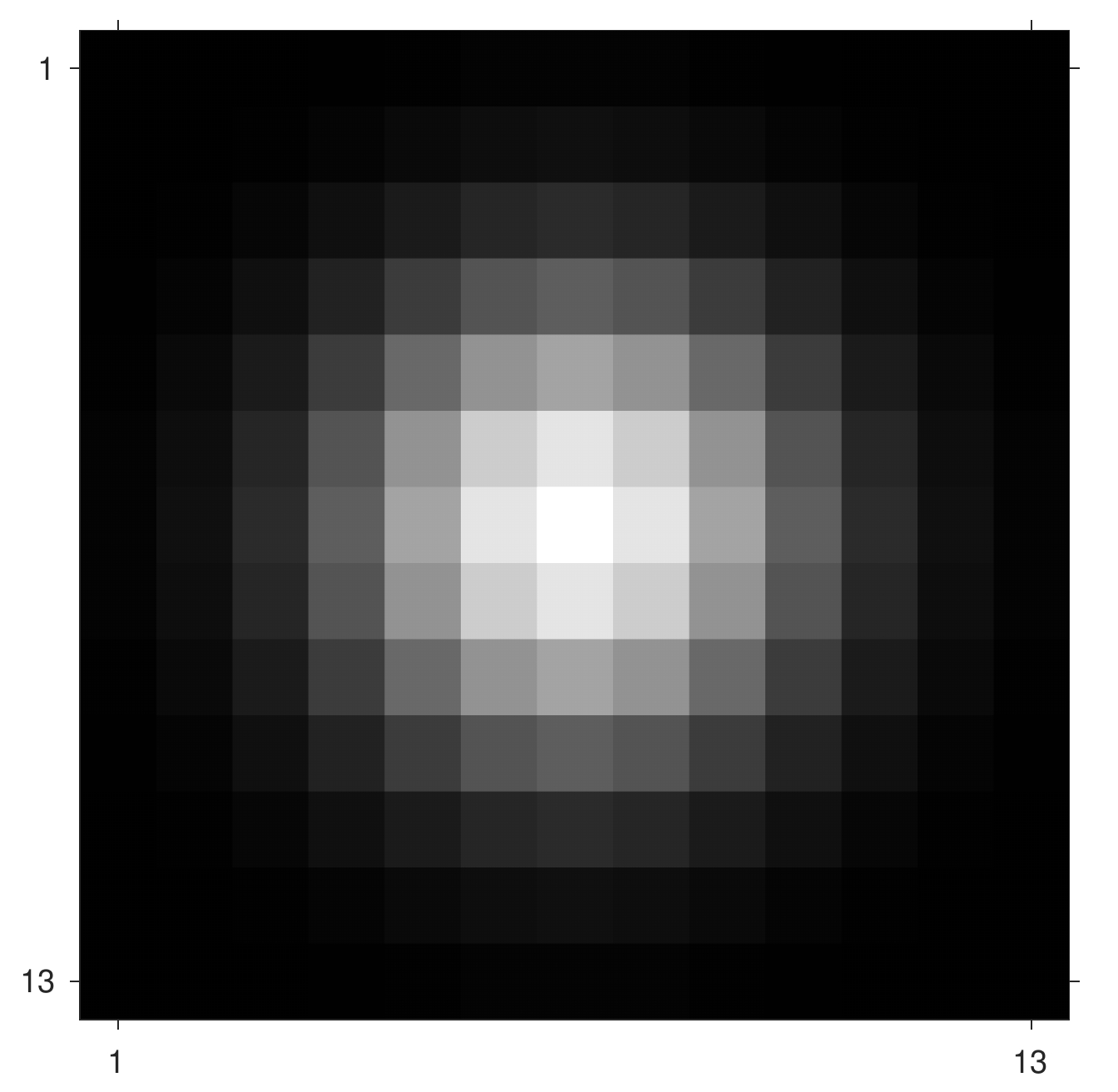}}
    \subfloat[38.4cm]{\includegraphics[width=0.25\linewidth]{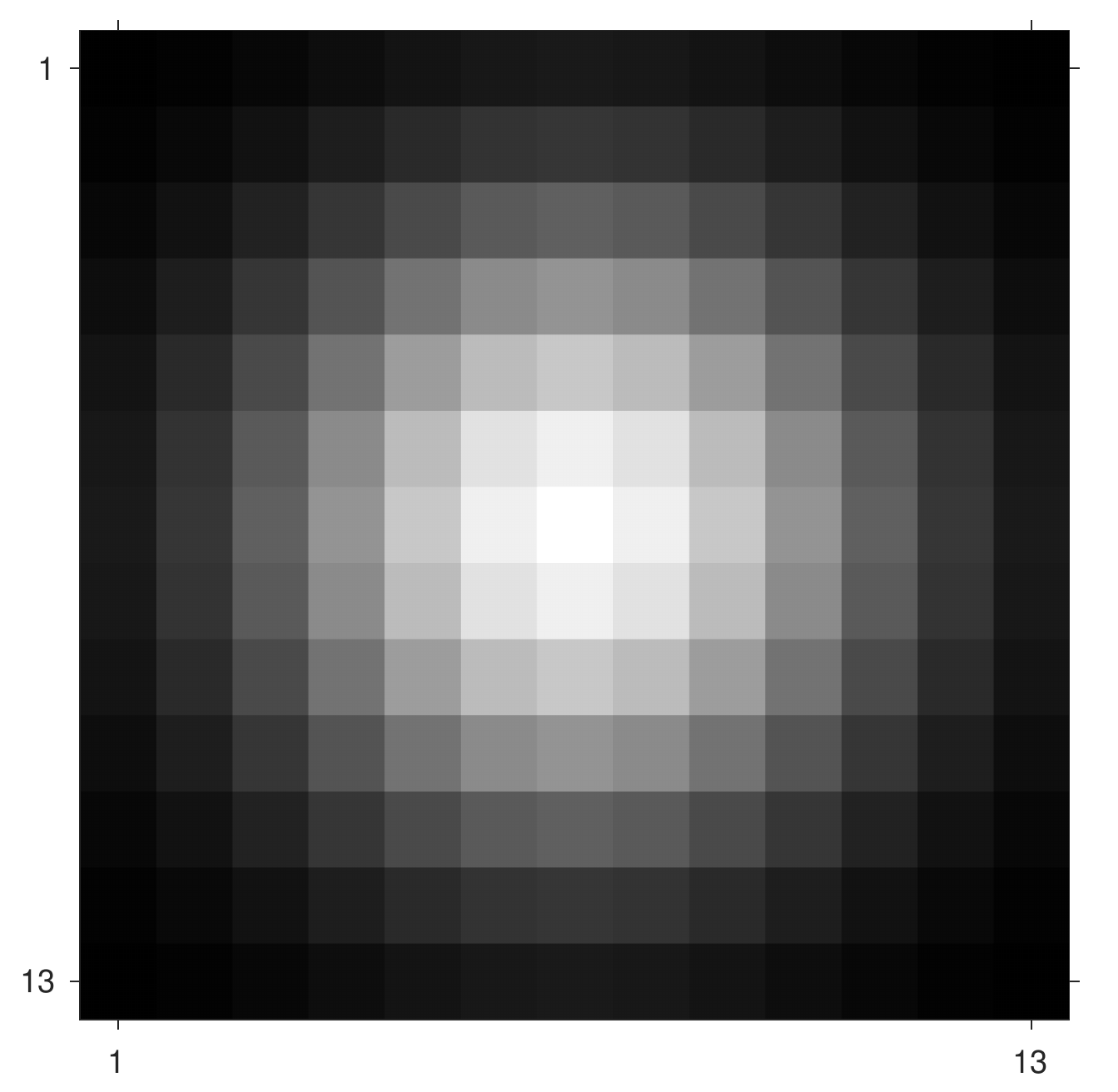}}
    \caption{Subfigure (a): A sample slice of simulated XCAT phantom's activity map; 
    (b): attenuation map;
    (c)-(f): point spread function at different depth locations.}
    \label{fig:xtrue,atten,psf}
\end{figure}
}

\subsubsection{Accuracy}

We first compared 
primary (no scatter events included)
projection images and profiles generated by 
our Julia projector with those from 
MC simulation and the Matlab projector.
For results of MC,
we ran two SIMIND simulations
for 1 billion histories
using \lu and \ytt as radionuclide source,
respectively.
Each simulation took about 
10 hours using 
a 3.2 GHz 16-Core Intel Xeon W
CPU on MacOS.
The Matlab projector
was originally implemented 
and compiled in C99
and then wrapped by a Matlab MEX file
as a part of the
Michigan Image Reconstruction Toolbox (MIRT)
\cite{fessler:21:mir}.
The physics modeling of the Matlab projector 
was the same as our Julia projector
except that it only implemented 3-pass 
1D linear interpolation for image rotation.
Unlike the memory-efficient Julia version,
the Matlab version pre-rotates the patient attenuation map
for all projection views.
This strategy saves time during EM iterations
for a single patient,
but uses considerable memory
and scales poorly
for DL training approaches
involving multiple patient datasets.

\begin{figure}[hbt!]
    \centering
    \subfloat[True activity]{\includegraphics[width=0.4\linewidth]{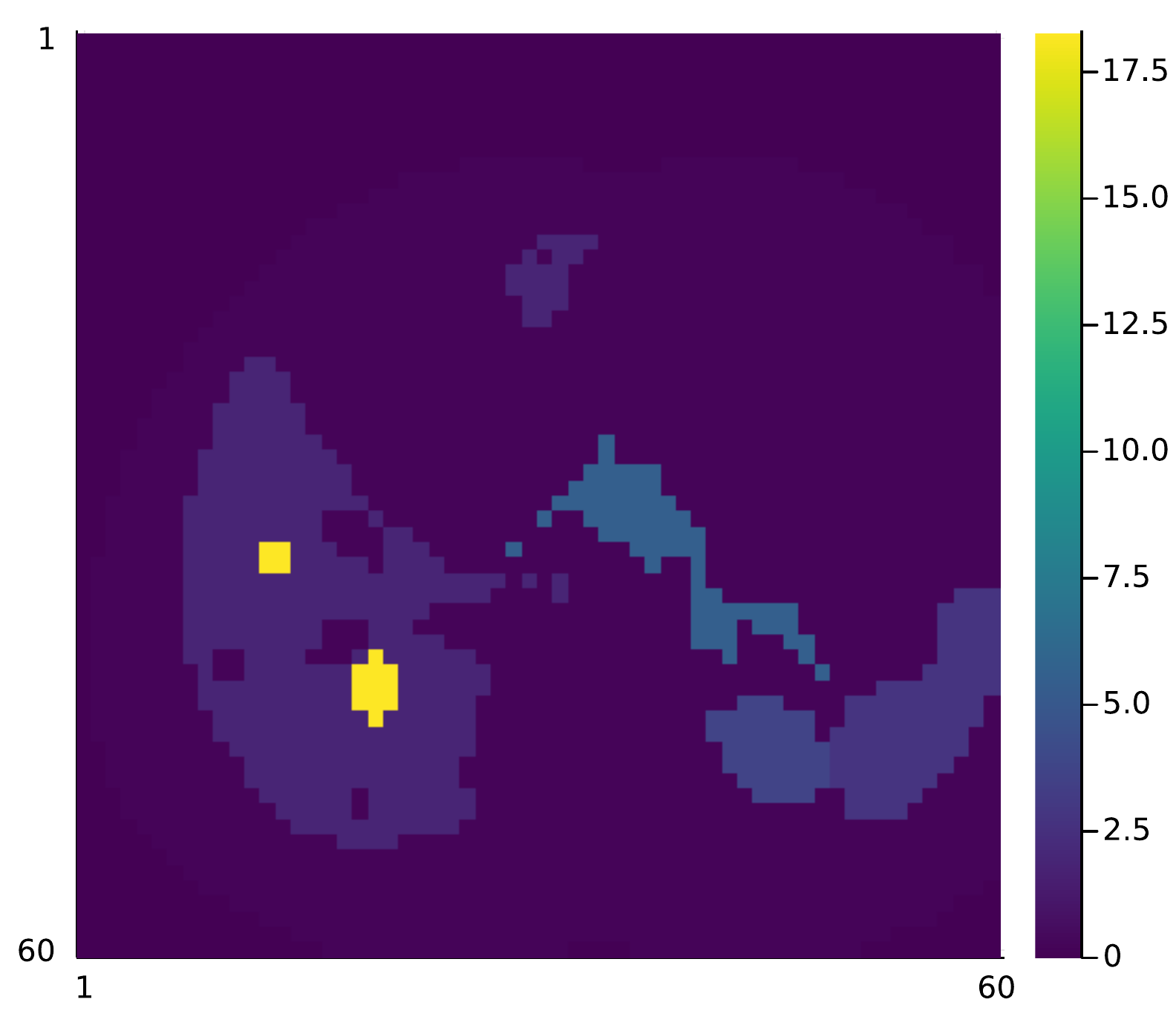}}
    \subfloat[OSEM-Matlab]{\includegraphics[width=0.4\linewidth]{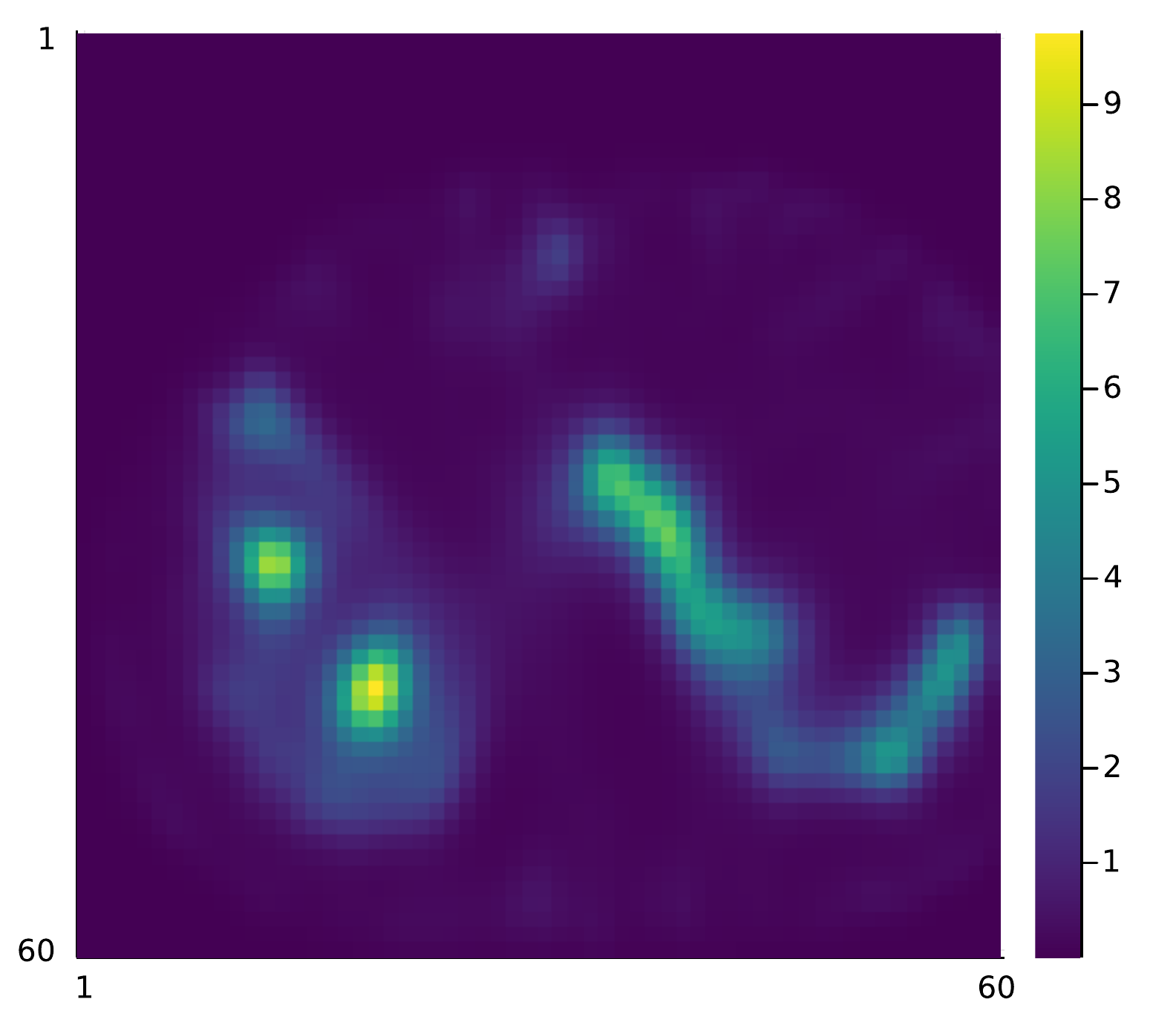}}
    \\
    \subfloat[OSEM-Julia 2D ]{\includegraphics[width=0.4\linewidth]{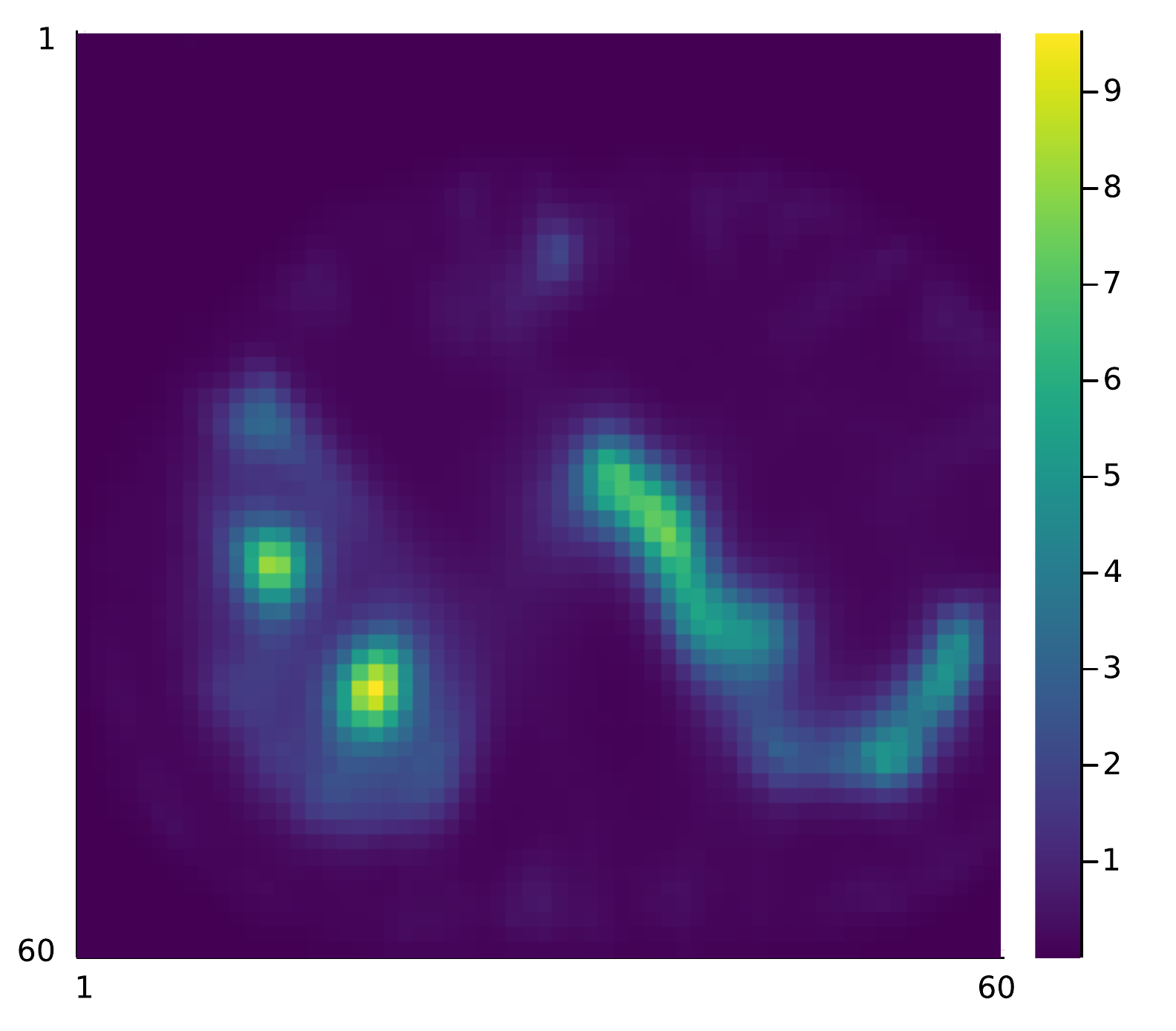}}
    \subfloat[Difference of (b) and (c)]{\includegraphics[width=0.4\linewidth]{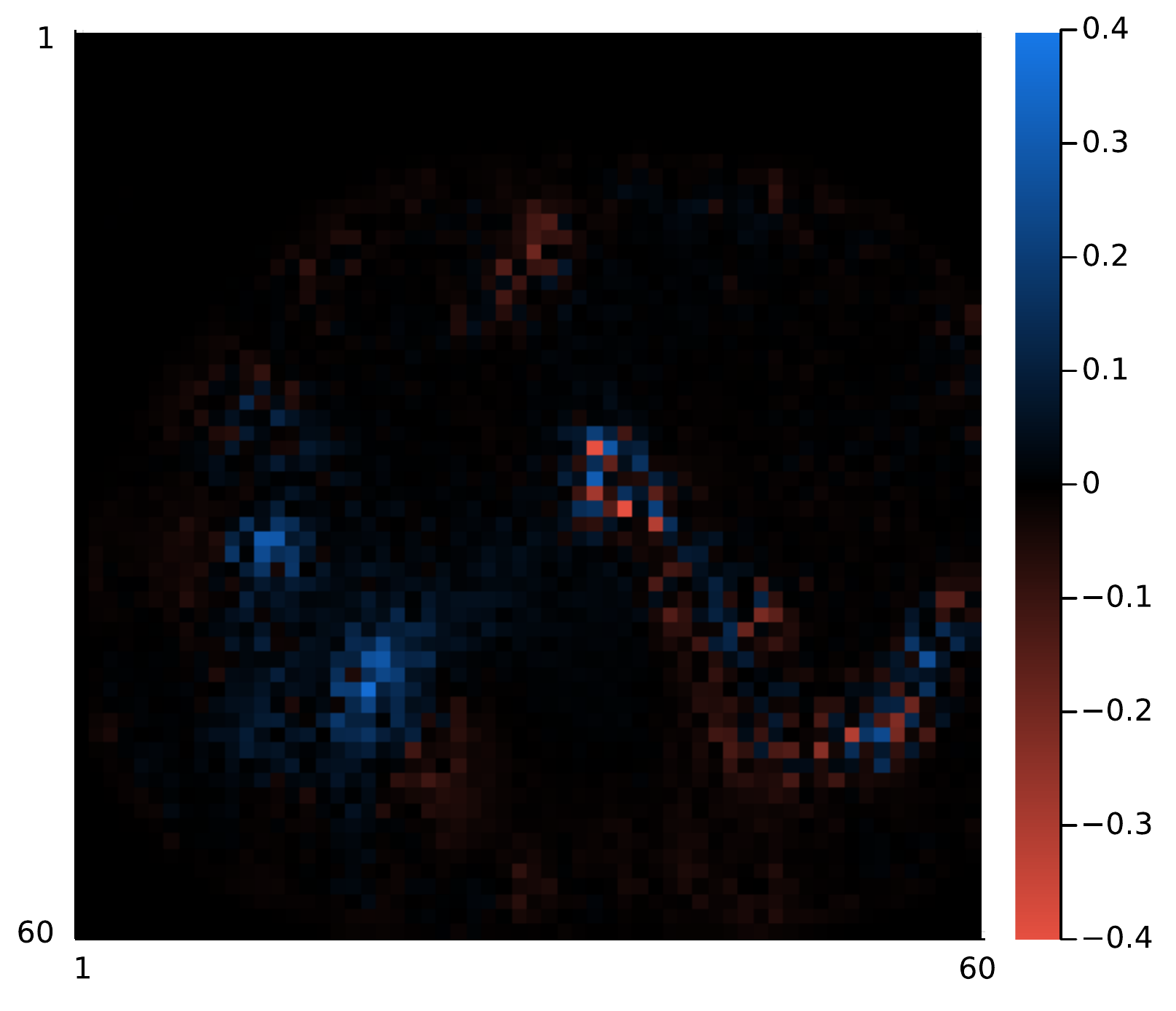}}
    \caption{Comparison of one slice of 
    the $128 \times 128 \times 80$
    OSEM reconstruction 
    (16 iterations, 4 subsets) 
    using Matlab and Julia (2D interpolation) projectors.}
    \label{fig:osem-recon-compare}
\end{figure}

\fref{fig:proj} compared the primary 
projections generated by different methods
without adding Poisson noise.
Visualizations of image slices 
and line profiles illustrate 
that our Julia projector
(with rotation based on 3-pass 1D interpolation)
is almost identical
to the Matlab projector,
while both give a reasonably good 
approximation to the MC.
Using MC as reference,
the NRMSE of Julia1D/Matlab/Julia2D projectors
were 7.9\%/7.9\%/7.6\% for \lu, respectively;
while the NRMSE were 
8.2\%/8.2\%/7.9\% for \ytt.
We also compared the OSEM reconstructed images
using Julia (2D) and Matlab projectors,
where we did not observe notable difference,
as shown in \fref{fig:osem-recon-compare}.
The overall NRMSD between Matlab and Julia (2D) projector
for the whole 3D OSEM reconstructed image
ranged from 2.5\% to 2.8\% across 3 noise realizations.  

\subsubsection{Speed and memory use}

We compared the memory use 
and compute times
between our Julia projector 
(with 2D bilinear interpolation)
and the Matlab projector
using different number of threads
when projecting a $128\times128\times80$ image.
\fref{fig:ap,vs,bfbp} shows that
our Julia projector has comparable computing time 
for a single projection with 128 view angles
using different number of CPU threads,
while using only a very small fraction of memory ($\sim$5\%)
and pre-allocation time ($\sim$1\%)
compared to the Matlab projector.

\begin{figure}[hbt!]
    \centering
    \includegraphics[width=\linewidth]{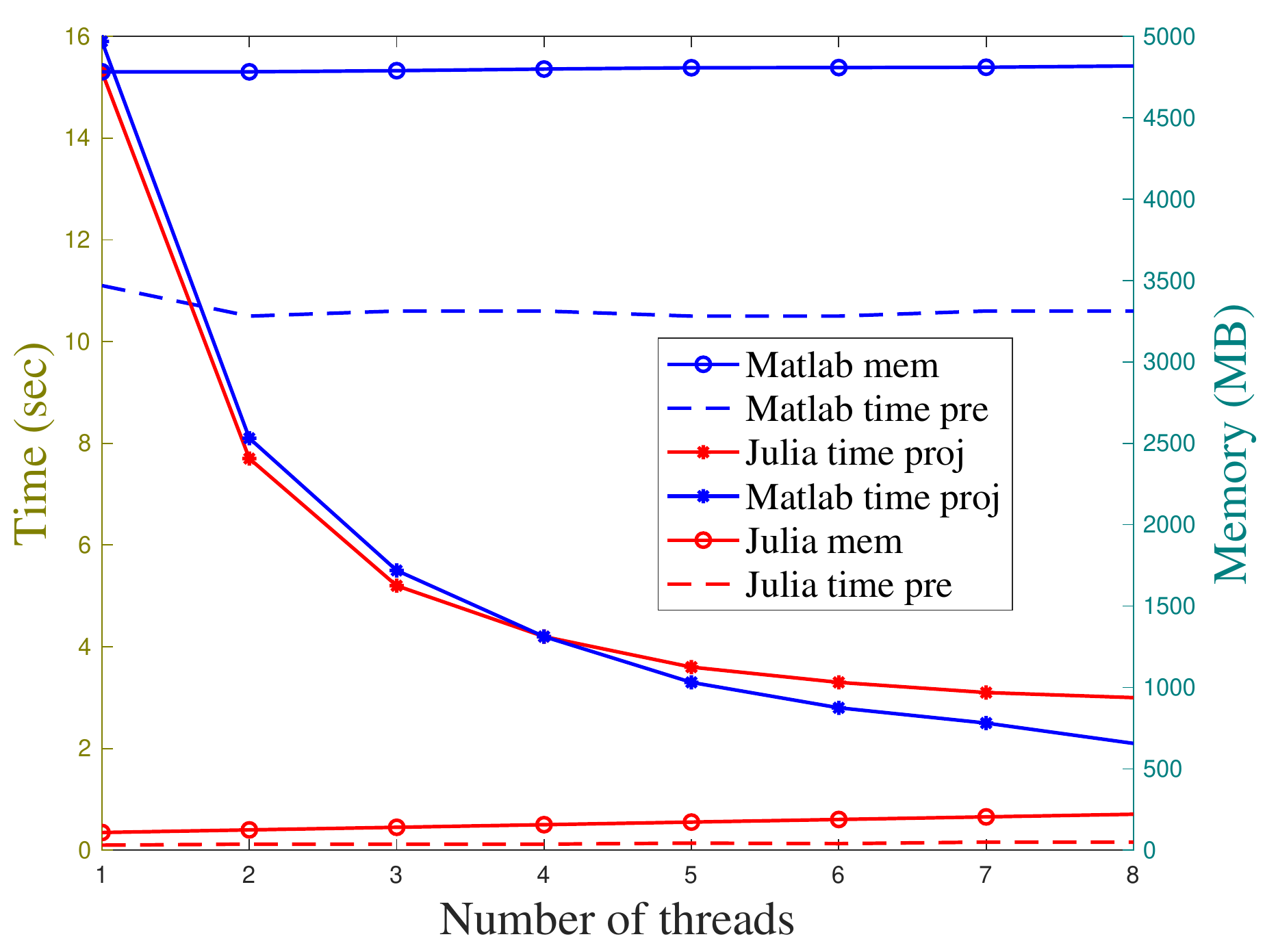}
    \caption{Time and memory comparison between Matlab projector and our Julia projector
    for projecting 128 view angles of a $128\times128\times80$ image.
    ``time pre" denotes the time cost for pre-allocating necessary arrays before projection;
    ``time proj" denotes the time cost
    for a single projection;
    ``mem" denotes the memory usage.
    All methods were tested on MacOS
    with a 3.8 GHz 8-Core Intel Core i7 CPU.}
    \label{fig:ap,vs,bfbp}
\end{figure}

\subsubsection{Adjoint of projector}

We generated a set of random numbers 
to verify that the backprojector 
is an exact adjoint of the forward projector.
Specifically, 
we generated the system matrix 
of size $(8\times 6 \times 7) \times (8\times 8 \times 6)$
using random (nonnegative) attenuation maps
and random (symmetric) PSF.
\fref{fig:adjoint,projector}
compares the transpose
of the forward projector
to the backprojector.
As shown in \fref{fig:adjoint,projector}(d),
the Frobenius norm error
of our backprojector 
agrees well with the regular transpose
within an accuracy of $10^{-6}$
across 100 different realizations,
as expected for 32-bit floating point calculations.
A more comprehensive comparison is 
available in the code tests
at \url{https://github.com/JuliaImageRecon/SPECTrecon.jl}.

\begin{figure}[hbt!]
    \centering
    \subfloat[Regular transpose]{\includegraphics[width=0.5\linewidth]{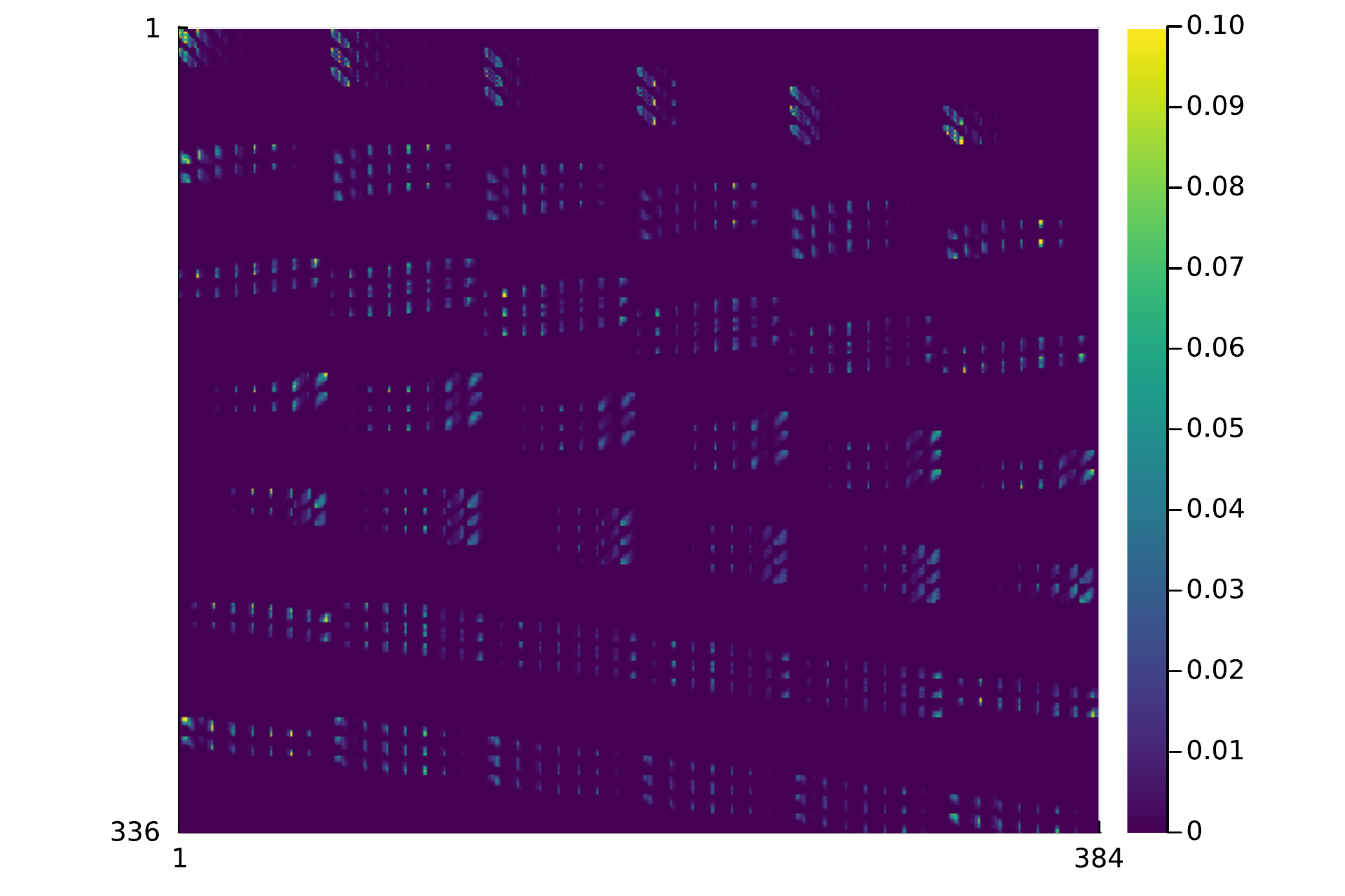}}
    \subfloat[Backprojector]{\includegraphics[width=0.5\linewidth]{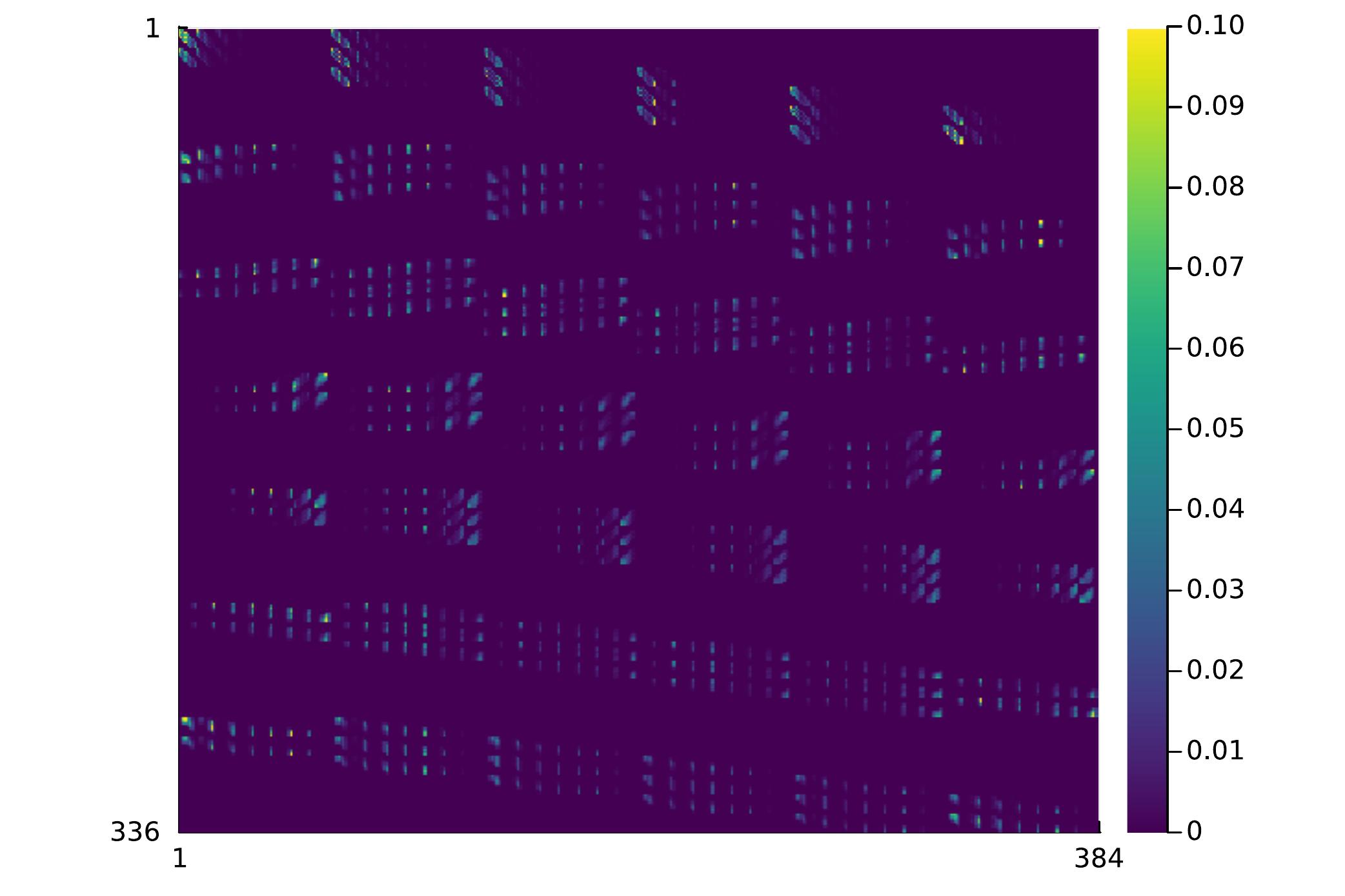}}
    \\
    \subfloat[Difference of (a) and (b) multiplied by $10^7$]{\includegraphics[width=0.45\linewidth]{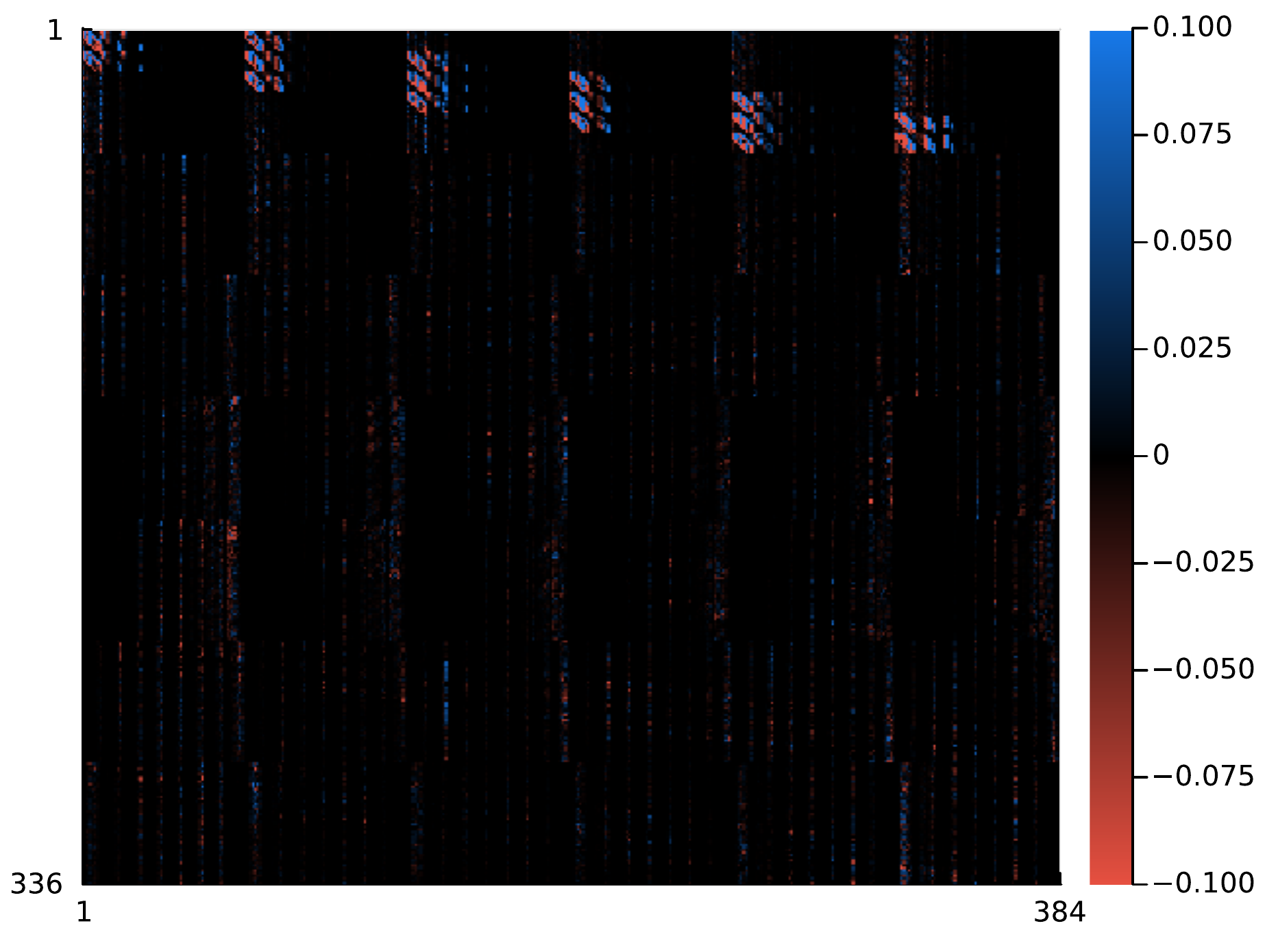}}
    \hspace{3mm}
    \subfloat[Frobenius norm error]{\includegraphics[width=0.45\linewidth]{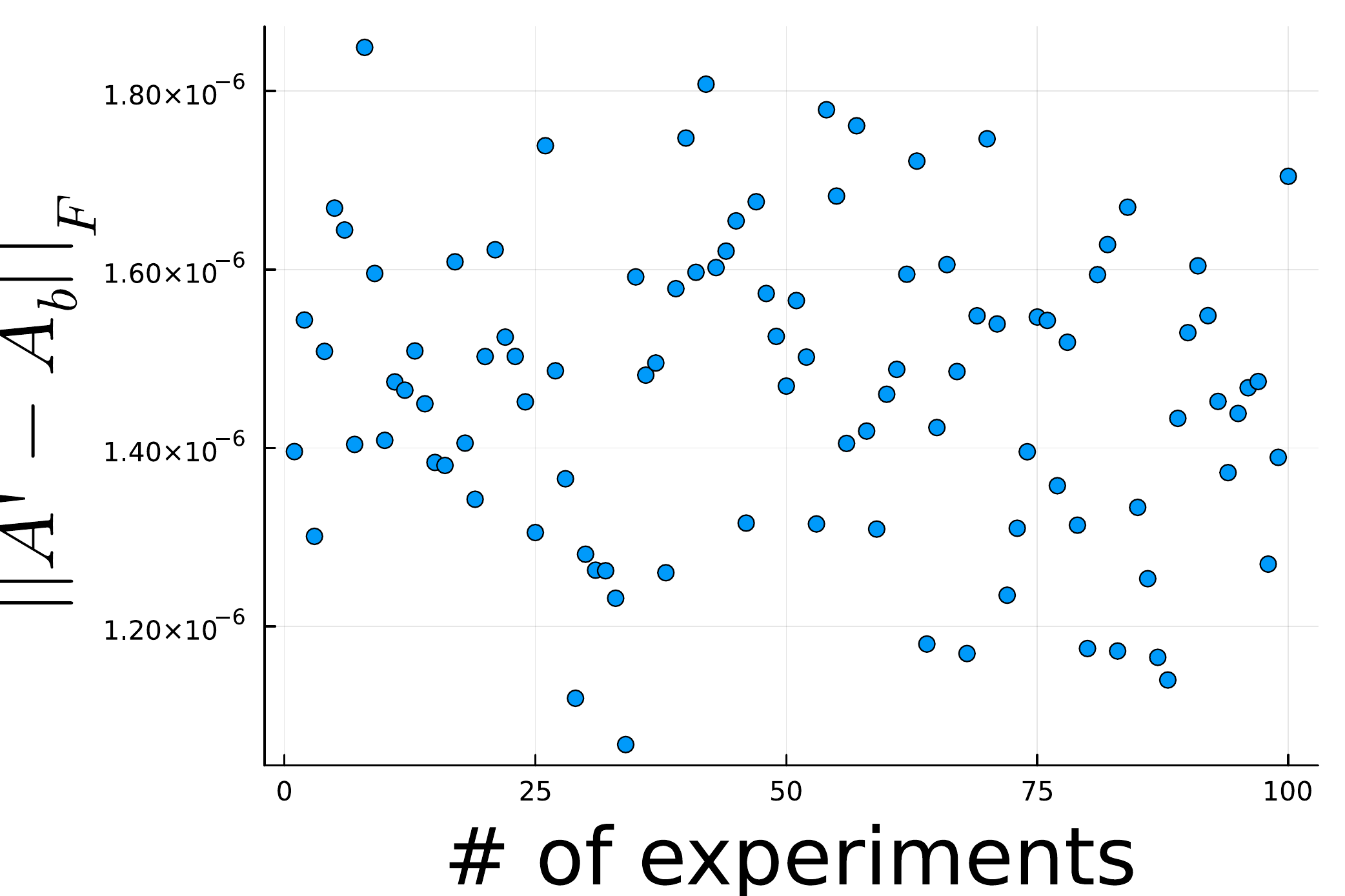}}
    \caption{Accuracy of the backprojector.
    In subfigure (d), $\A'$ denotes regular transpose of $\A$; $\A_b$ denotes the backprojector.}
    \label{fig:adjoint,projector}
\end{figure}

\subsection{Comparison of CNN-regularized EM using different training methods}

This section compares end-to-end training
with other training methods
that have been used previously for SPECT image reconstruction,
namely the gradient truncation and sequential training.
The training targets were simulated activity maps
on \lu XCAT phantoms
and \lu \& \ytt virtual patient phantoms.
We implemented an 
unrolled CNN-regularized EM algorithm
with 3 outer iterations,
each of which had
one inner iteration.
Only 3 outer iterations were used 
(compared to previous works such as \cite{mehranian:19:mbd})
because we used the
16-iteration 4-subset OSEM reconstructed image
as a warm start for all reconstruction algorithms.
We set the regularization parameter 
(defined in \eqref{e,reg}) as
$\beta=1$.
The regularizer
was a 3-layer 3D CNN,
where each layer had a 
$3\times3\times3$ 
convolutional filter
followed by
ReLU activation (except the last layer),
and hence had 657 trainable parameters in total.
We added the input image $\xk$ to the output of CNN
following
the common residual learning strategy
\cite{he:16:drl}.
End-to-end training
and gradient truncation
could also work with
a shared weights CNN approach,
but were not included here
for fair comparison purpose,
since the sequential training
only works with non-shared weights CNN.
All the neural networks
were initialized  
with the same parameters
(drawn from a Gaussian distribution)
and trained on an Nvidia RTX 3090 GPU
for 600 epochs
by minimizing mean square error (loss)
using AdamW optimizer \cite{loshchilov:19:dwd}
with a constant learning rate 0.002.

Besides line profiles for qualitative comparison,
we also used mean activity error (MAE)
and normalized root mean square error (NRMSE)
as quantitative evaluation metrics,
where MAE is defined as 
\begin{equation}\label{e,mae}
\mathrm{MAE} \defequ     
\Bigg|
1 - 
\frac{\frac{1}{n_p}\sum_{j \in \mathrm{VOI}} \hat{\x}[j]}{
\frac{1}{n_p}\sum_{j \in \mathrm{VOI}} 
\xtrue[j] 
}
\Bigg|
\times 100\%
,
\end{equation}
where $n_p$ denotes number of voxels
in the voxels of interest (VOI).
$\hat{\x}$ and
$\xtrue$ denote the 
reconstructed image and 
the true activity map, respectively.
The NRMSE is defined as
\begin{equation}\label{e,nrmse}
\mathrm{NRMSE} \defequ 
\frac{\sqrt{\frac{1}{n_p}
\sum_{j \in \mathrm{VOI}}
\paren{\hat{\x}[j] - \xtrue[j]}^2
}}{\sqrt{\frac{1}{n_p}
\paren{\sum_{j \in \mathrm{VOI}}\xtrue[j]}^2
}}
\times 100\%
.
\end{equation}
All activity images were scaled
by a factor that normalized 
the whole activity to 1 MBq 
per field of view (FOV) before comparison.
All quantitative results (\tref{tab:xcat}, 
\tref{tab:mae,lu177}, \tref{tab:mae,y90}) 
were averaged across
3 different noise realizations.

\subsubsection{Loss function, computing time and memory use}

\begin{figure}[hbt!]
    \centering
    \subfloat[Training loss]{\includegraphics[width=0.5\linewidth]{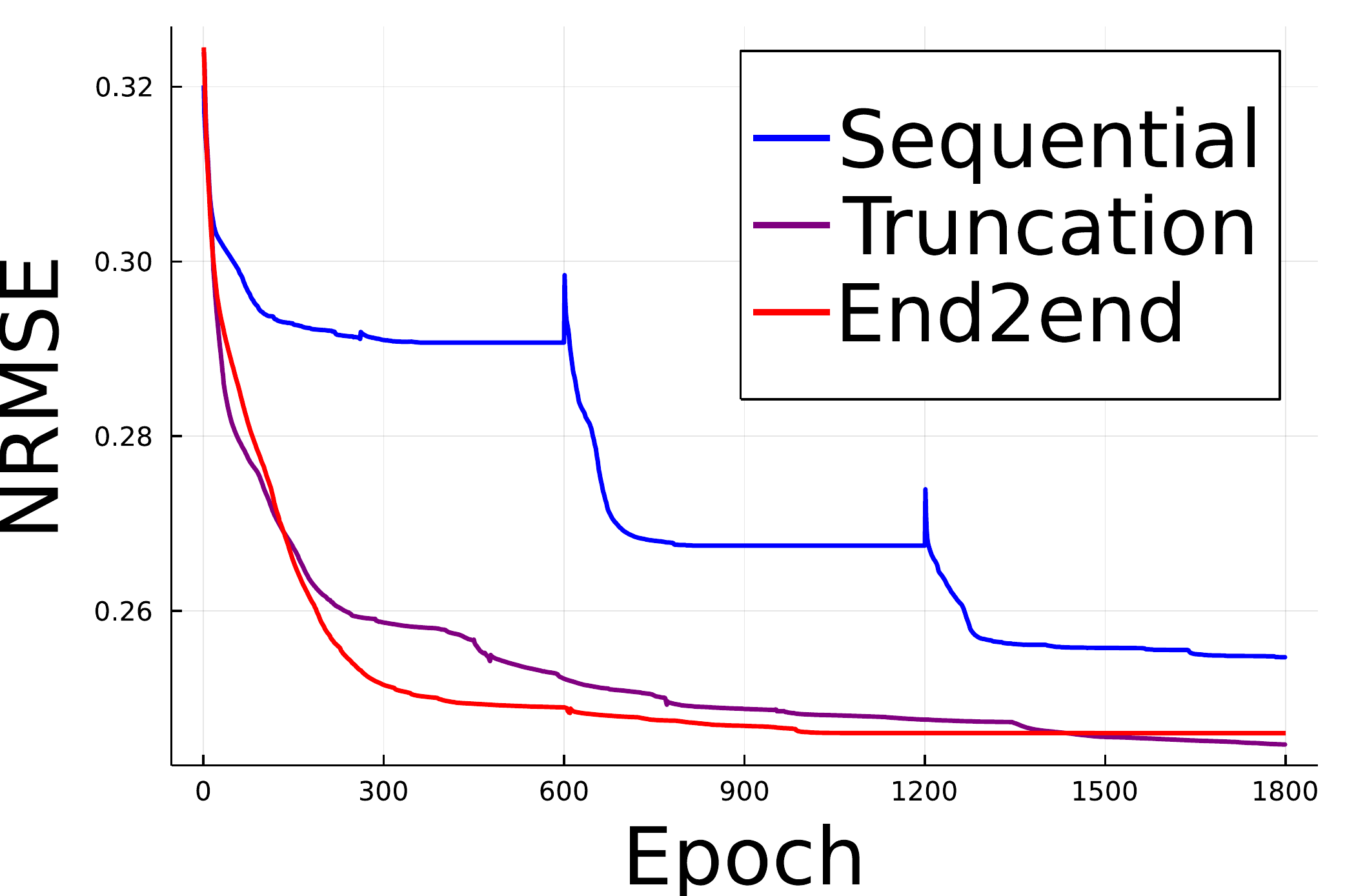}}
    \subfloat[Validation loss]{\includegraphics[width=0.5\linewidth]{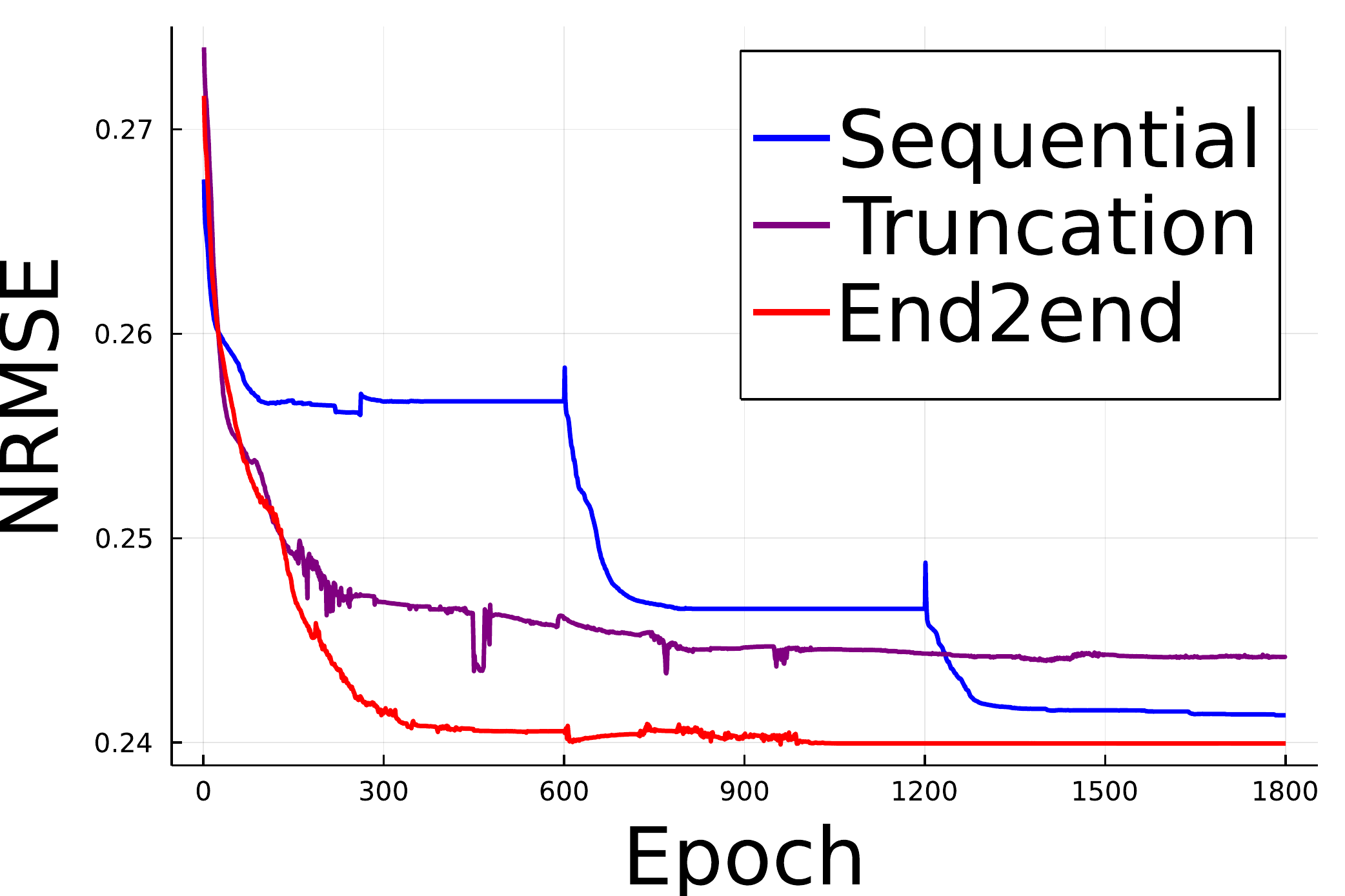}}
    \caption{Training and validation loss of three backpropagation methods.}
    \label{fig:train-valid-loss}
\end{figure}

We compared the training and validation loss 
using sequential training, gradient truncation 
and end-to-end training.
We ran 1800 epochs for each method on \lu XCAT phantoms
with the AdamW optimizer
\cite{loshchilov:19:dwd}.
\fref{fig:train-valid-loss}
shows that
the end-to-end training achieved the lowest validation loss
while it had comparable training loss with the gradient truncation
(which became lower at around 1400 epochs).
For visualization,
we concatenated the first 600 epochs of each outer iteration
for the sequential training method,
as shown by the spikes in sequential training curve.
We ran 600 epochs for each algorithm for subsequent experiments 
because the validation losses were 
pretty much settled at around 600 epochs.

We also compared the computing time of each training method.
We found that for
MLEM with 3 outer iterations and 1 inner iteration,
where each outer iteration had 
a 3-layer convolutional neural network, 
sequential training took 48.6 seconds to complete a training epoch; 
while gradient truncation took 327.1 seconds and
end-to-end training took 336.3 seconds.
Under the same experiment settings, 
we found sequential training
took less than 1GB of memory to backpropagate through 
one outer iteration;
compared to approximately 6GB used in gradient truncation and end-to-end training
that backpropagated through three outer iterations.

\subsubsection{Results on \texorpdfstring{\lu}{Lu-177} XCAT phantoms}

We evaluated the CNN-regularized EM algorithm
with three training methods
on 4 \lu XCAT phantoms we simulated.
We generated the primary projections
by calling forward operation of our Julia projector 
and then added uniform scatters
with
10\% of the primary counts
before adding Poisson noise.
Of the 4 phantoms,
we used 2 for training, 1 for validation and 1 for testing.

\comment{
\subsubsection{Convergence loss}
\fref{fig:train,valid,loss}
shows that the end-to-end
training achieved the lowest
training and validation NRMSE
compared to other methods.
As a reminder, 
the sequential training 
only had one iteration of CNN
at each time of training 
and hence had a different starting NRMSE.
The ``jump" of loss curve for 
sequential training at 200 and 400 epochs 
results from starting the 
next outer iteration
with an untrained CNN.
The gradient truncation method,
although decreases the training 
NRMSE faster than
end-to-end at the first 100 epochs,
slowed down decreasing after 200 epochs,
presumably due to the approximated gradients
leading to a suboptimal minimizer.

\begin{figure}[hbt!]
    \centering
    \includegraphics[width=\linewidth]{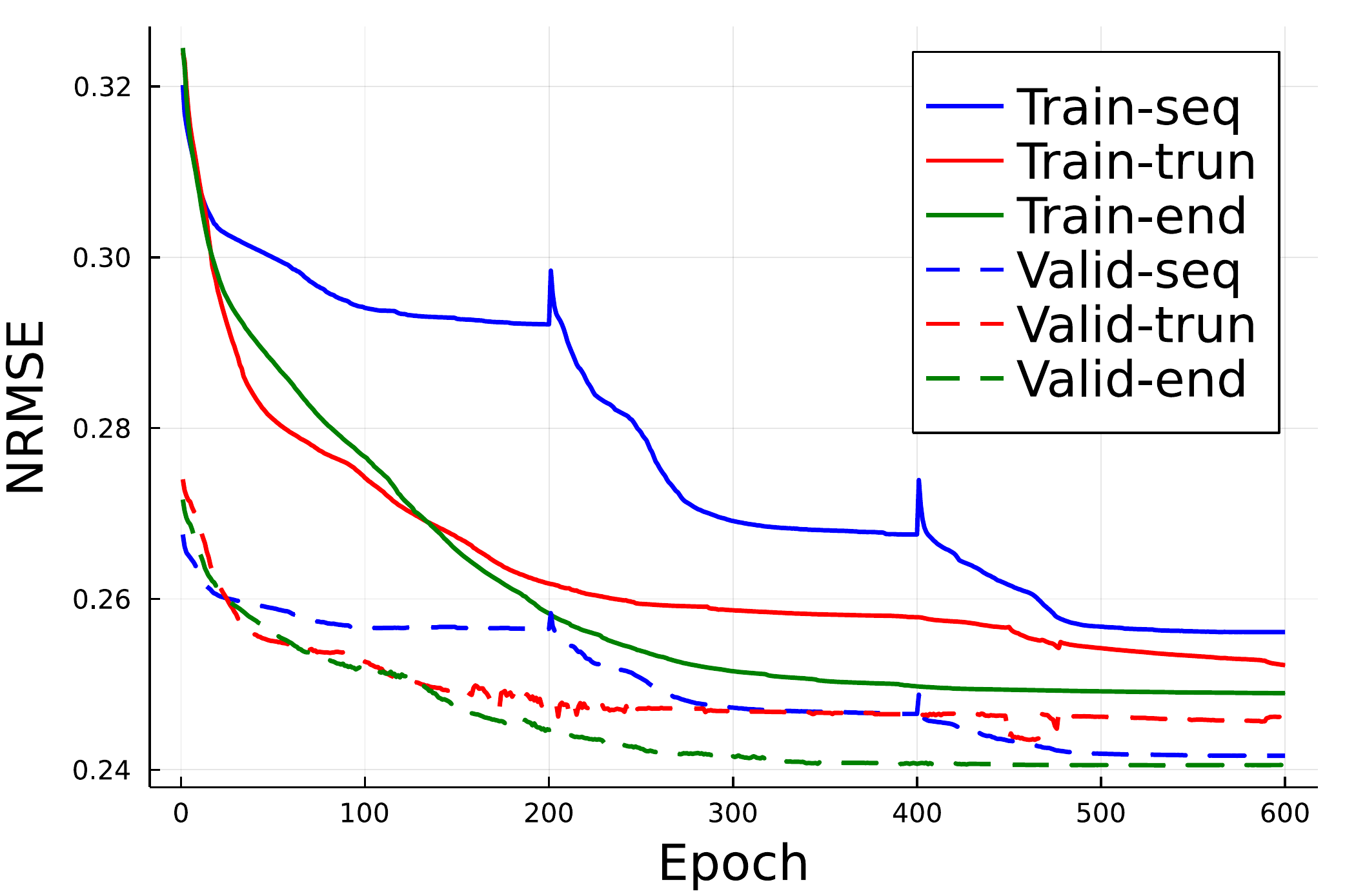}
    \caption{Training and validation loss for
    different training methods.
    ``seq" denotes sequential training,
    ``trun" denotes gradient truncation,
    ``end" denotes end-to-end training.
    For sequential training, 
    we concatenated the first 200 epochs
    (already converged) 
    of each outer iteration
    along the ``epoch" axis in the plot.}
    \label{fig:train,valid,loss}
\end{figure}

} 

\fref{fig:xcat7,slice32} 
shows that
the end-to-end training 
yielded incrementally better
reconstruction of the tumor
in the liver center
over OSEM, sequential training 
and gradient truncation.
\fref{fig:xcat7,slice32}(g)
also illustrates
this improvement
by the line profile across the tumor.
For the tumor at the top-right corner
of the liver,
all methods had comparable performance;
this can be attributed to 
the small tumor size (5mL)
for which
partial volume (PV) effects
associated with SPECT resolution are higher;
and hence its recovery
is even more challenging.

\comment{
For the spleen,
all DL methods 
(\fref{fig:xcat7,slice32}(d)-(f))
visually 
showed higher resolution and better recovery
compared to OSEM (\fref{fig:xcat7,slice32} (c)).
}

\begin{figure}[hbt!]
    \centering
    \subfloat[True activity]{\includegraphics[width=0.33\linewidth]{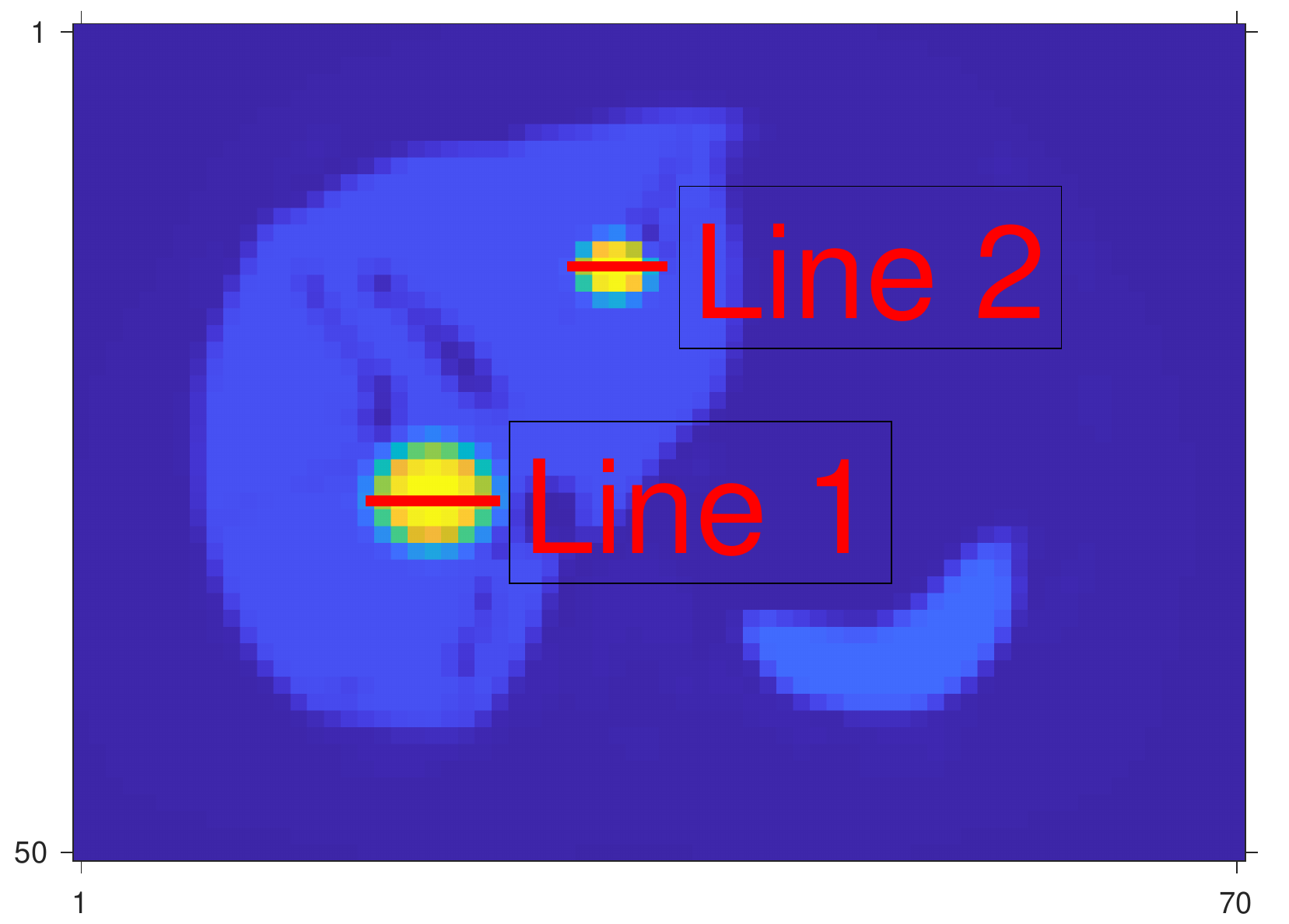}}
    \subfloat[Attenuation]{\includegraphics[width=0.33\linewidth]{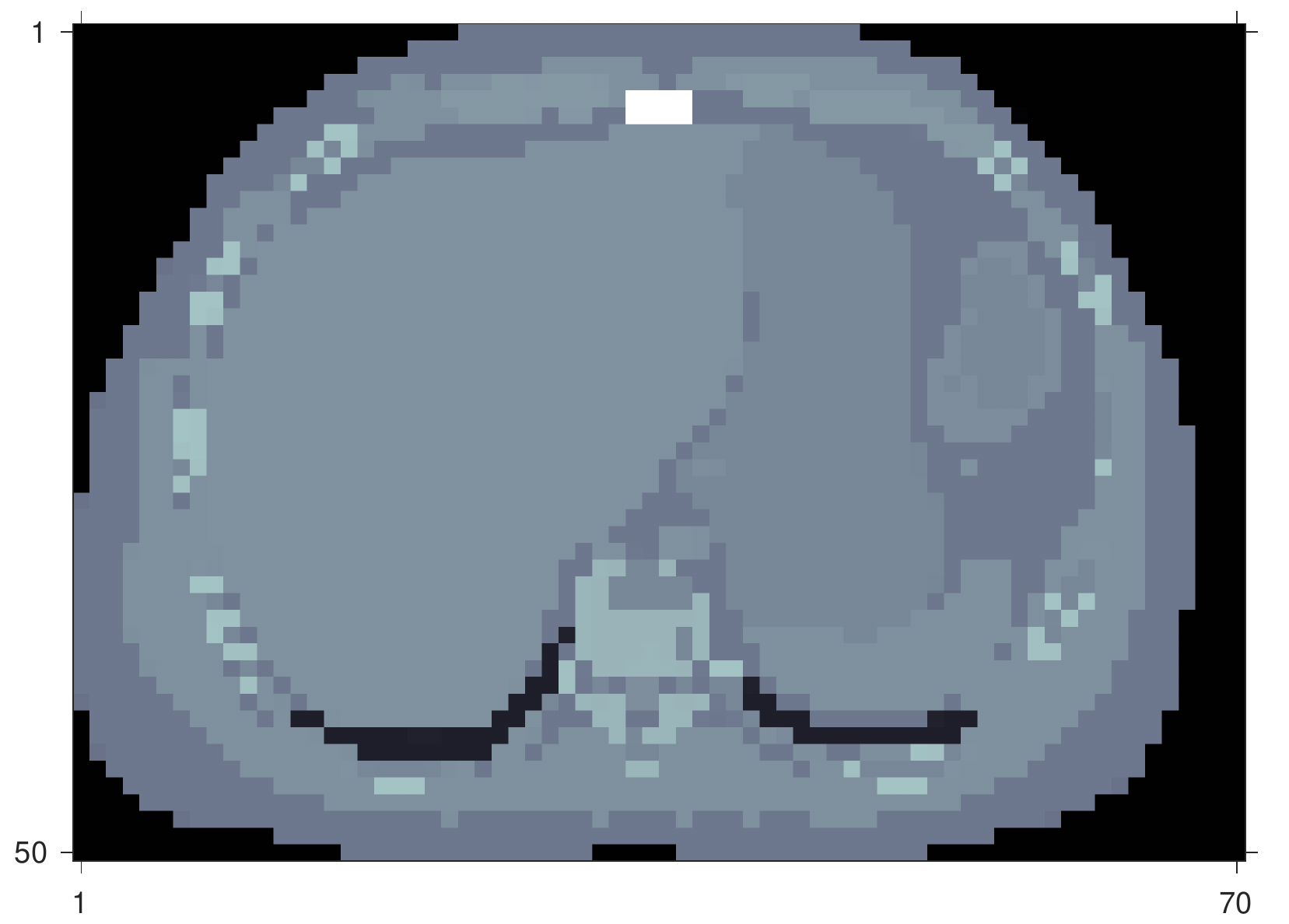}}
    \subfloat[OSEM]{\includegraphics[width=0.33\linewidth]{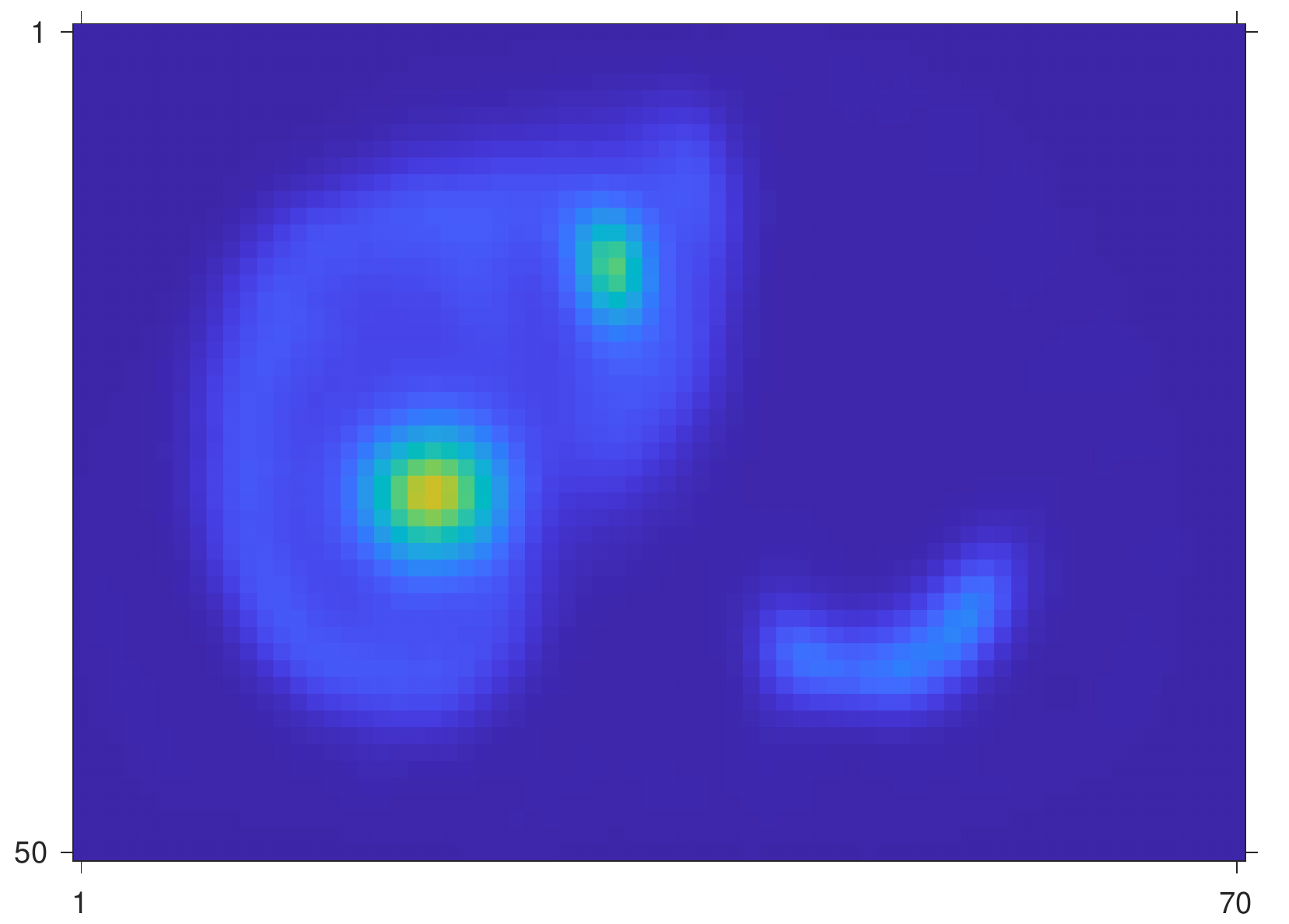}}
    \\
    \subfloat[Sequential]{\includegraphics[width=0.33\linewidth]{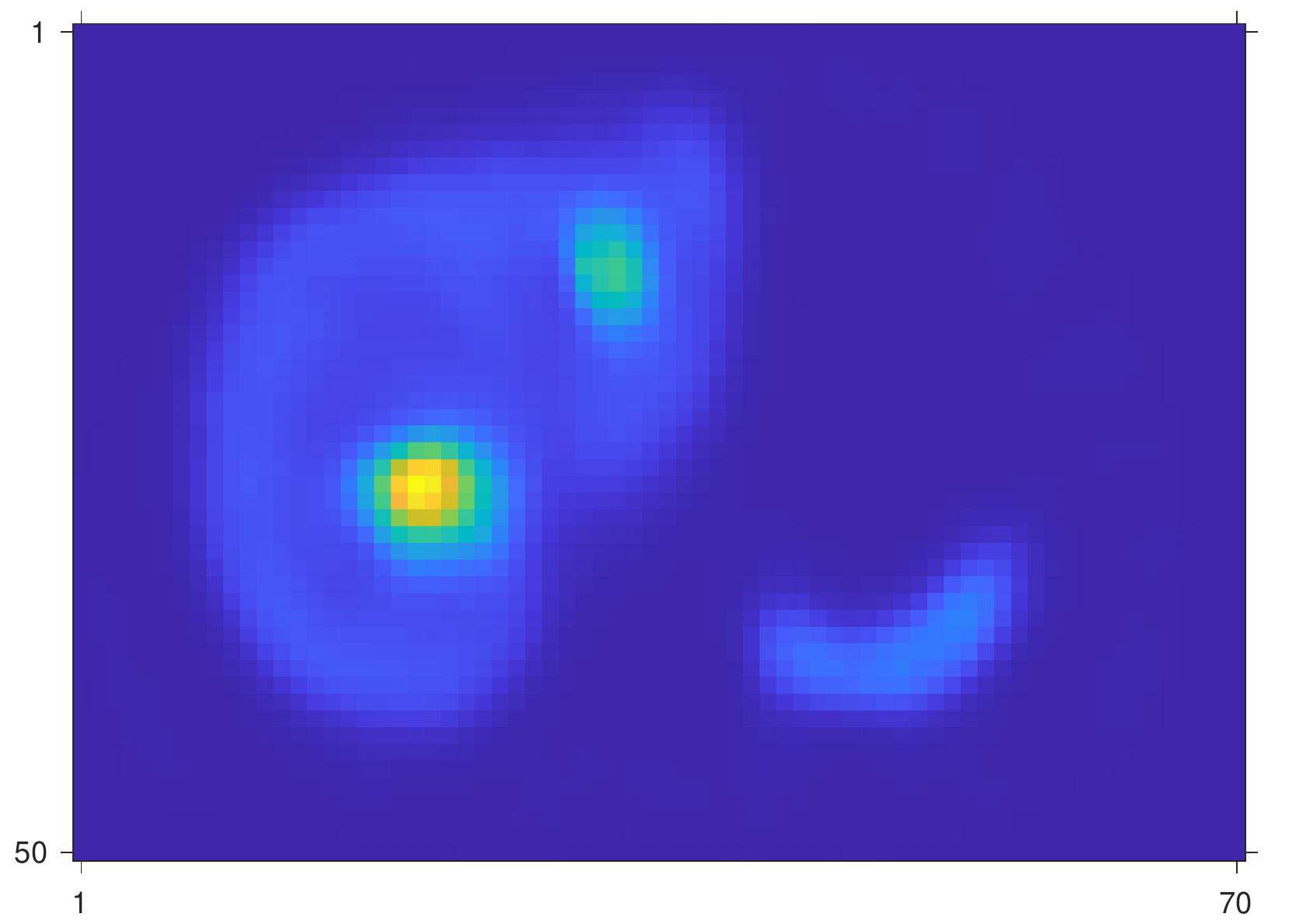}}
    \subfloat[Truncation]{\includegraphics[width=0.33\linewidth]{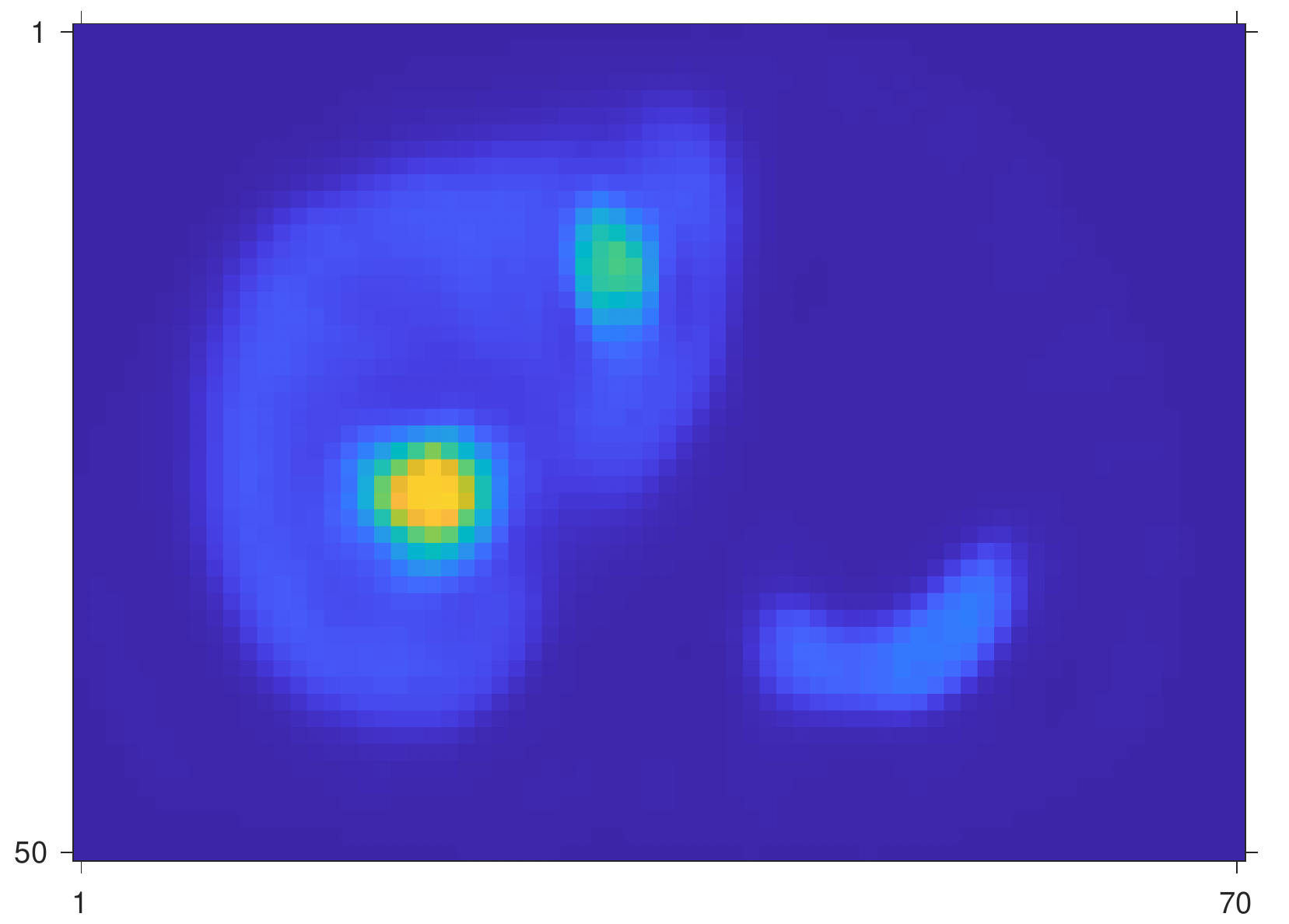}}
    \subfloat[End2end]{\includegraphics[width=0.33\linewidth]{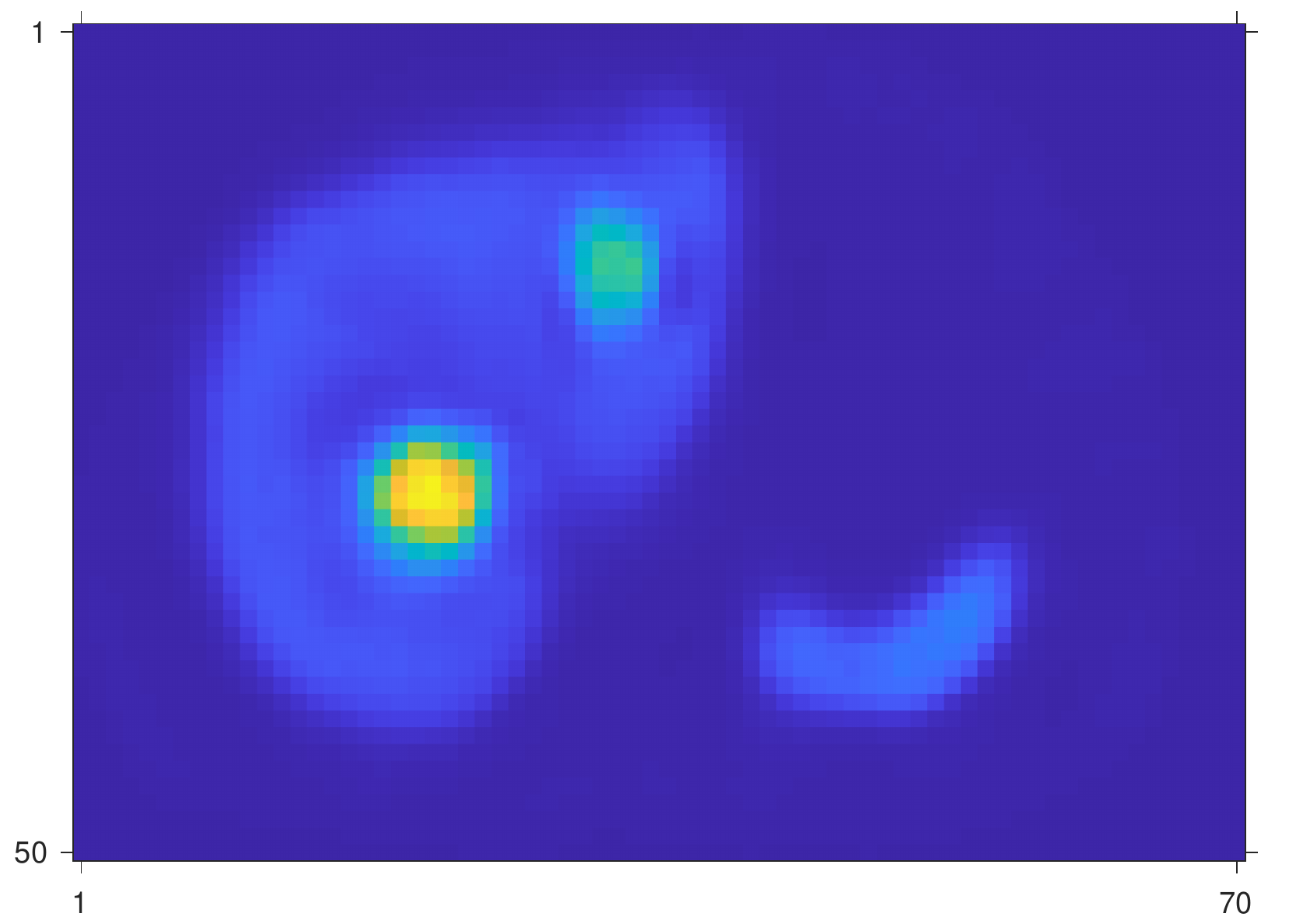}}
    \\
    \subfloat[Line 1 profile]{\includegraphics[width=0.5\linewidth]{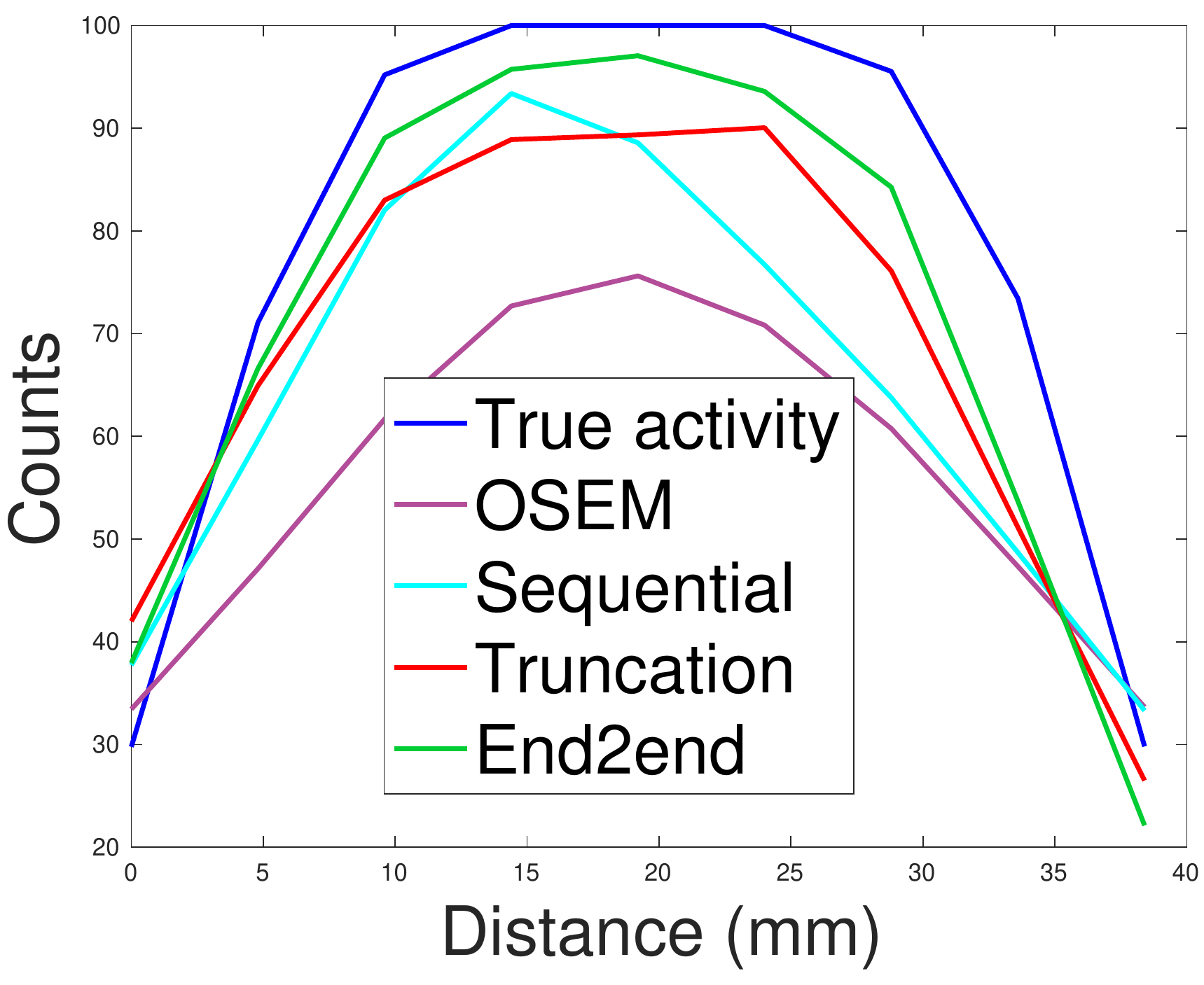}}
    \subfloat[Line 2 profile]{\includegraphics[width=0.5\linewidth]{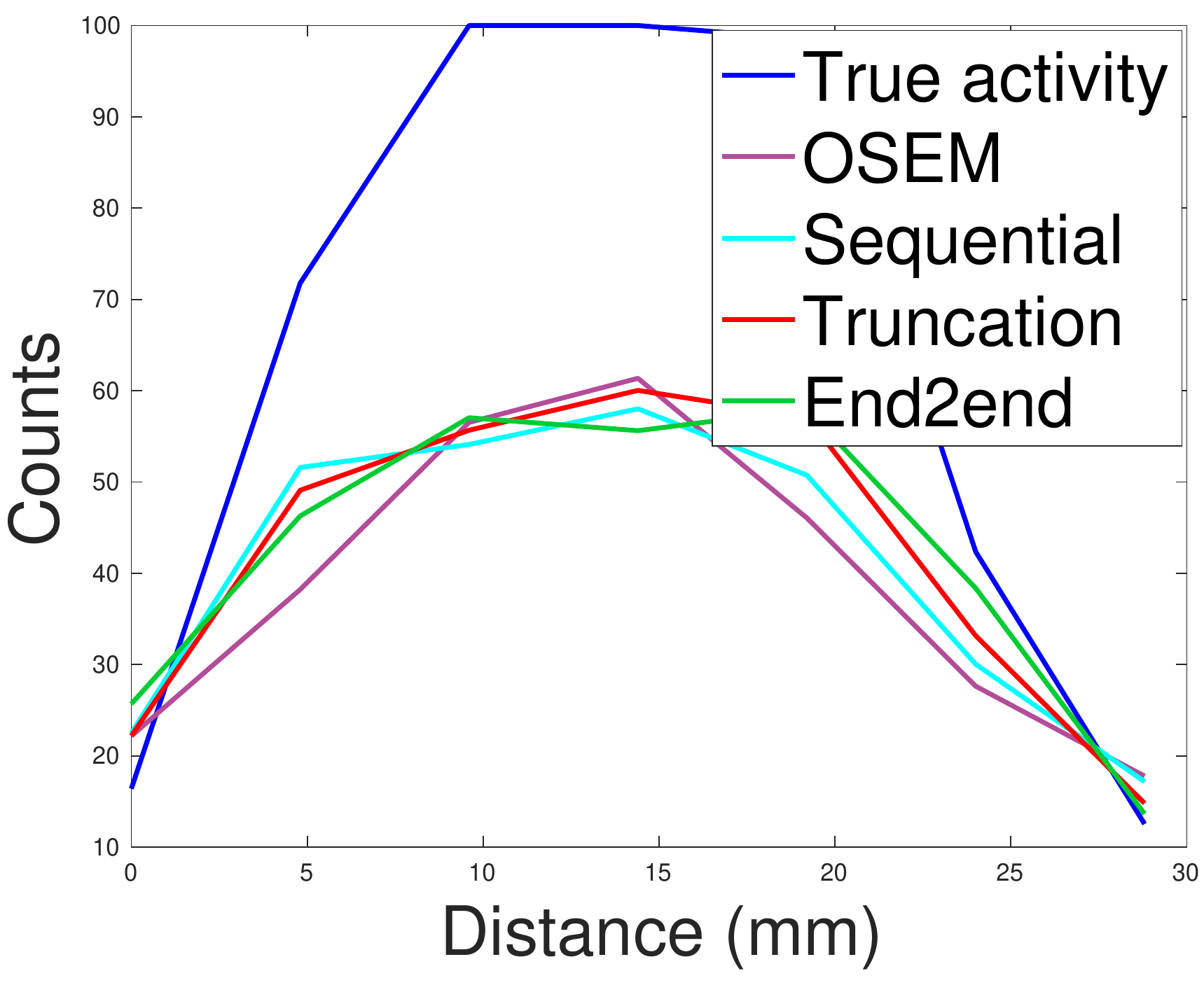}}
    \caption{Qualitative comparison of 
    different training methods and OSEM tested on \lu XCAT phantoms.
    Subfigure (a)-(c): true activity map, attenuation map and OSEM reconstruction
    (16 iterations and 4 subsets);
    (d)-(f): regularized EM using sequential training, gradient truncation, end-to-end training, respectively;
    (g) and (h): line profiles in (a).}
    \label{fig:xcat7,slice32}
\end{figure}

\tref{tab:xcat}
demonstrates that the CNN-regularized EM algorithm
with all training methods 
(sequential training,
gradient truncation
and end-to-end training)
consistently had lower reconstruction error
than the OSEM method.
Among all training methods,
the proposed end-to-end training  
had lower MAE 
over nearly all lesions and organs
than other training methods.
The relative reduction in MAE
by the end-to-end training
was up to 32\% (for lesion 3)
compared to sequential training.
End-to-end training
also had lower NRMSE
for most lesions and organs,
and was otherwise comparable
to other training methods.
The relative improvement
compared to sequential training 
was up to 29\% (for lesion 3).

\begin{table}[hbt!]
\caption{
The average($\pm$standard deviation) MAE(\%) and NRMSE(\%)
across 3 noise realizations of \lu XCAT phantoms.}
    \centering
    \resizebox{\linewidth}{!}{
    \begin{tabular}{|c|c|c|c|c|}
    \hline
    \multicolumn{5}{|c|}{MAE(\%)}
    \\
    \hline
    Lesion/Organ & OSEM & Sequential & Truncation & End2end\\
    \hline
    Lesion 1 (67mL) 
    & 
    12.5 $\pm$ 0.6
    & 
    6.7 $\pm$ 1.8
    & 
    2.8 $\pm$ 0.9
    & 
    \textbf{2.1} $\pm$ 1.1 
    \\
    \hline
    Lesion 2 (10mL) 
    & 
    20.2 $\pm$ 0.9
    & 
    11.5 $\pm$ 4.1
    & 
    10.8 $\pm$ 0.9
    & 
    \textbf{9.7} $\pm$ 1.1
    \\
    \hline
    Lesion 3 (9mL) 
    & 
    25.6 $\pm$ 0.6
    & 
    18.8 $\pm$ 0.4
    & 
    15.2 $\pm$ 0.9
    & 
    \textbf{12.8} $\pm$ 1.0
    \\
    \hline
    Lesion 4 (5mL) 
    & 
    43.0 $\pm$ 0.6
    & 
    40.0 $\pm$ 1.2
    & 
    38.8 $\pm$ 0.8
    & \textbf{38.7} $\pm$ 0.7
    \\
    \hline
    Liver 
    & 
    6.4 $\pm$ 0.7
    & 
    6.2 $\pm$ 1.5
    & 
    4.6 $\pm$ 1.1
    & 
    \textbf{3.7} $\pm$ 1.2
    \\
    \hline
    Lung 
    & 
    2.4 $\pm$ 0.7
    & 
    2.2 $\pm$ 0.4
    & 
    \textbf{0.7} $\pm$ 0.6
    & 
    0.9 $\pm$ 0.5
    \\
    \hline
    Spleen 
    & 
    14.2 $\pm$ 0.9
    & 
    12.6 $\pm$ 2.4
    & 
    \textbf{8.9} $\pm$ 0.7
    & 
    9.3 $\pm$ 1.5
    \\
    \hline
    Kidney 
    & 
    15.9 $\pm$ 1.0
    & 
    15.1 $\pm$ 1.2
    & 
    14.4 $\pm$ 1.4
    & 
    \textbf{13.6} $\pm$ 1.6
    \\
    \hline
    \multicolumn{5}{|c|}{NRMSE(\%)}
    \\
    \hline
    Lesion/Organ & OSEM & Sequential & Truncation & End2end\\
    \hline
    Lesion 1 (67mL) 
    & 
    27.3 $\pm$ 0.3
    & 
    21.7 $\pm$ 1.3
    & 
    18.9 $\pm$ 0.6
    & 
    \textbf{18.3}  $\pm$ 0.6
    \\
    \hline
    Lesion 2 (10mL) 
    & 
    26.8 $\pm$ 0.6
    & 
    19.2 $\pm$ 2.2
    & 
    16.4 $\pm$ 0.4
    & 
    \textbf{16.3} $\pm$ 0.8
    \\
    \hline
    Lesion 3 (9mL) 
    & 
    28.4 $\pm$ 0.4
    & 
    22.8 $\pm$ 0.8
    & 
    18.3 $\pm$ 0.7
    & 
    \textbf{16.3} $\pm$ 0.7
    \\
    \hline
    Lesion 4 (5mL) 
    & 
    43.5 $\pm$ 0.5
    & 
    41.1 $\pm$ 1.3
    & 
    \textbf{40.0} $\pm$ 0.7
    & 
    40.2 $\pm$ 0.6
    \\
    \hline
    Liver 
    & 
    28.5 $\pm$ 0.1
    & 
    25.0 $\pm$ 0.8
    & 
    \textbf{24.3} $\pm$ 0.3
    & 
    24.5 $\pm$ 0.3
    \\
    \hline
    Lung 
    & 
    32.1 $\pm$ 0.1
    & 
    31.2 $\pm$ 1.1
    & 
    \textbf{29.5} $\pm$ 0.3
    & 
    30.4 $\pm$ 0.4
    \\
    \hline
    Spleen 
    & 
    25.7 $\pm$ 0.3
    & 
    22.8 $\pm$ 1.1
    & 
    20.4 $\pm$ 0.4
    & 
    \textbf{19.9} $\pm$ 0.6
    \\
    \hline
    Kidney 
    & 
    40.8 $\pm$ 0.3
    & 
    39.7 $\pm$ 0.4
    & 
    39.7 $\pm$ 0.2
    & 
    \textbf{39.2} $\pm$ 0.3
    \\
    \hline
    \end{tabular}}
    \label{tab:xcat}
\end{table}

\subsubsection{Results on \texorpdfstring{\lu}{Lu-177} VP phantoms}

Next we present test results 
on 8 \lu virtual patient phantoms.
Out of 8 \lu phantoms, we used 4 for training,
1 for validation and 3 for testing.

\begin{figure}[hbt!]
    \centering
    \subfloat[True activity]{\includegraphics[width=0.33\linewidth]{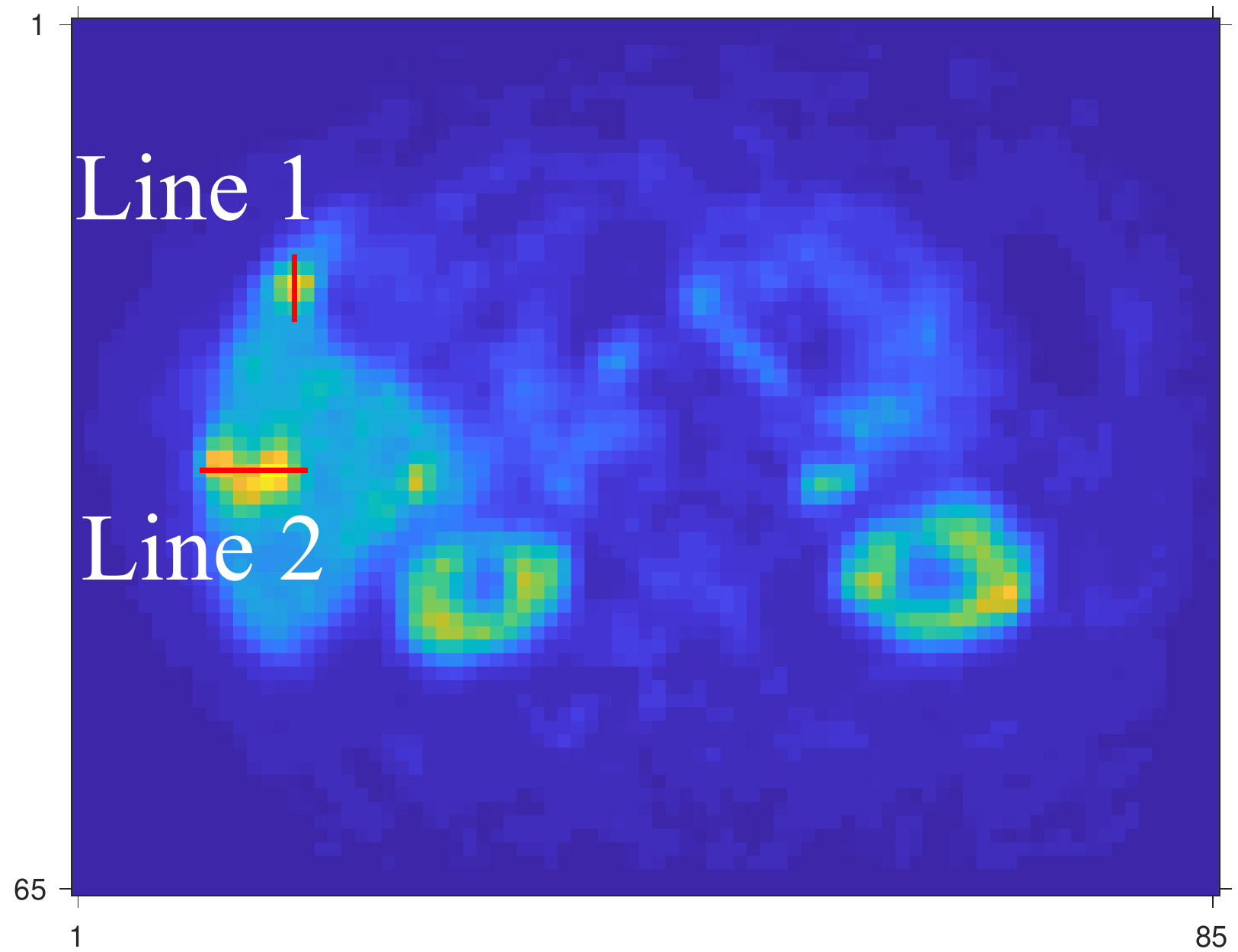}}
    \subfloat[Attenuation map]{\includegraphics[width=0.33\linewidth]{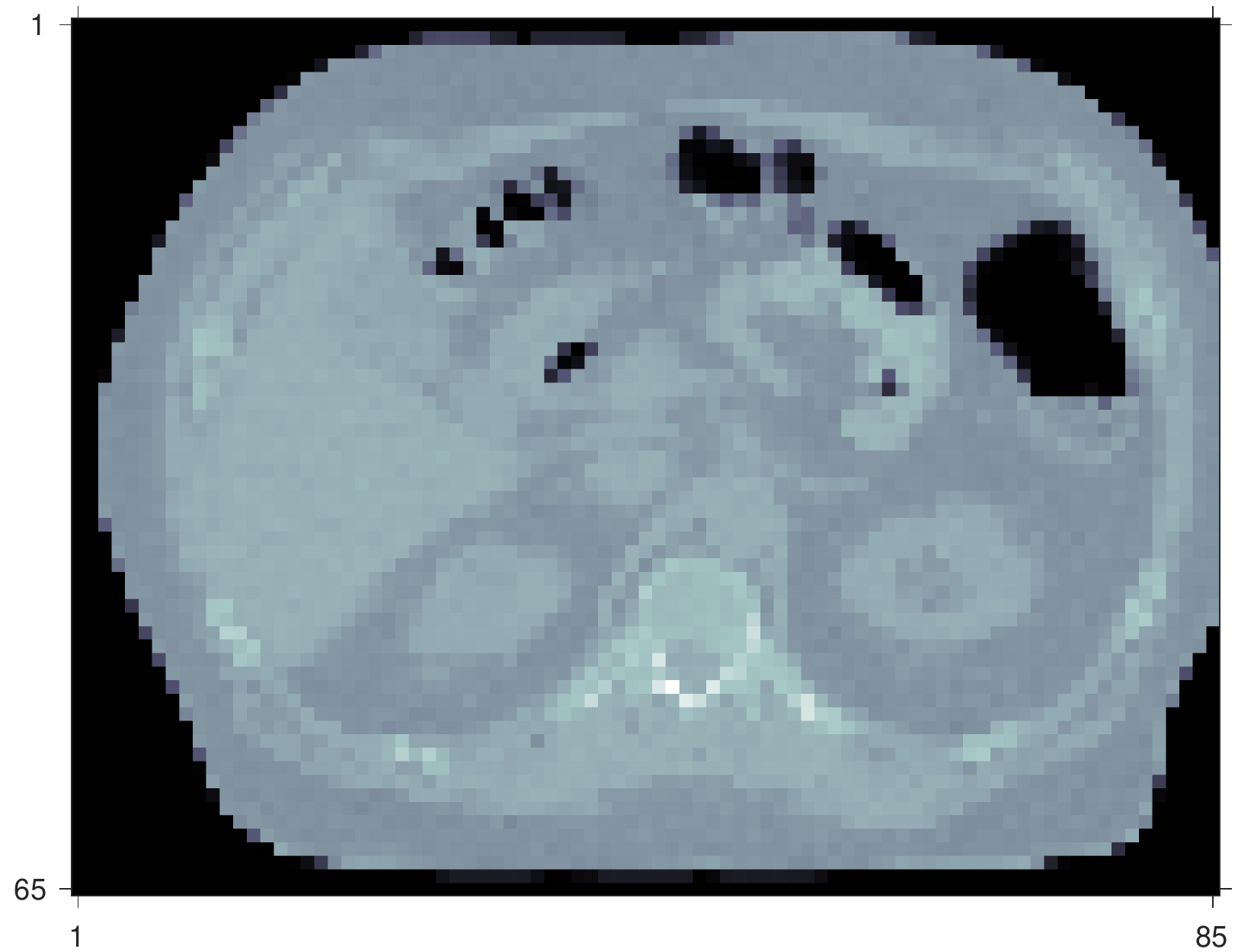}}
    \subfloat[OSEM]{\includegraphics[width=0.33\linewidth]{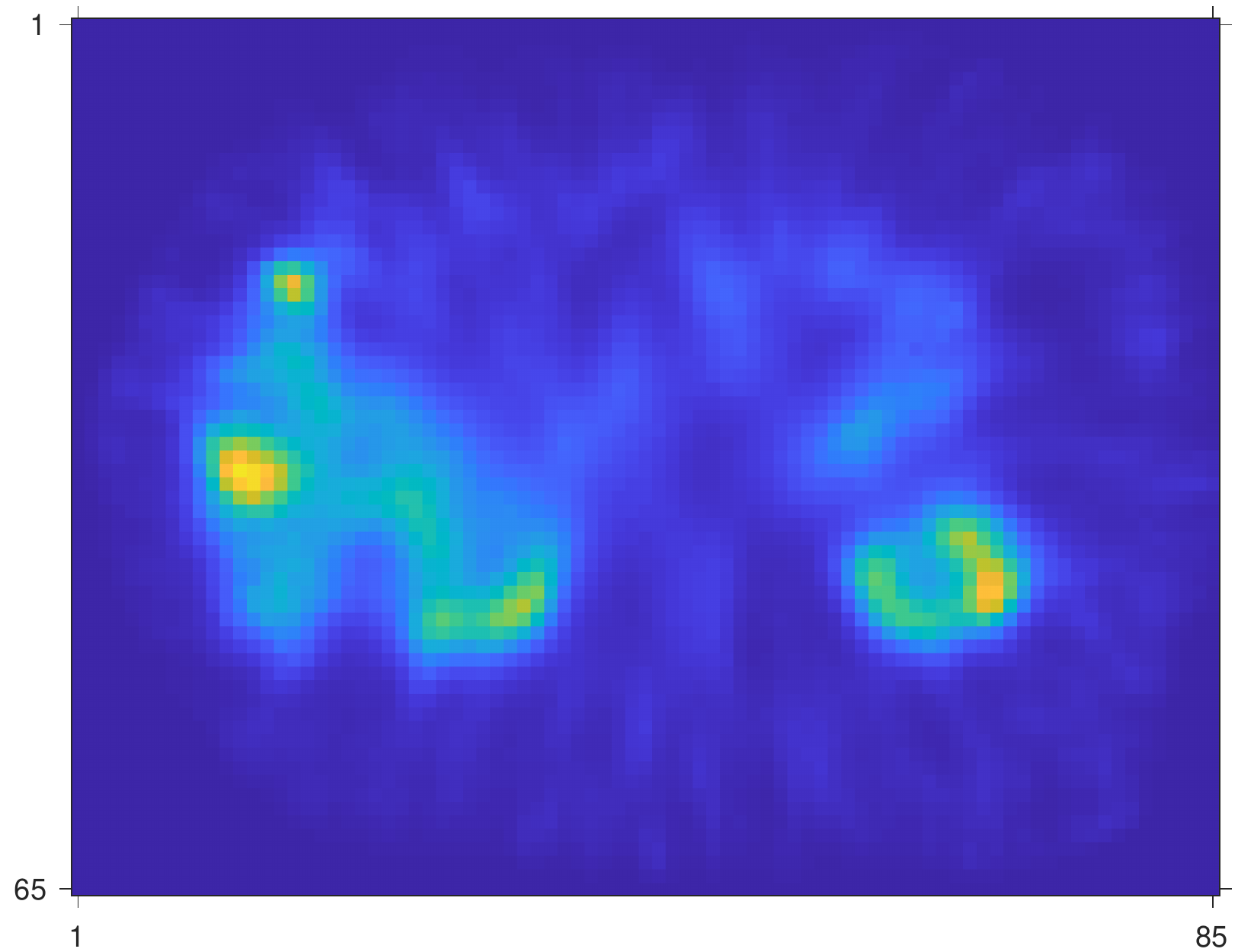}}
    \\
    \subfloat[Sequential]{\includegraphics[width=0.33\linewidth]{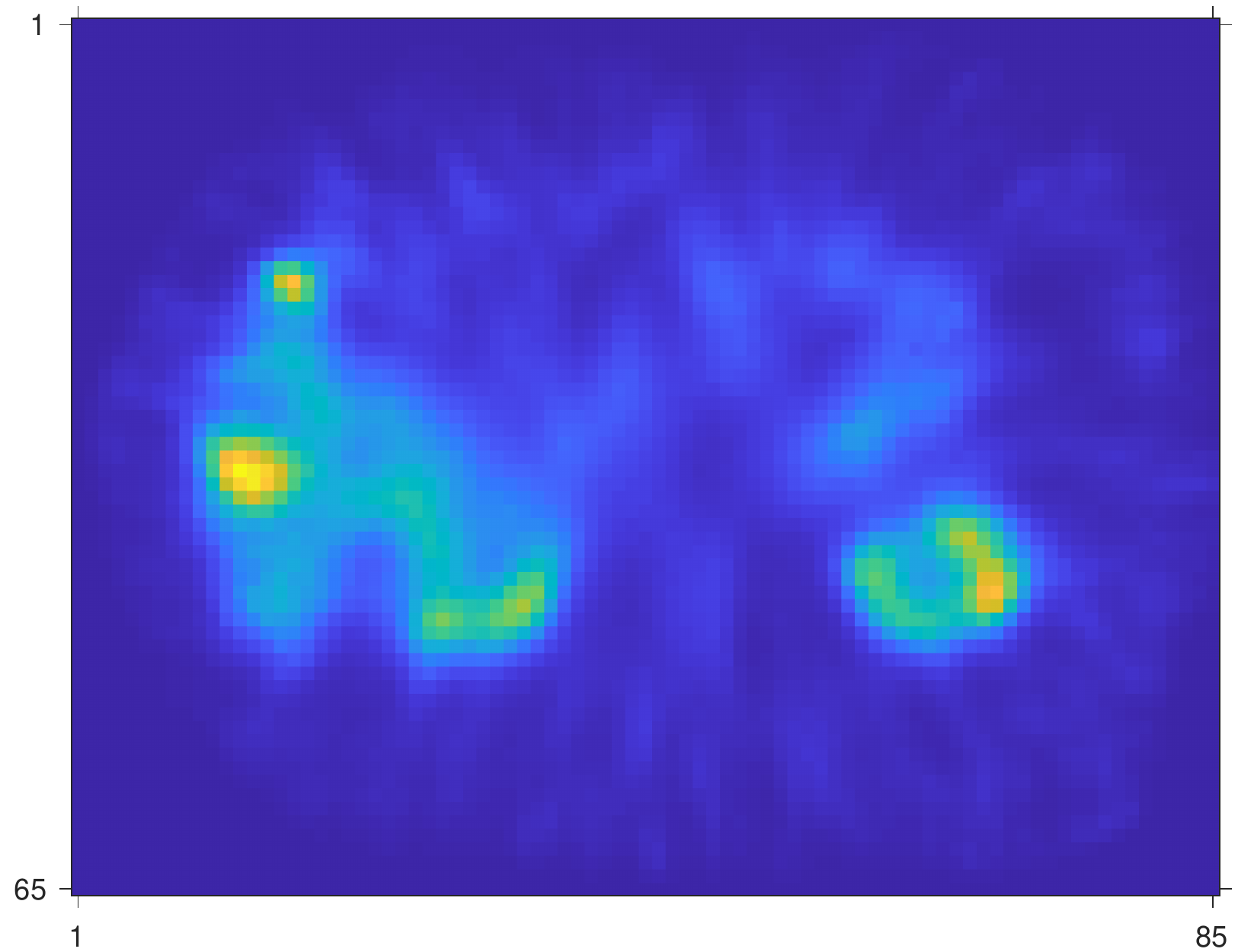}}
    \subfloat[Truncation]{\includegraphics[width=0.33\linewidth]{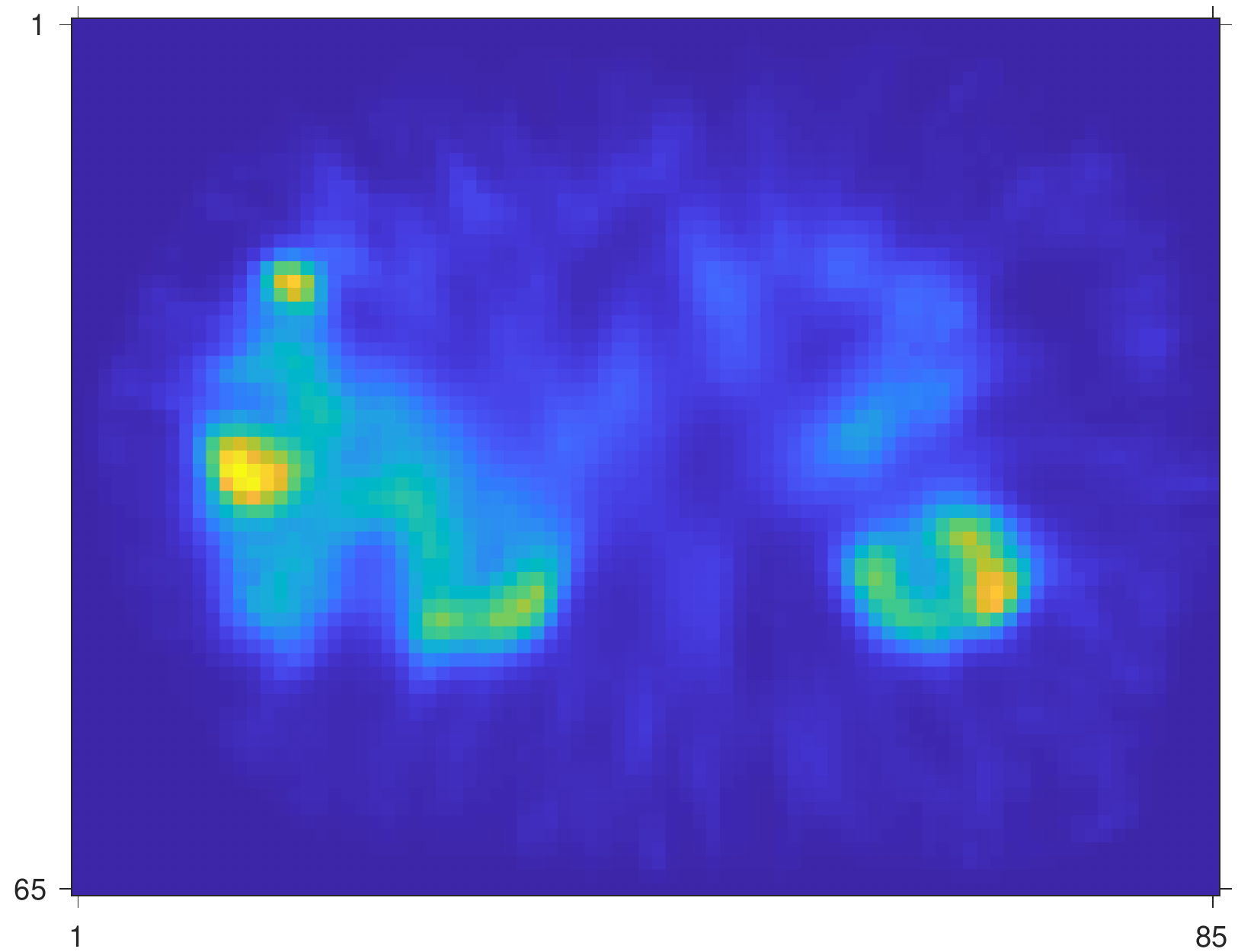}}
    \subfloat[End2end]{\includegraphics[width=0.33\linewidth]{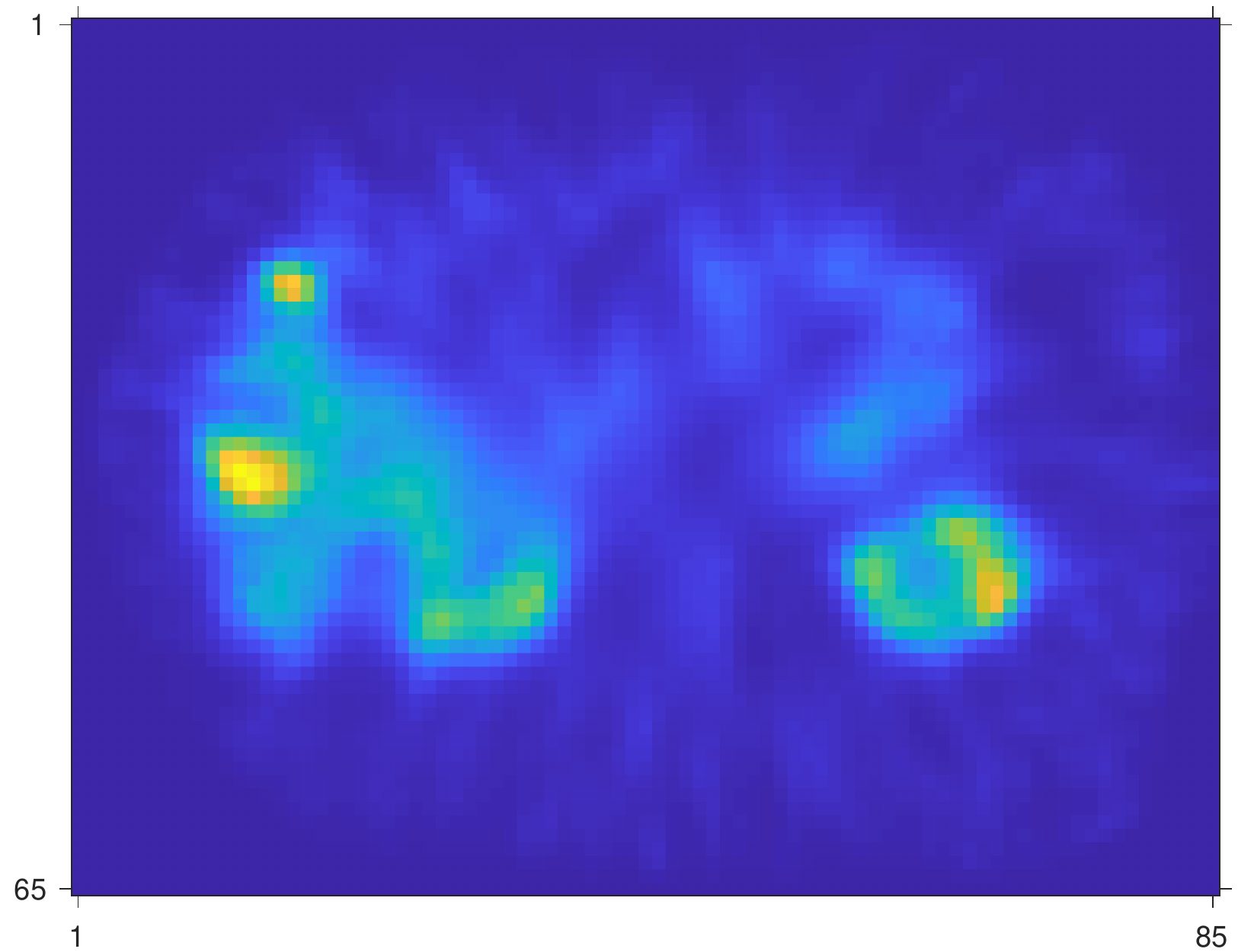}}
    \\
    \subfloat[Line 1 profile]{\includegraphics[width=0.5\linewidth]{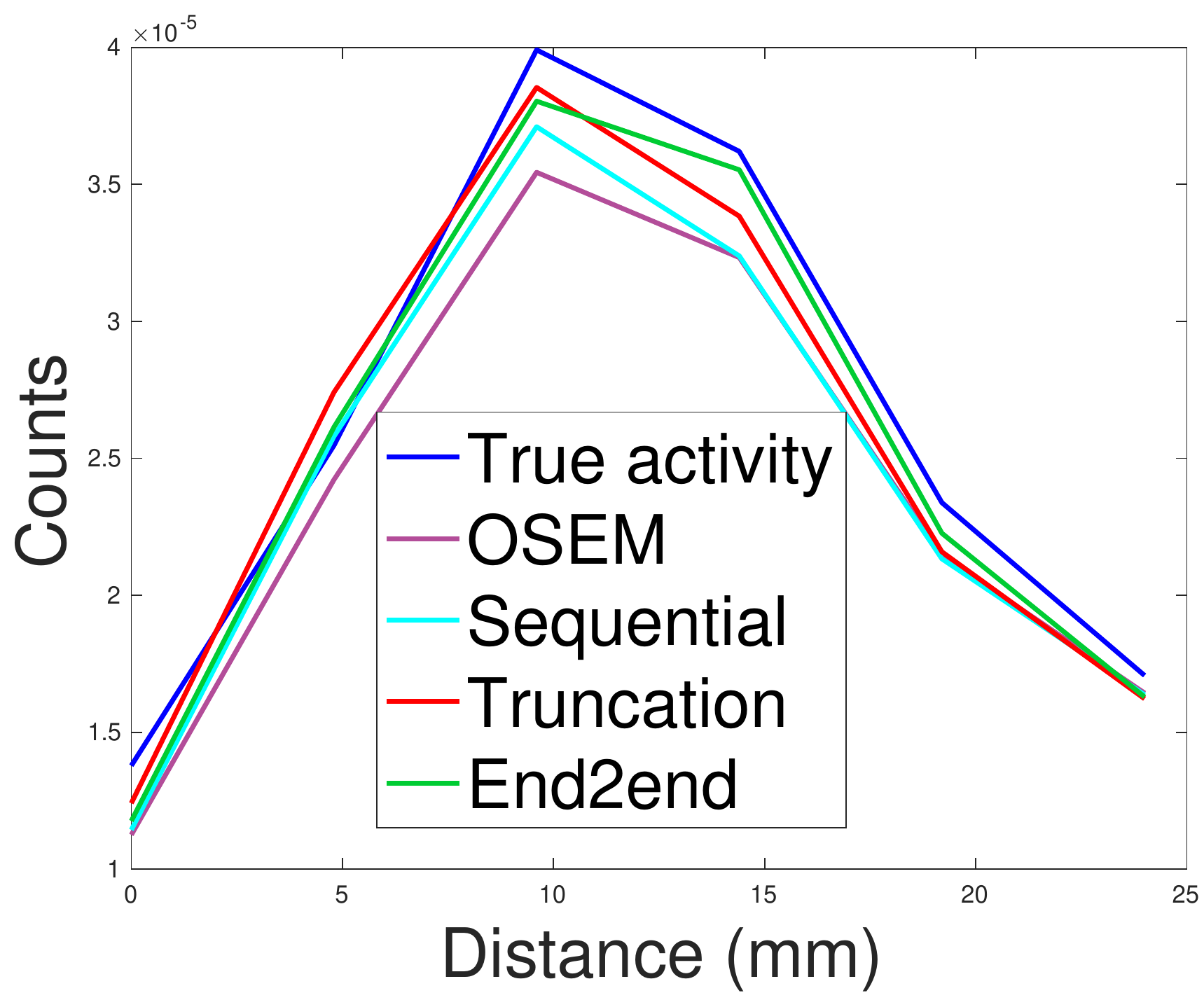}}
    \subfloat[Line 2 profile]{\includegraphics[width=0.5\linewidth]{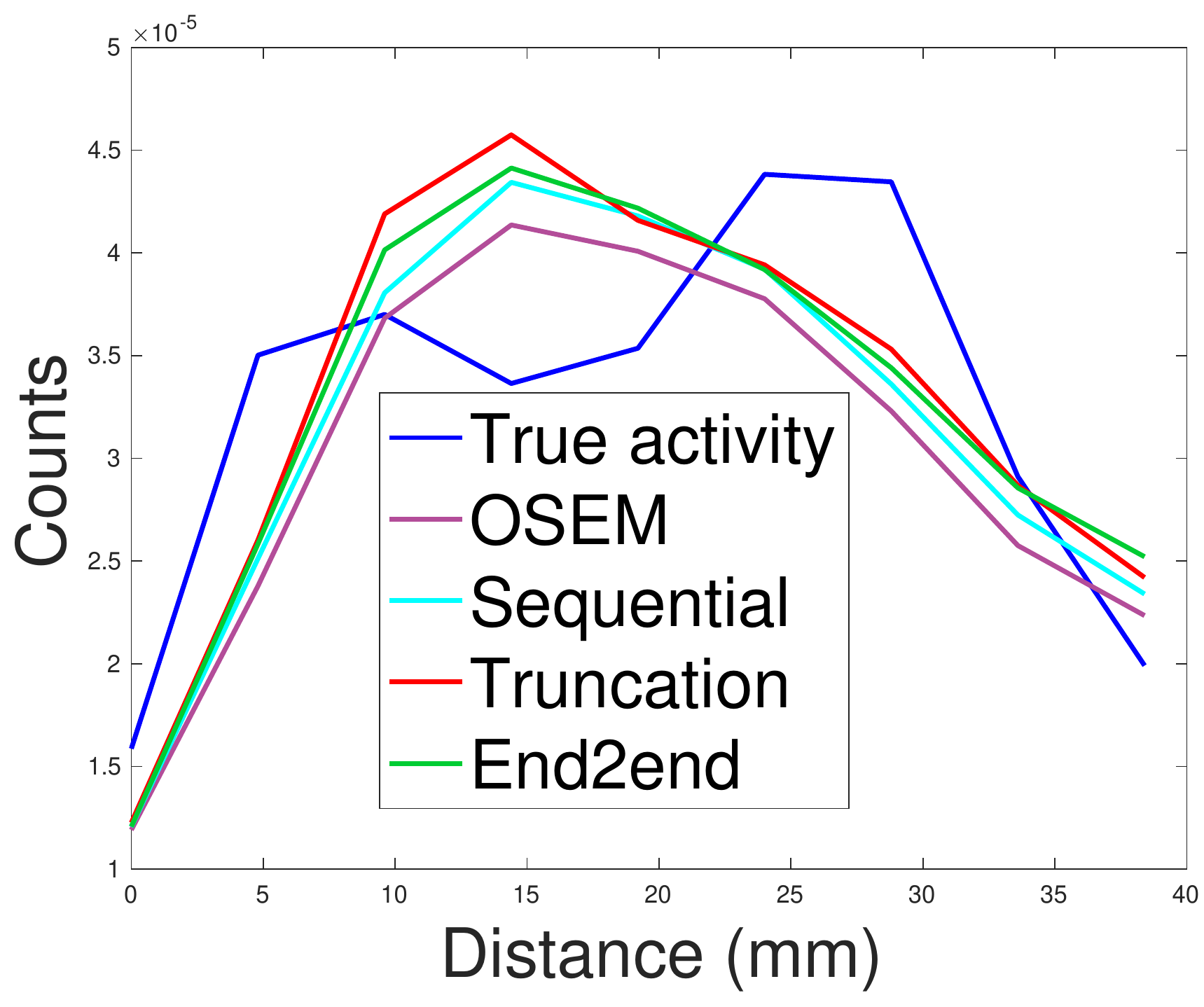}}
    \caption{Qualitative comparison of 
    different training methods and OSEM 
    tested on $^{177}$Lu VP phantoms.
    Subfigure (g) and (h) correspond to
    line profiles marked in (a).}
    \label{fig:lu177}
\end{figure}

\fref{fig:lu177} shows that
the improvement of 
all learning-based methods 
was limited compared to OSEM,
which was also evident from line profiles.
For example, in \fref{fig:lu177}(g),
where the line profile was drawn on a small tumor.
We found that 
OSEM yielded a fairly accurate estimate already,
and we did not observe as much improvement as 
we had seen on \lu XCAT phantoms
for end-to-end training or even learning-based methods.
\tref{tab:mae,lu177} also demonstrates 
this observation.
The OSEM method had substantially lower MAE and NRMSE
compared to the errors shown
for \lu XCAT data (cf \tref{tab:xcat}).
Moreover, the end-to-end training method
had comparable accuracy
with gradient truncation.
For example,
gradient truncation was the best
on lesion, liver and lung in terms of MAE;
end-to-end training had the lowest NRMSE
on lesion, liver, lung, kidney and spleen.
Perhaps this could be due to 
the loss function used for training,
\ie, MSE loss was used in our experiments 
so that end-to-end training might yield lower NRMSE.
A more comprehensive study
would be needed 
to verify this conjecture.

\begin{table}[hbt!]
\caption{The average($\pm$standard deviation) MAE(\%) and NRMSE(\%)
across 3 noise realizations of \lu VP phantoms.}
    \centering
    \resizebox{\linewidth}{!}{
    \begin{tabular}{|c|c|c|c|c|}
    \hline
    \multicolumn{5}{|c|}{MAE(\%)}
    \\
    \hline
    Lesion/Organ & OSEM & Sequential & Truncation & End2end\\
    \hline
    Lesion (6-152mL) & 11.1 $\pm$ 2.5 
    & 9.4 $\pm$ 3.2
    & \textbf{6.7} $\pm$ 2.4 
    & 7.3 $\pm$ 2.8 \\
    \hline
    Liver 
    & 
    4.8 $\pm$ 0.1
    & 
    4.5 $\pm$ 0.2
    & 
    \textbf{3.4} $\pm$ 0.6
    & 
    4.0 $\pm$ 0.2
    \\
    \hline
    Healthy liver 
    & 
    4.1 $\pm$ 0.1
    & 
    4.1 $\pm$ 0.1
    & 
    \textbf{3.5} $\pm$ 0.6
    & 
    4.1 $\pm$ 0.2
    \\
    \hline
    Lung 
    & 
    3.4 $\pm$ 0.1
    & 
    3.0 $\pm$ 0.2
    & 
    \textbf{2.4} $\pm$ 0.7
    & 
    3.0 $\pm$ 0.5
    \\
    \hline
    Kidney 
    & 
    5.2 $\pm$ 0.3
    & 
    4.3 $\pm$ 0.1
    & 
    2.6 $\pm$ 0.1
    & 
    \textbf{2.3} $\pm$ 0.2
    \\
    \hline
    Spleen 
    & 
    0.8 $\pm$ 0.2
    & 
    \textbf{0.6} $\pm$ 0.1
    & 
    1.3 $\pm$ 0.6
    & 
    1.2 $\pm$ 0.4
    \\
    \hline
    \multicolumn{5}{|c|}{NRMSE(\%)}
    \\
    \hline
    Lesion/Organ & OSEM & Sequential & Truncation & End2end\\
    \hline
    Lesion (6-152mL) 
    & 
    16.1 $\pm$ 2.2
    & 
    14.9 $\pm$ 2.4
    & 
    14.3 $\pm$ 1.7
    & 
    \textbf{14.2} $\pm$ 2.1
    \\
    \hline
    Liver 
    & 
    15.9 $\pm$ 0.2
    & 
    \textbf{15.3} $\pm$ 0.1
    & 
    15.5 $\pm$ 0.6
    & 
    \textbf{15.3} $\pm$ 0.1
    \\
    \hline
    Healthy liver 
    & 
    16.8 $\pm$ 0.1
    & 
    \textbf{16.6} $\pm$ 0.1
    & 
    17.3 $\pm$ 0.5
    & 
    17.1 $\pm$ 0.3
    \\
    \hline
    Lung 
    & 
    22.3 $\pm$ 0.3
    & 
    22.1 $\pm$ 0.4 
    & 
    22.0 $\pm$ 0.4
    & 
    \textbf{21.9} $\pm$ 0.5
    \\
    \hline
    Kidney 
    & 
    17.4 $\pm$ 0.1
    & 
    16.8 $\pm$ 0.1
    & 
    16.4 $\pm$ 0.3
    & 
    \textbf{16.3} $\pm$ 0.5
    \\
    \hline
    Spleen 
    & 
    13.5 $\pm$ 0.2
    & 
    12.4 $\pm$ 0.3
    & 
    \textbf{12.3} $\pm$ 0.7
    & 
    \textbf{12.3} $\pm$ 0.5
    \\
    \hline
    \end{tabular}}
    \label{tab:mae,lu177}
\end{table}

\subsubsection{Results on \texorpdfstring{\ytt}{Y-90} VP phantoms}

We also tested with 8 \ytt virtual patient phantoms.
Of the 8 phantoms, we used 4 for training, 
1 for validation and 3 for testing.
\comment{Then we generated density maps using patients' CT images
according to the following formula that
was determined from a prior calibration
measurement using a phantom 
with 16 tissue equivalent rods:
\begin{equation}\label{ct2den}
    \rho(HU) = 
    \left\{
    \begin{array}{ll}
    0.00108 \cdot HU + 1.02351, & HU \le 0 \\
    0.00069 \cdot HU + 1.03107, & HU > 0
    . \end{array}
    \right .
\end{equation}
}
\comment{out of 128 in total,
with each phantom's activity map and density map as input.
The simulated energy window was set 
from 105 keV to 195 keV
to account for $^{90}$Y large fraction of scatters.
Poisson noise was added after 
the 128 projection views 
were scaled to a 10 million total counts, 
which corresponds to a typical 
average count-level in \blue{\ytt bremstrahlung SPECT imaging}.
The SPECT reconstruction used the OSEM algorithm
(implemented by our BFBP) with CT-based 
attenuation correction, 
\blue{scatter correction using the true scatter from SIMIND}
and collimator-detector response modelling
(4 subsets and 16 iterations, 
$128\times128\times80$ matrix
with voxel size
$4.8\times4.8\times4.8$~mm$^3$
and
no Gaussian smoothing).
Out of 8 \ytt virtual patient phantoms, 
we randomly selected 4 for training, 
1 for validation and 3 for testing.}

\begin{figure}[hbt!]
    \centering
    \subfloat[True activity]{\includegraphics[width=0.33\linewidth]{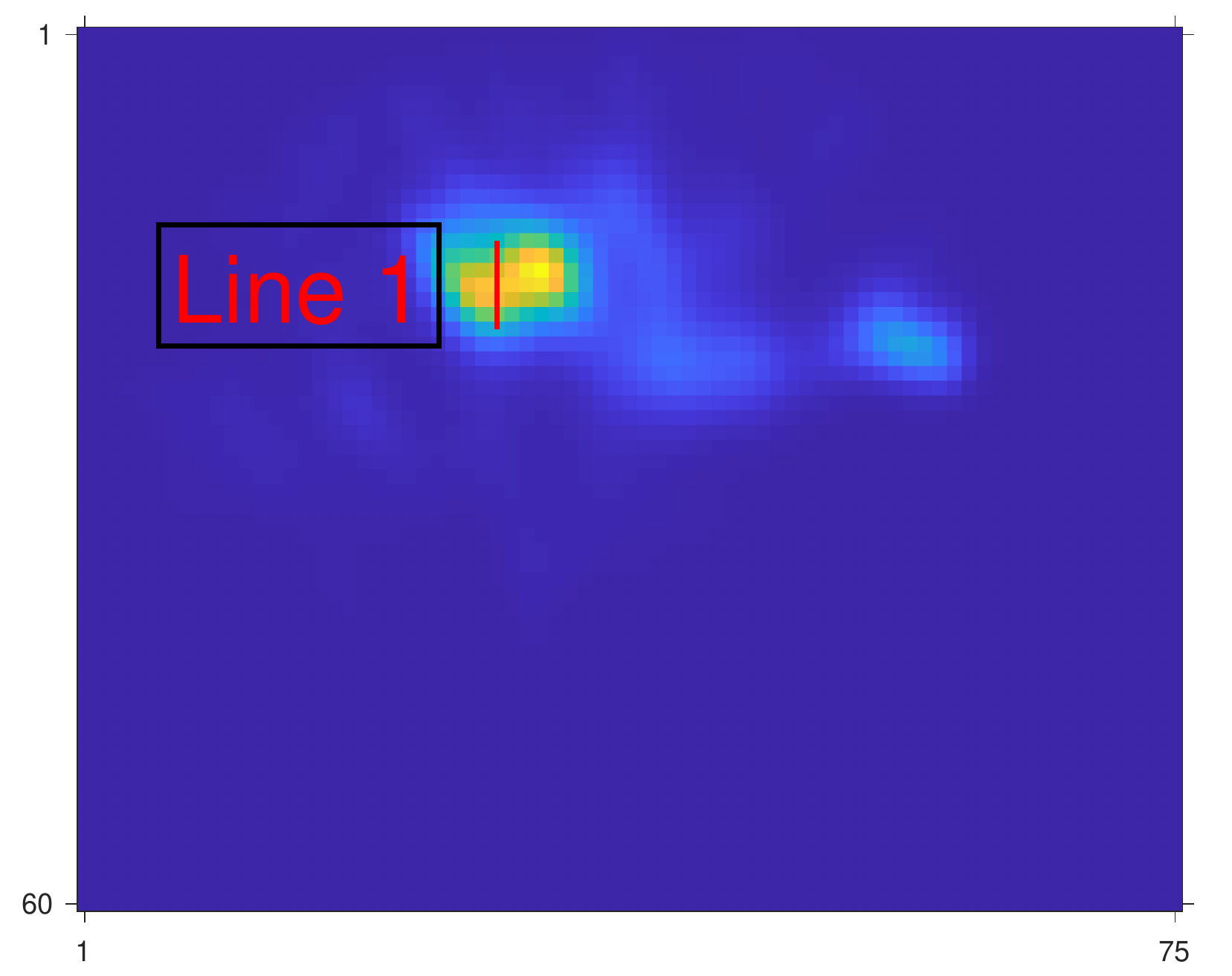}}
    \subfloat[Attenuation map]{\includegraphics[width=0.33\linewidth]{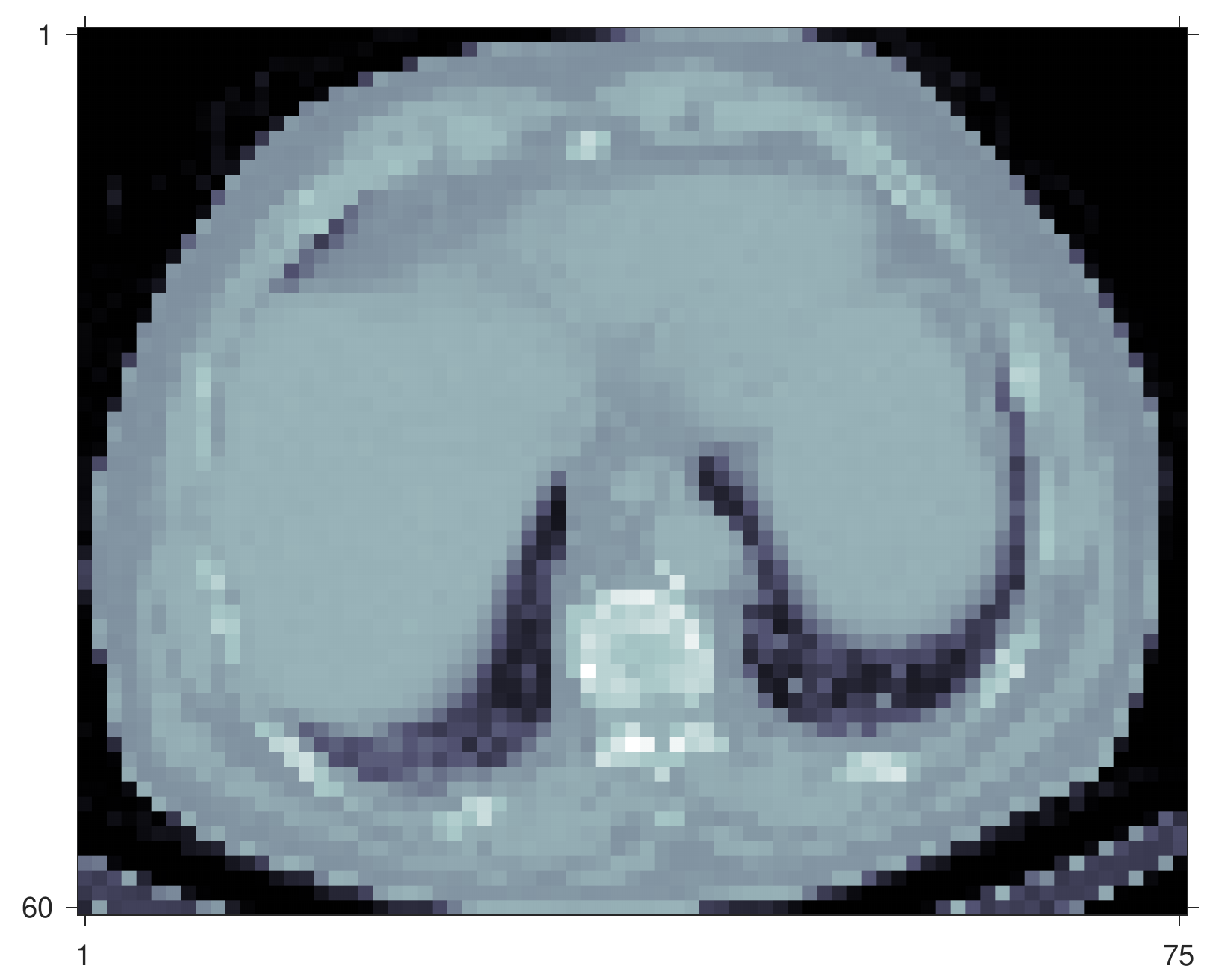}}
    \subfloat[OSEM]{\includegraphics[width=0.33\linewidth]{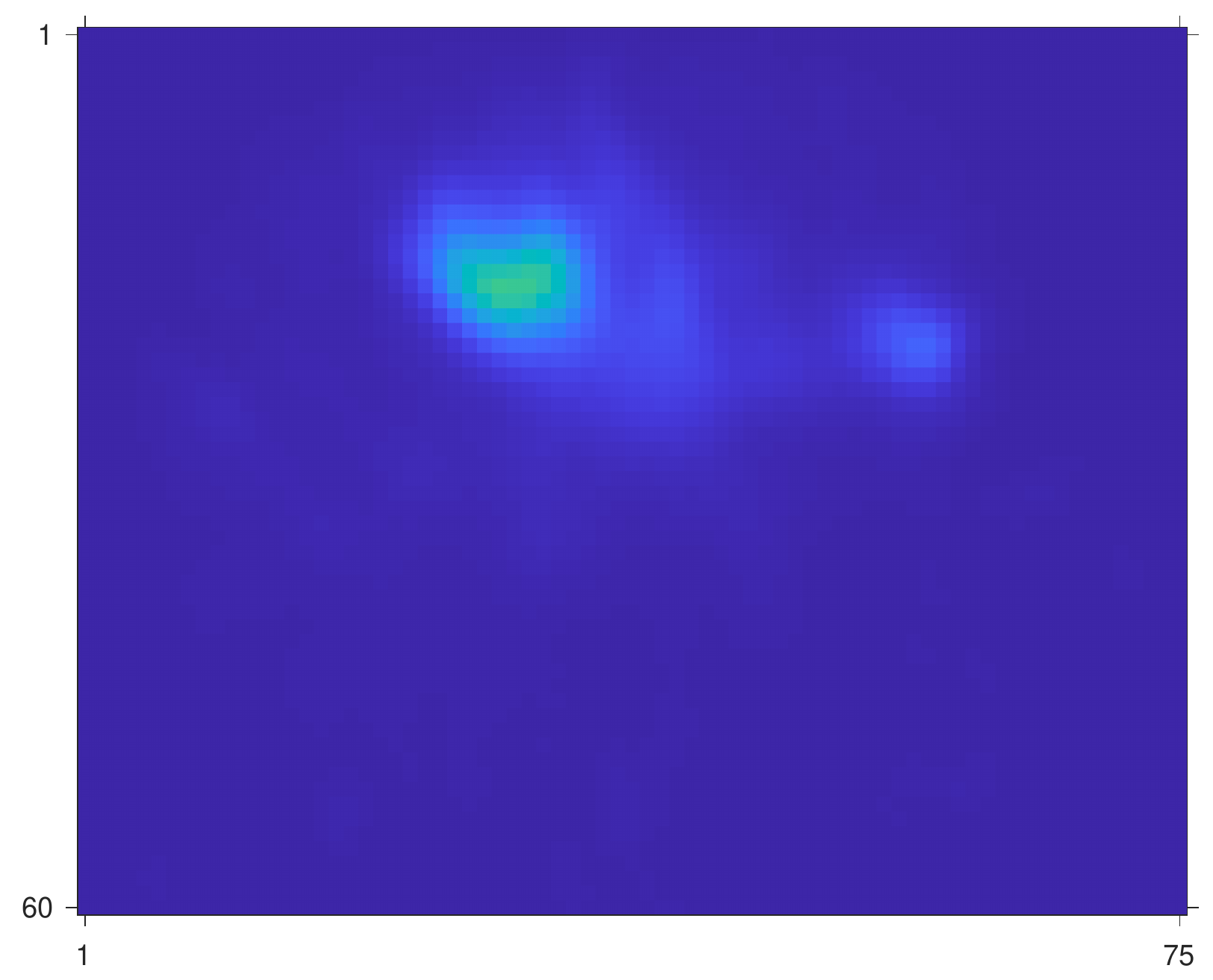}}
    \\
    \subfloat[Sequential]{\includegraphics[width=0.33\linewidth]{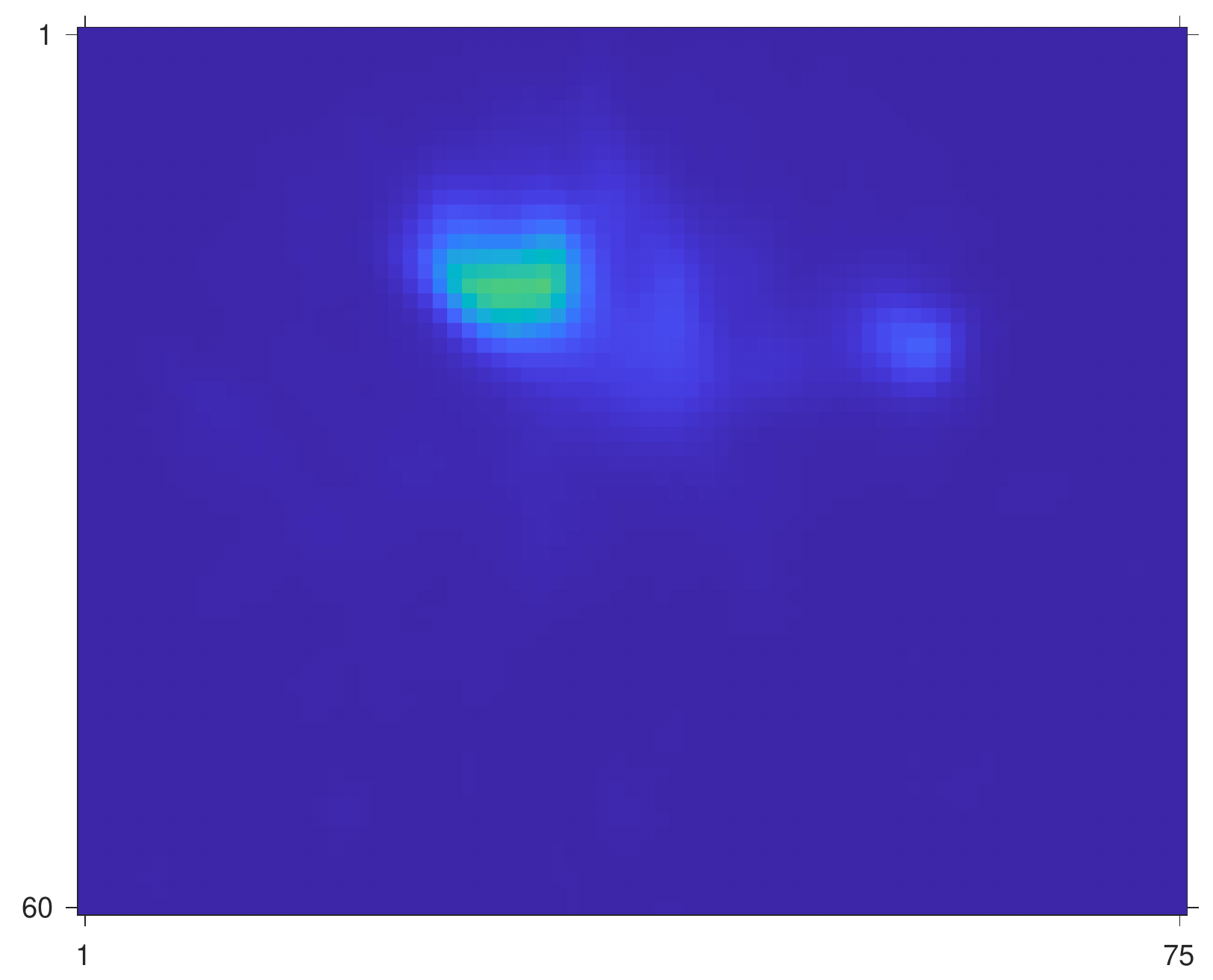}}
    \subfloat[Truncation]{\includegraphics[width=0.33\linewidth]{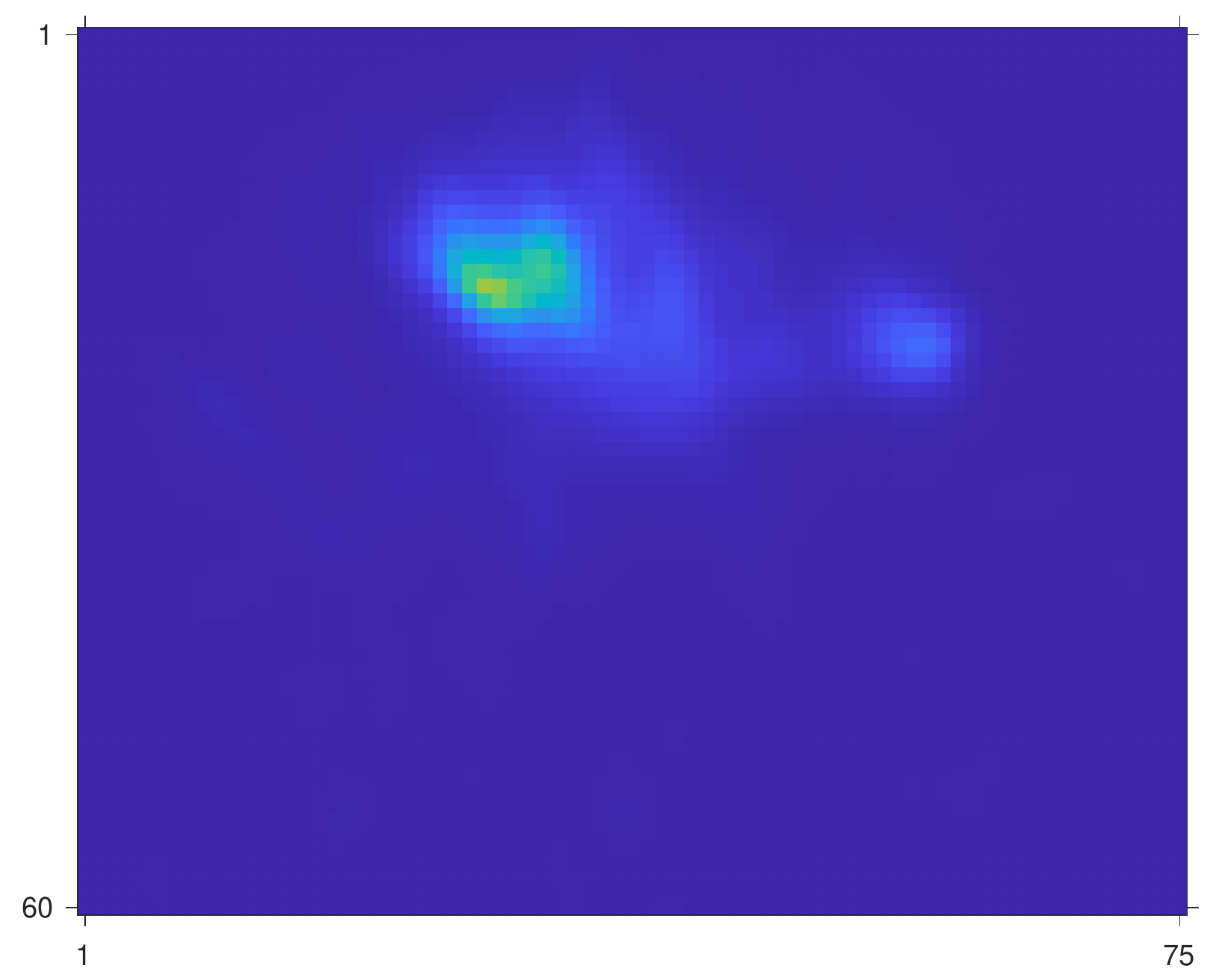}}
    \subfloat[End2end]{\includegraphics[width=0.33\linewidth]{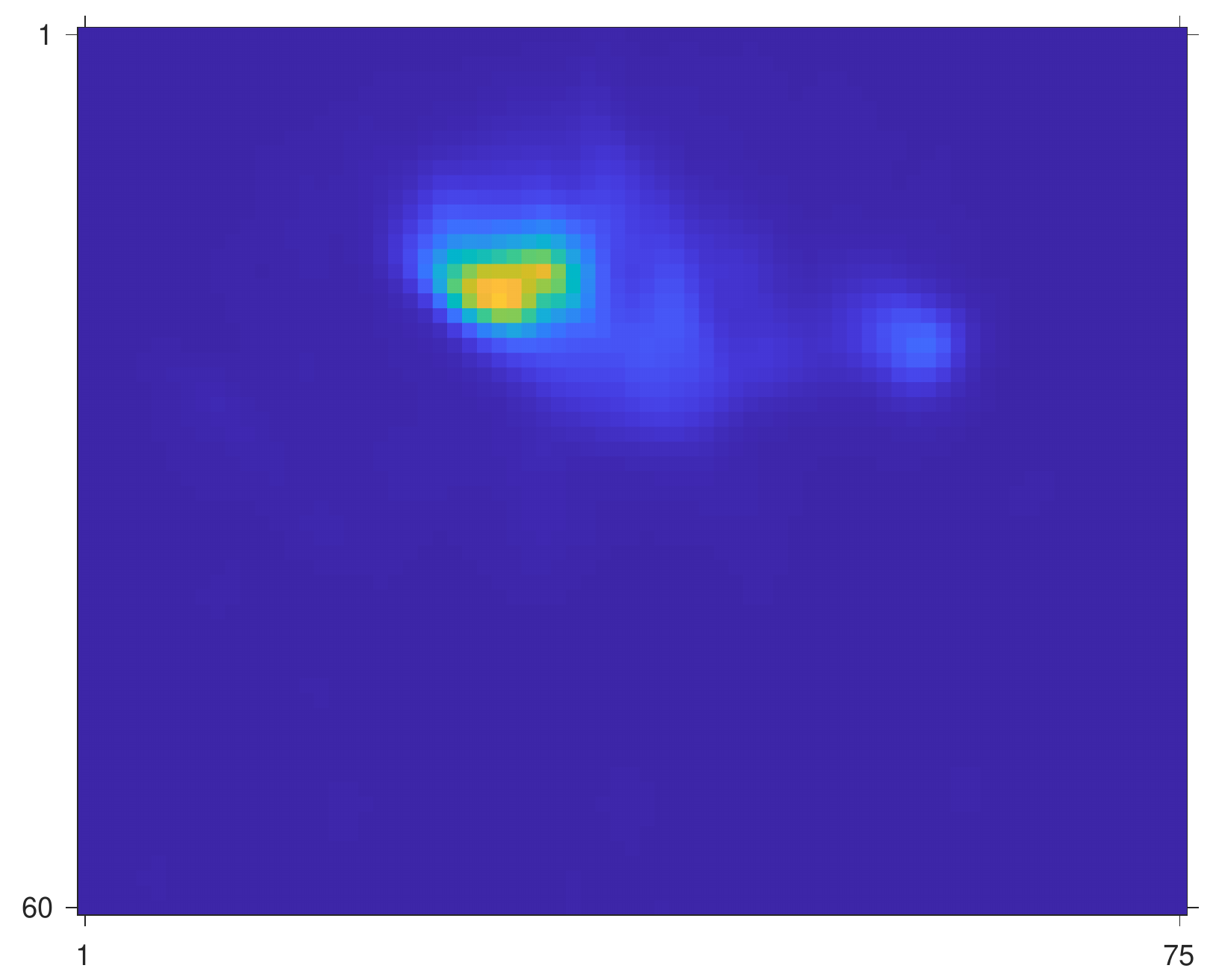}}
    \\
    \subfloat[True activity]{\includegraphics[width=0.33\linewidth]{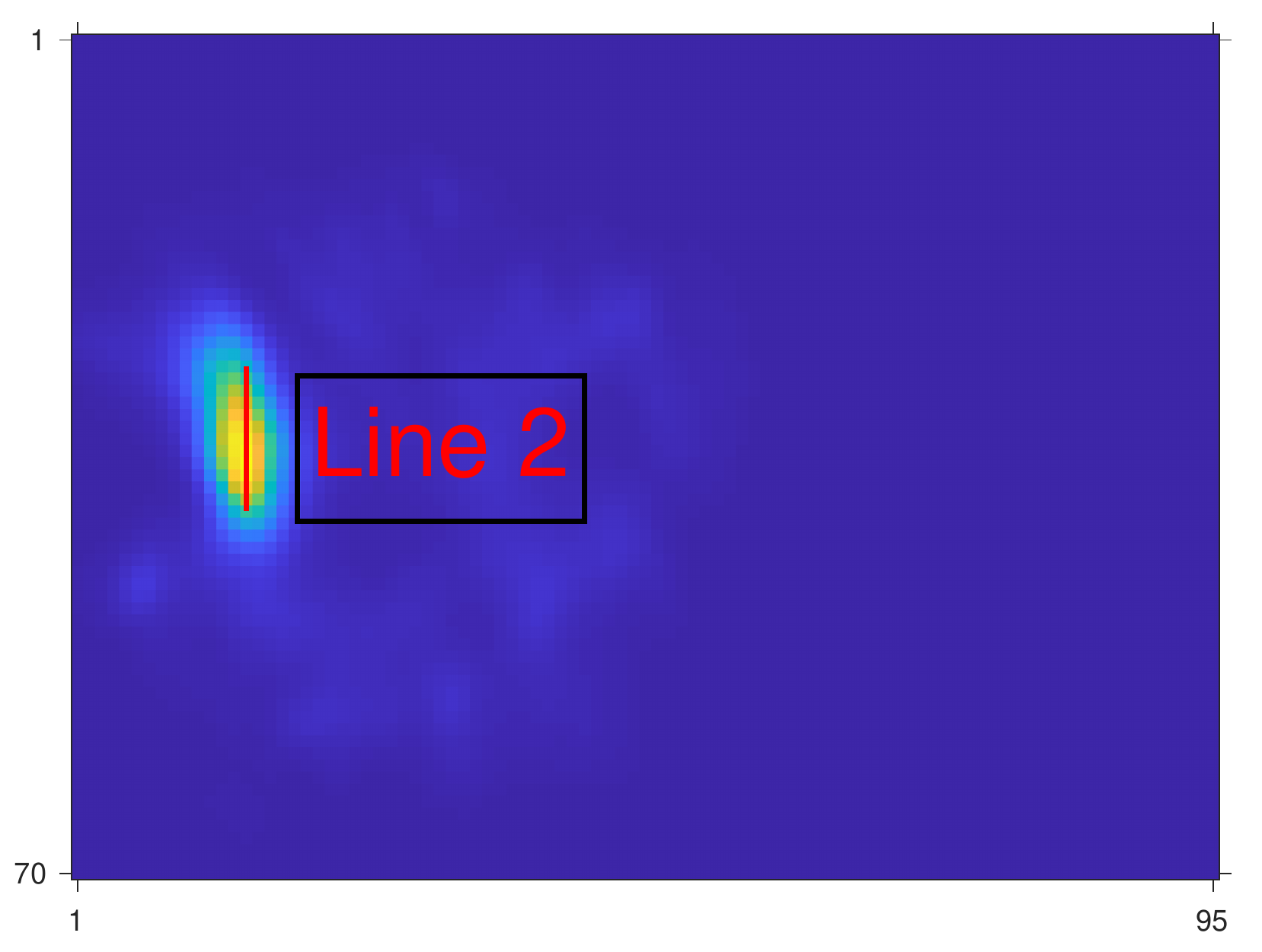}}
    \subfloat[Attenuation map]{\includegraphics[width=0.33\linewidth]{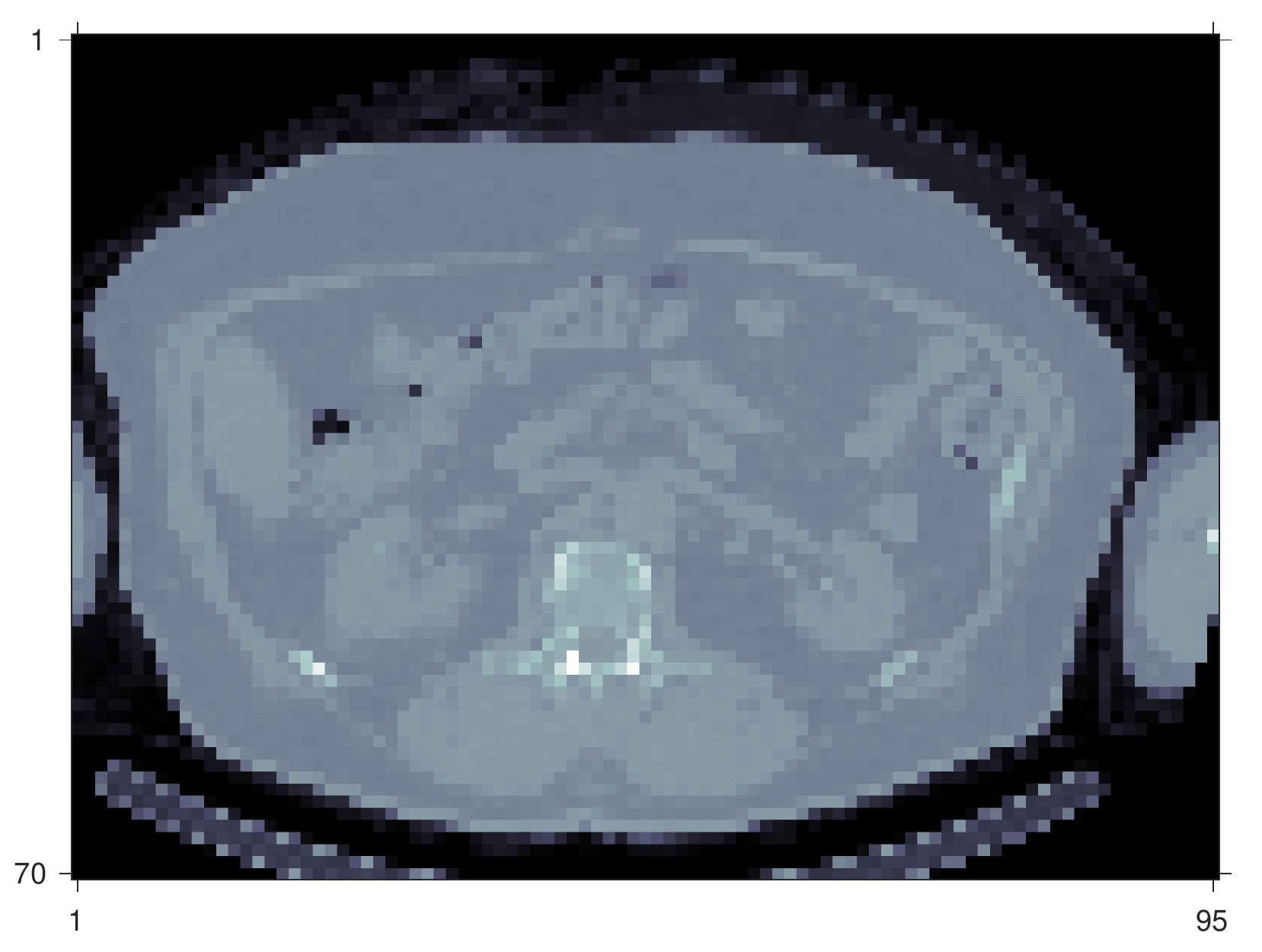}}
    \subfloat[OSEM]{\includegraphics[width=0.33\linewidth]{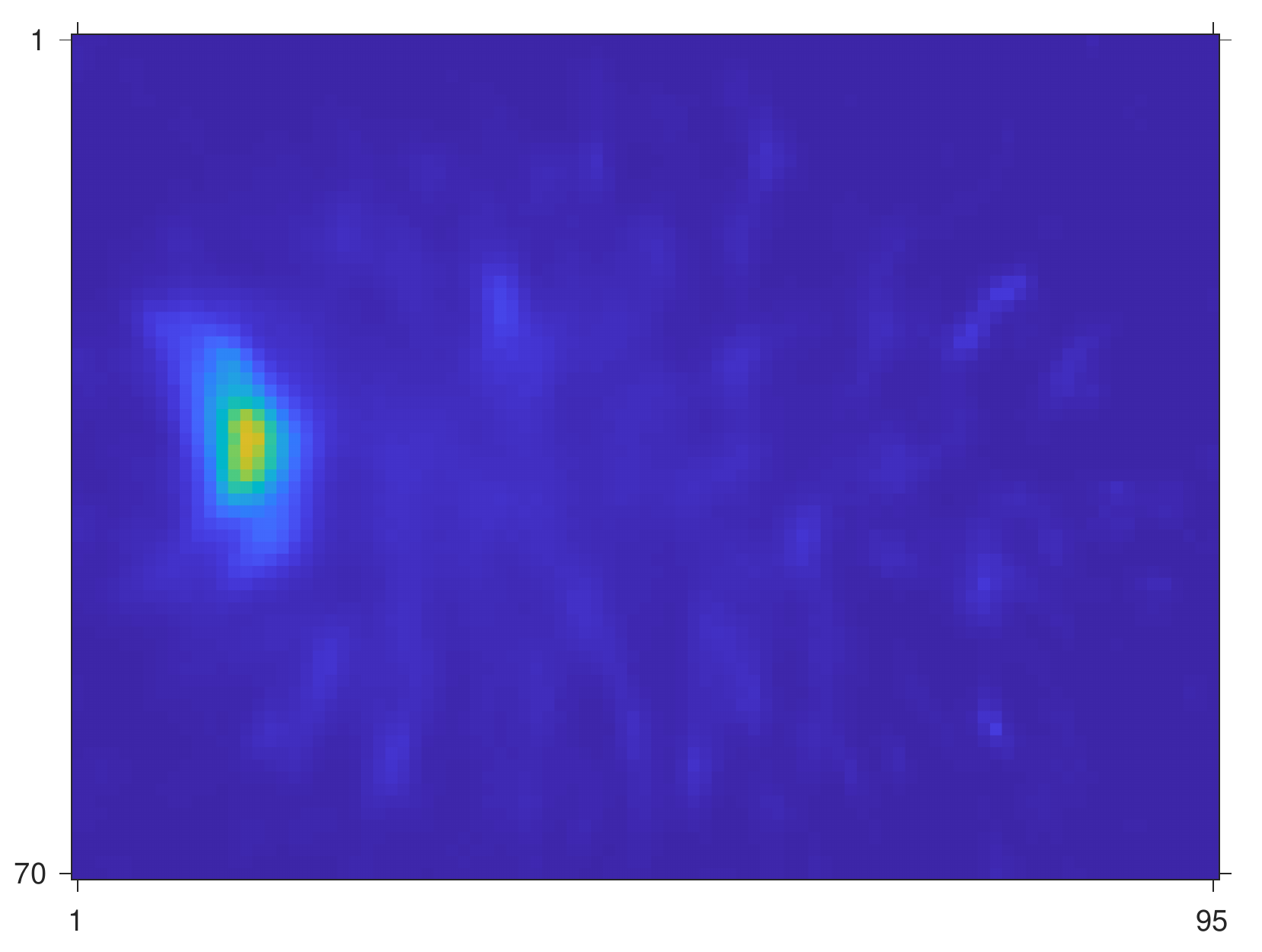}}
    \\
    \subfloat[Sequential]{\includegraphics[width=0.33\linewidth]{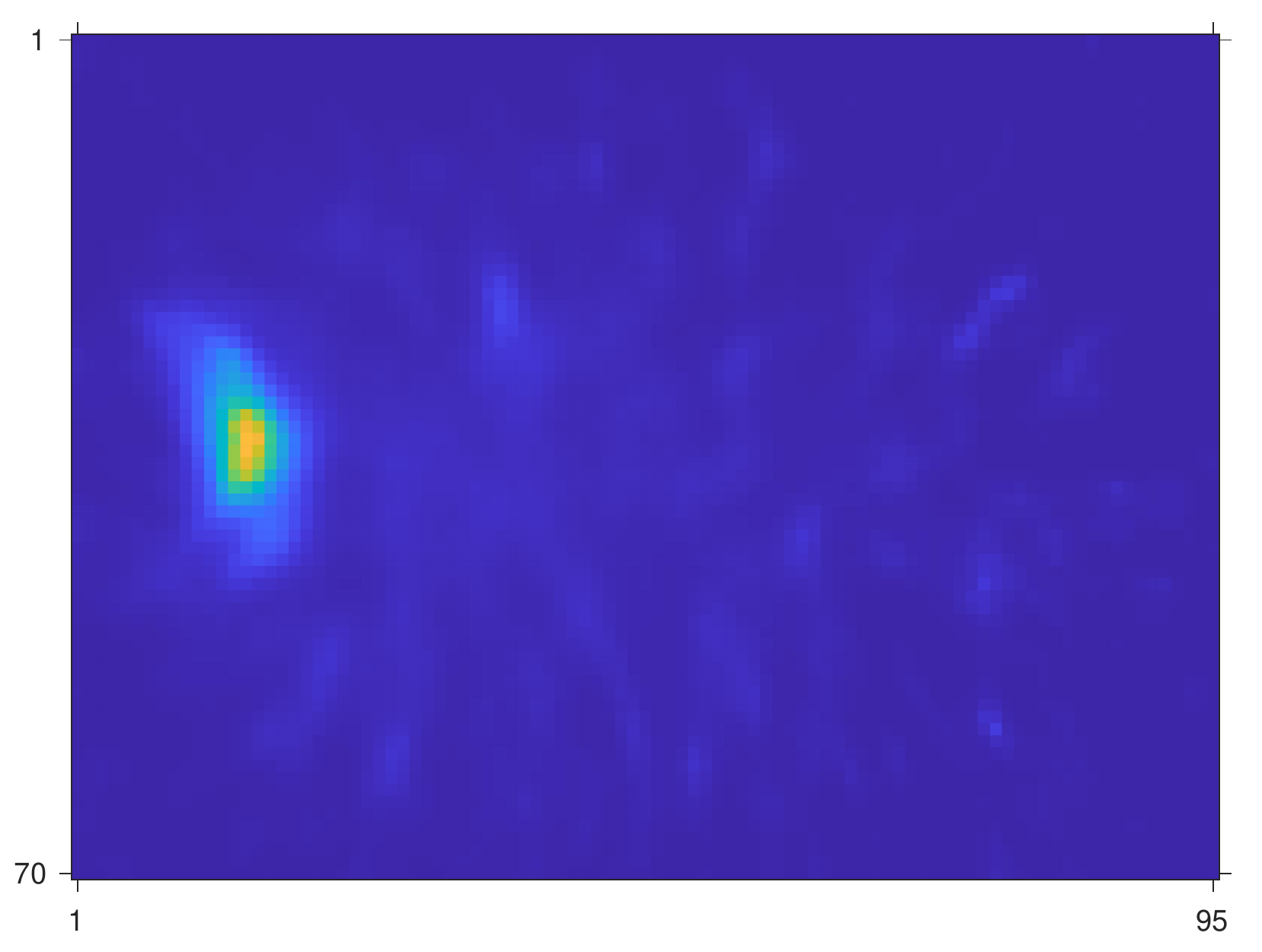}}
    \subfloat[Truncation]{\includegraphics[width=0.33\linewidth]{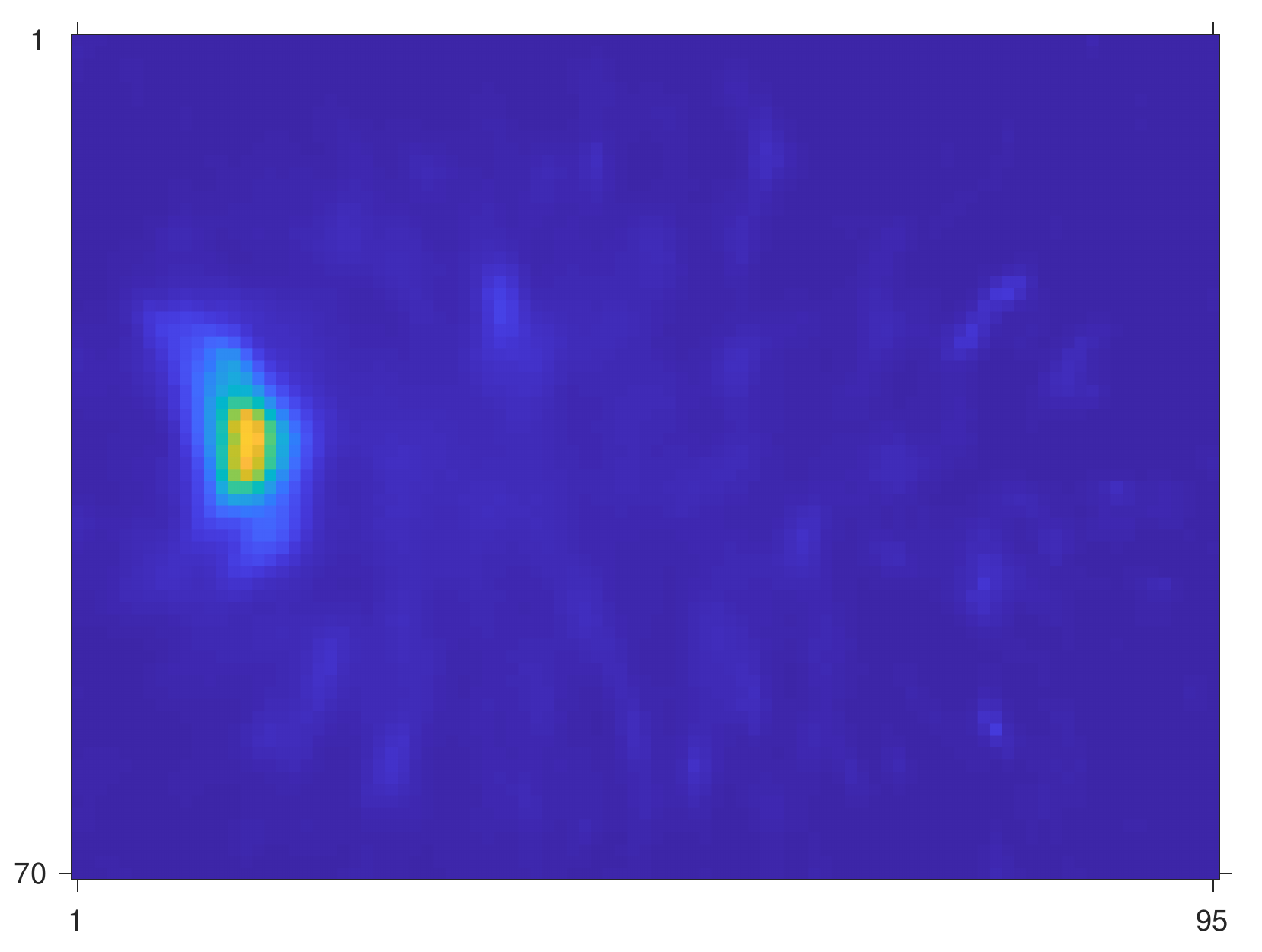}}
    \subfloat[End2end]{\includegraphics[width=0.33\linewidth]{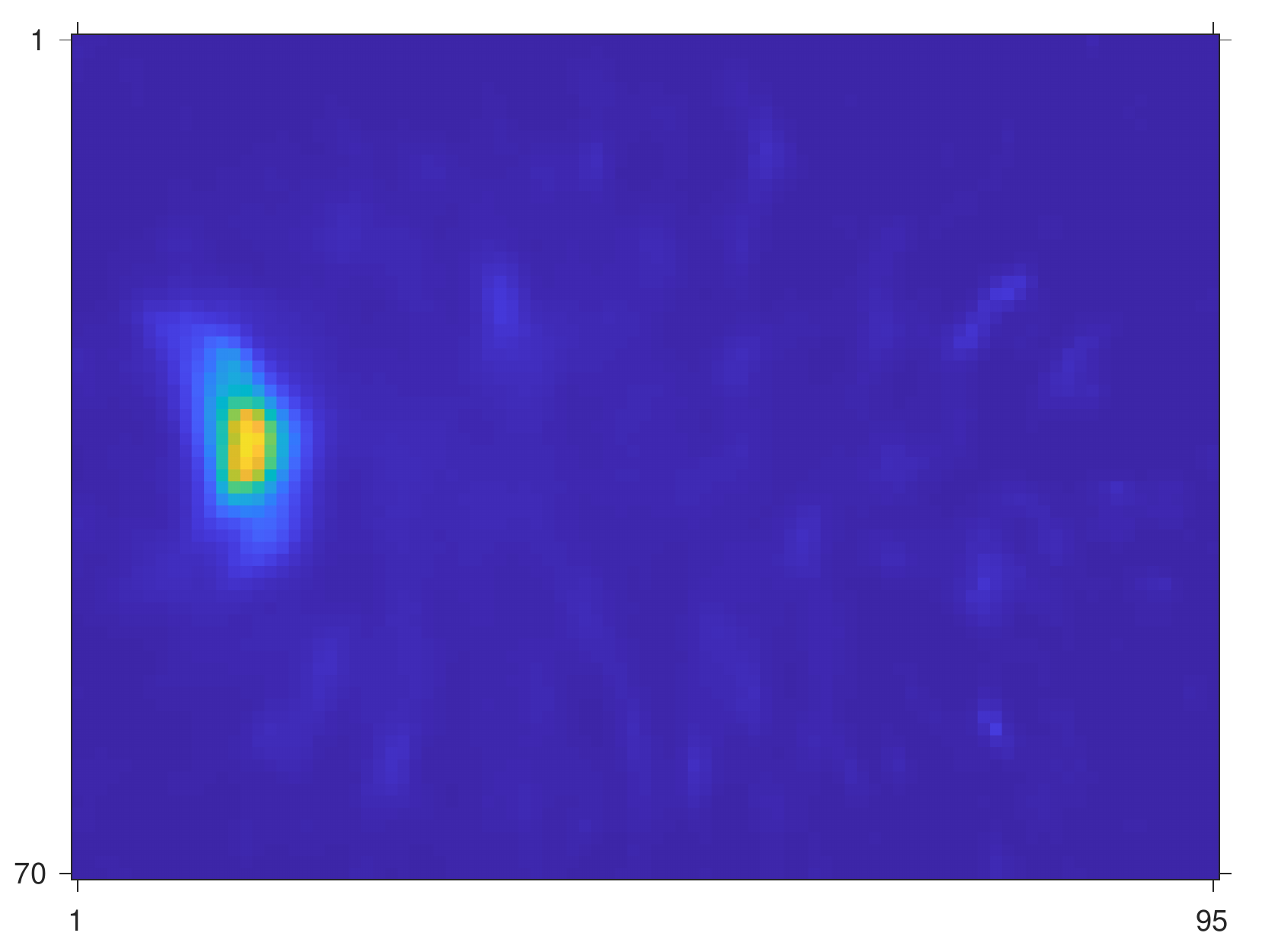}}
    \\
    \subfloat[Line 1 profile]{\includegraphics[width=0.5\linewidth]{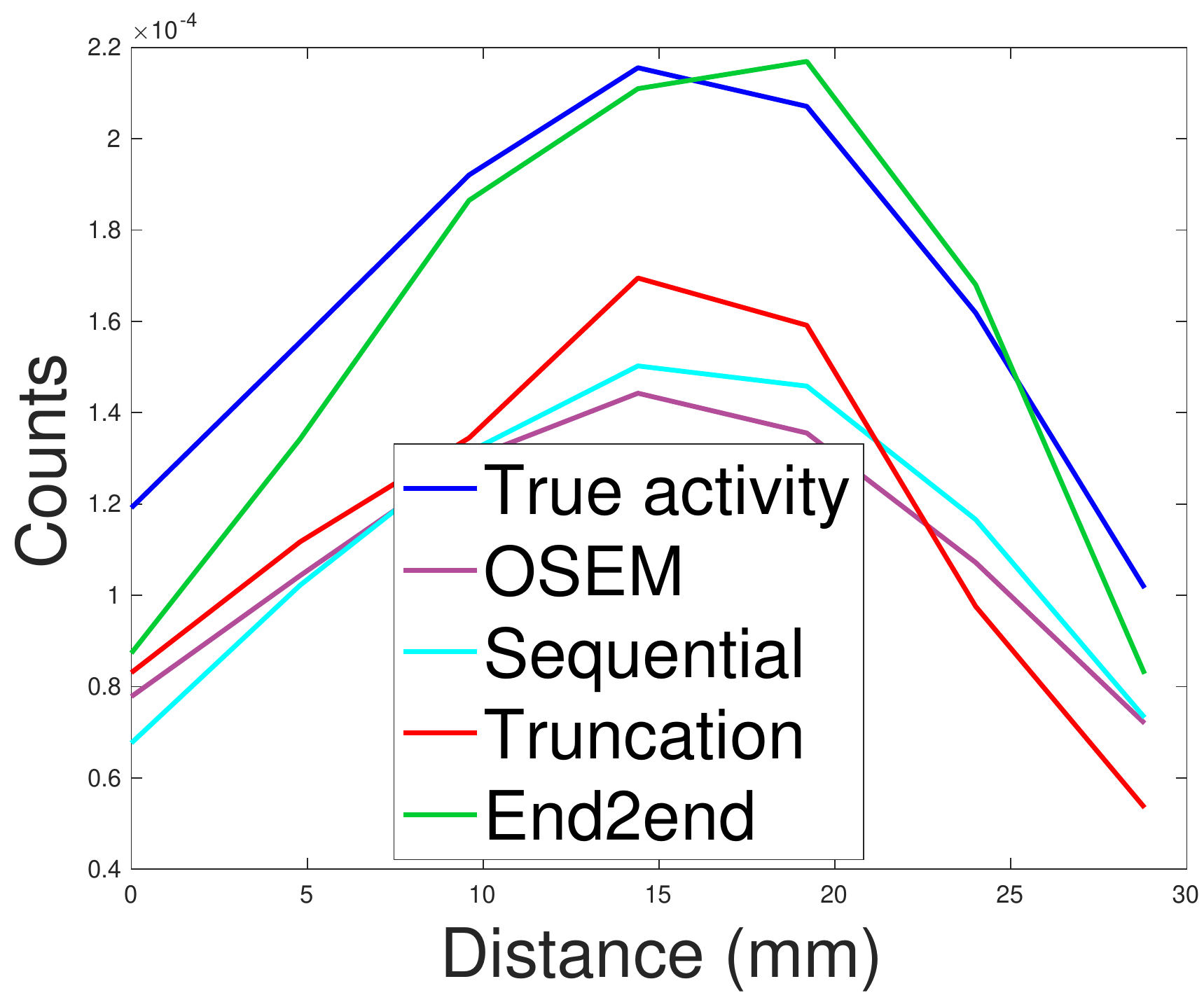}}
    \subfloat[Line 2 profile]{\includegraphics[width=0.5\linewidth]{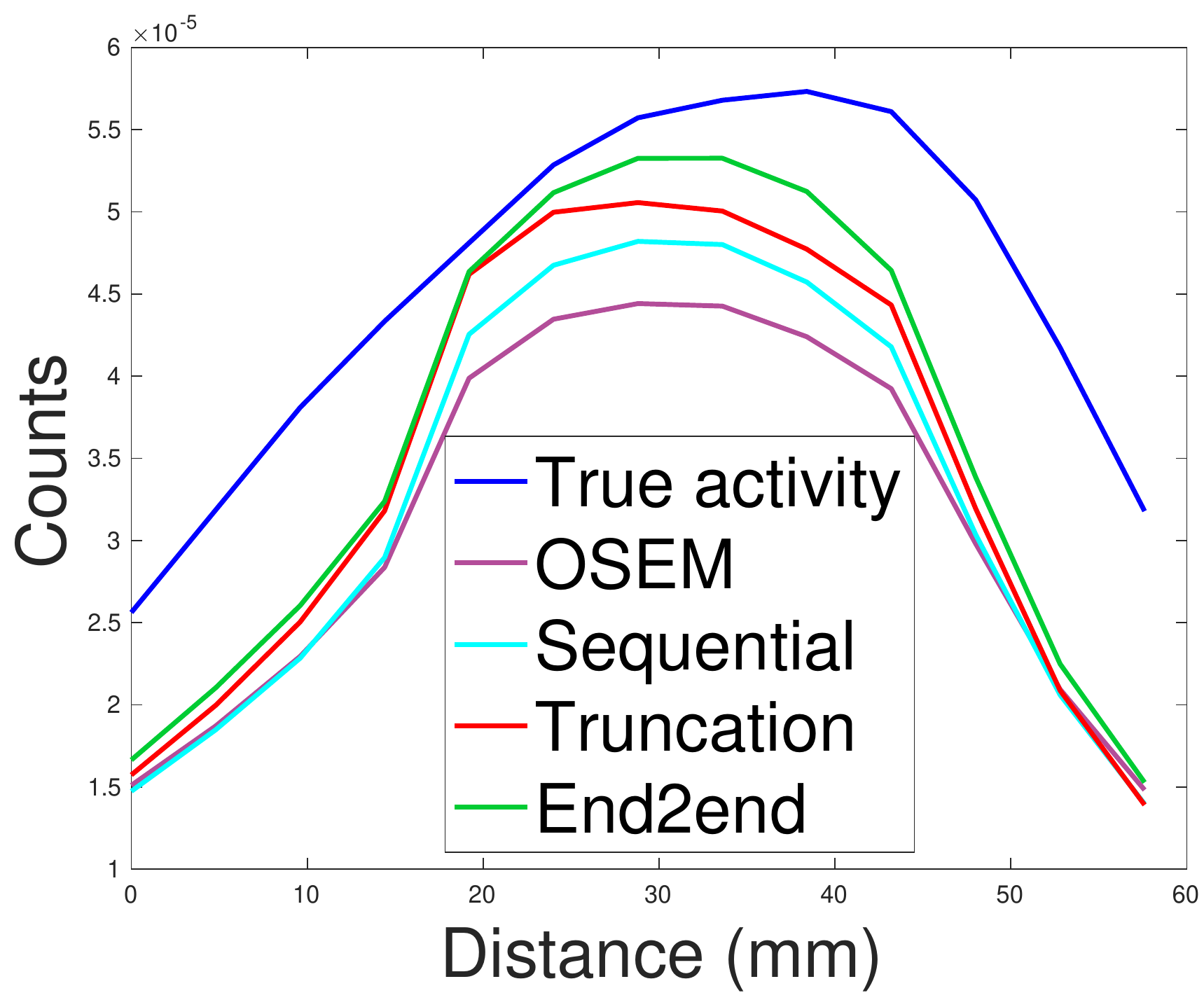}}
    \caption{Qualitative comparison of 
    different training methods and OSEM 
    tested on \ytt VP phantoms.
    Subfigure (a)-(f) and (g)-(l) show two slices
    from two testing phantoms.
    Subfigure (m) and (n) correspond to
    line profiles in (a) and (g), respectively.
    }
    \label{fig:y90_result}
\end{figure}

\fref{fig:y90_result} compares the 
reconstruction quality between
OSEM and
CNN-regularized EM algorithm
using sequential training, 
gradient truncation and end-to-end training.
Visually, the end-to-end training reconstruction 
yields the closest estimate to the true activity.
This is also evident through the line profiles
(subfigure (m) and (n))
across the tumor and the liver.

\tref{tab:mae,y90} reports the mean activity error (MAE)
and NRMSE for lesions and organs 
across all testing phantoms.
Similar to the qualitative assessment (\fref{fig:y90_result}),
the end-to-end training also 
produced lower errors
consistently across all testing lesions and organs.
For instance, compared to 
sequential training/gradient truncation,
the end-to-end training
relatively reduced MAE on average by
8.7\%/7.2\%, 18.5\%/11.0\% and 24.7\%/16.1\%
for lesion, healthy liver and lung, respectively.
The NRMSE was also relatively reduced by
6.1\%/3.8\%,
7.2\%/4.1\% and 6.1\%/3.0\%
for lesion, healthy liver and lung, respectively.
All learning-based methods
consistently had lower errors
than the OSEM method.

\begin{table}[hbt!]
\caption{The average($\pm$standard deviation) MAE(\%) and NRMSE(\%)
across 3 noise realizations of $^{90}$Y VP phantoms.}
    \centering
    \resizebox{\linewidth}{!}{
    \begin{tabular}{|c|c|c|c|c|}
    \hline
    \multicolumn{5}{|c|}{MAE(\%)}
    \\
    \hline
    Lesion/Organ & OSEM & Sequential & Truncation & End2end\\
    \hline
    Lesion (3-356mL) 
    & 
    32.5 $\pm$ 1.3
    & 
    25.3 $\pm$ 1.3
    & 
    24.9 $\pm$ 1.0
    & 
    \textbf{23.1} $\pm$ 1.8
    \\
    \hline
    Liver 
    & 
    25.0 $\pm$ 0.1
    & 
    18.7 $\pm$ 0.1
    & 
    17.8 $\pm$ 1.3
    & 
    \textbf{15.6} $\pm$ 3.6
    \\
    \hline
    Healthy liver 
    & 
    25.1 $\pm$ 0.2
    & 
    23.8 $\pm$ 0.5
    & 
    21.8 $\pm$ 1.2
    & 
    \textbf{19.4} $\pm$ 3.1
    \\
    \hline
    Lung 
    & 
    88.4 $\pm$ 2.1
    & 
    64.9 $\pm$ 1.6
    & 
    58.3 $\pm$ 6.6
    & 
    \textbf{48.9} $\pm$ 8.4
    \\
    \hline
    \multicolumn{5}{|c|}{NRMSE(\%)}
    \\
    \hline
    Lesion/Organ & OSEM & Sequential & Truncation & End2end\\
    \hline
    Lesion (3-356mL) 
    & 
    35.3 $\pm$ 1.5
    & 
    29.6 $\pm$ 1.4
    & 
    28.9 $\pm$ 1.1
    & 
    \textbf{27.8} $\pm$ 1.2
    \\
    \hline
    Liver 
    & 
    29.9 $\pm$ 0.4
    & 
    22.7 $\pm$ 0.1
    & 
    22.1 $\pm$ 0.9
    & 
    \textbf{21.2} $\pm$ 1.5
    \\
    \hline
    Healthy liver 
    & 
    31.6 $\pm$ 0.4
    & 
    27.9 $\pm$ 0.3
    & 
    27.0 $\pm$ 0.9
    & 
    \textbf{25.9} $\pm$ 2.0
    \\
    \hline
    Lung 
    & 
    62.4 $\pm$ 1.3
    & 
    59.2 $\pm$ 1.1
    & 
    57.3 $\pm$ 3.0
    & 
    \textbf{55.6} $\pm$ 4.6
    \\
    \hline
    \end{tabular}}
    \label{tab:mae,y90}
\end{table}

\subsection{Results at intermediate iterations}

\begin{figure}[hbt!]
\centering
\subfloat[True activity]{\includegraphics[width=0.33\linewidth]{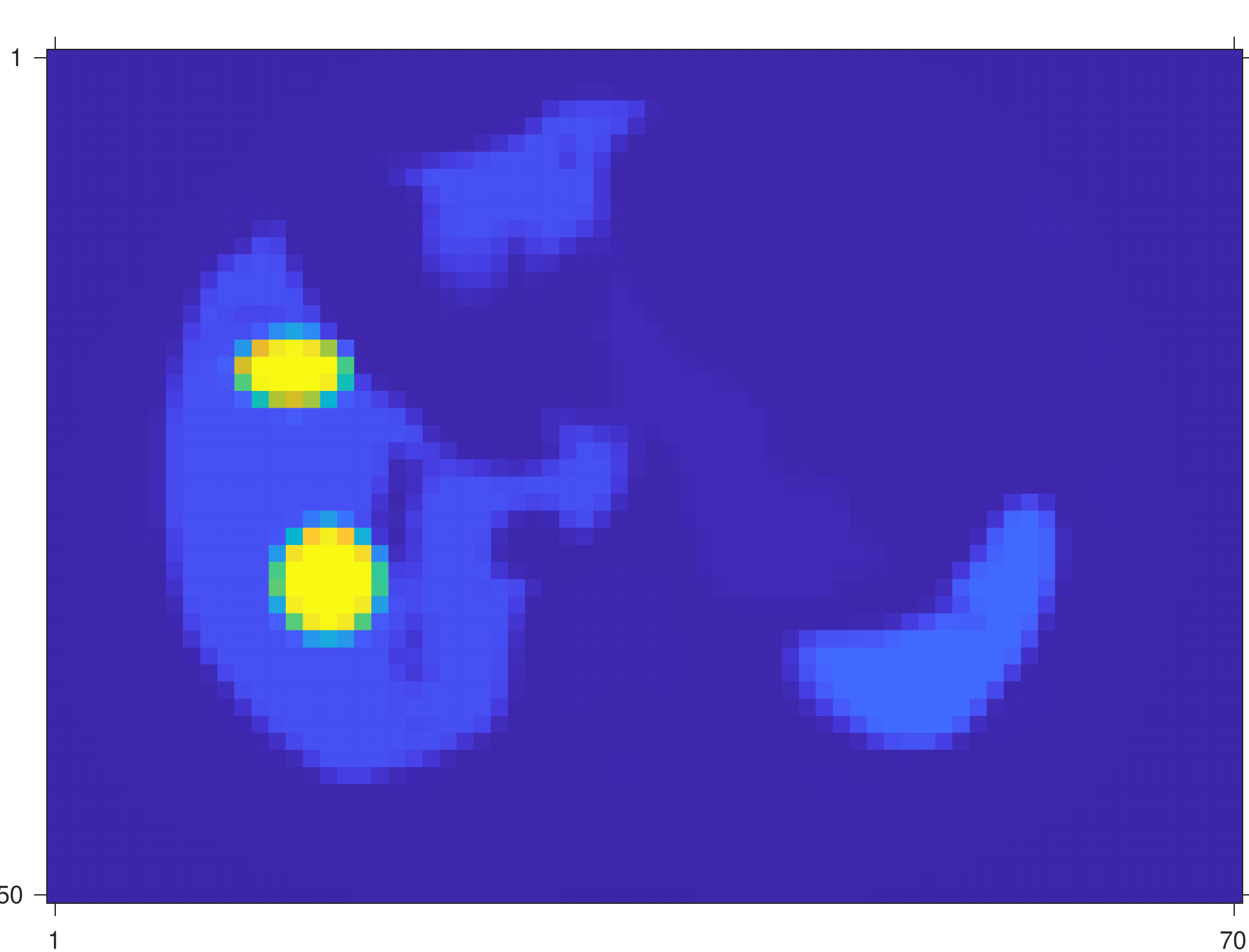}}
\subfloat[Attenuation map]{\includegraphics[width=0.33\linewidth]{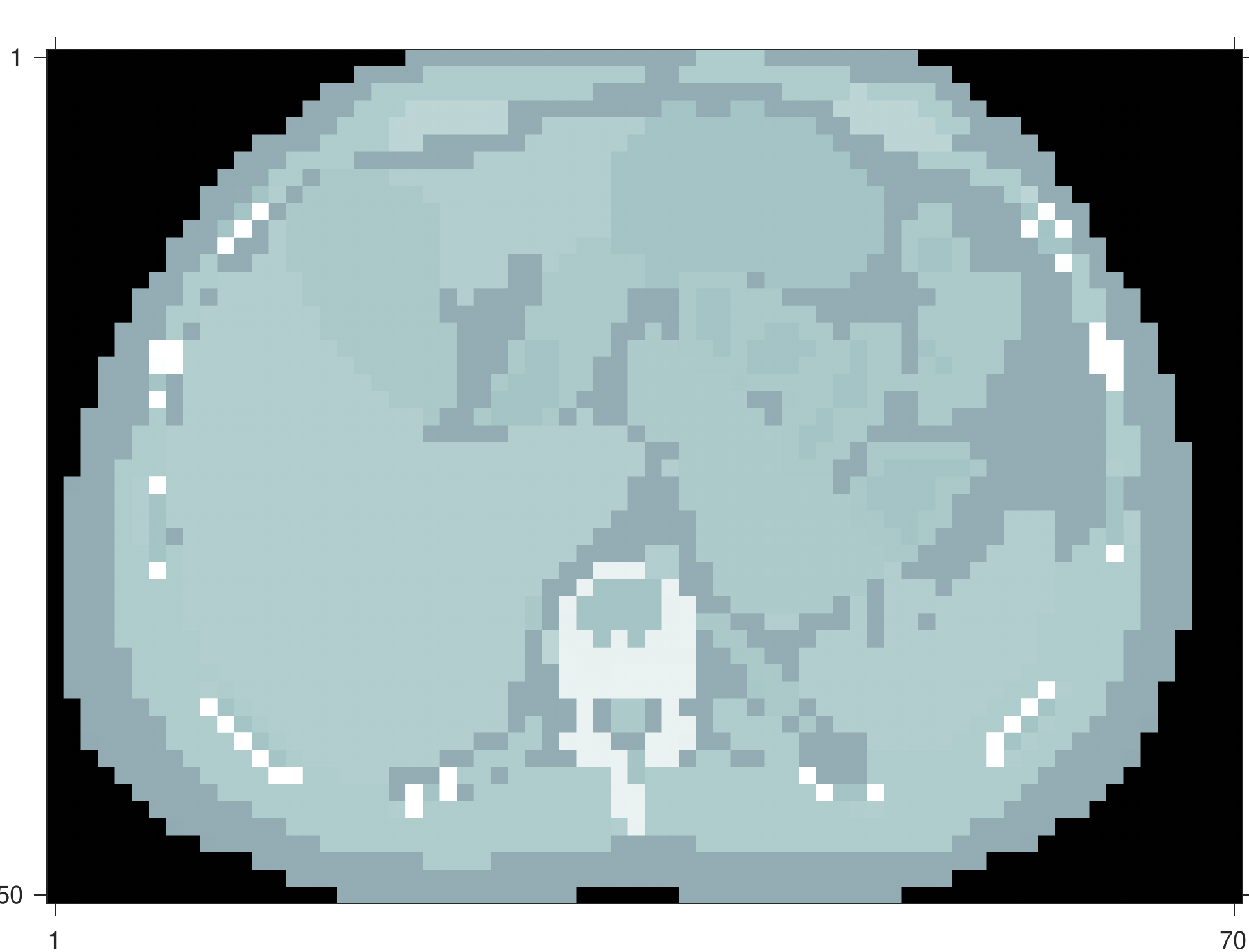}}
\subfloat[OSEM]{\includegraphics[width=0.33\linewidth]{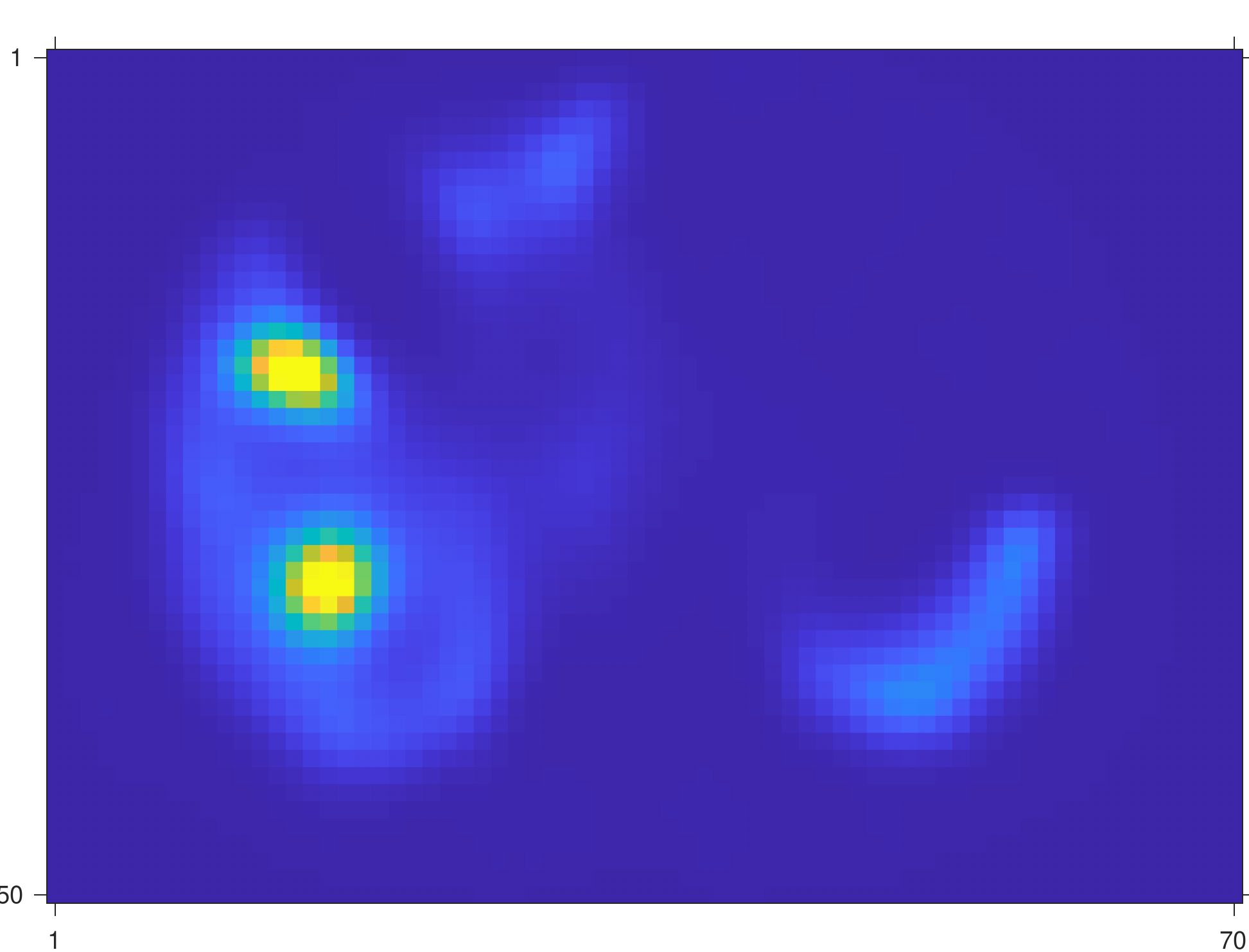}}
\\
\subfloat[Seq-iter1]{\includegraphics[width=0.33\linewidth]{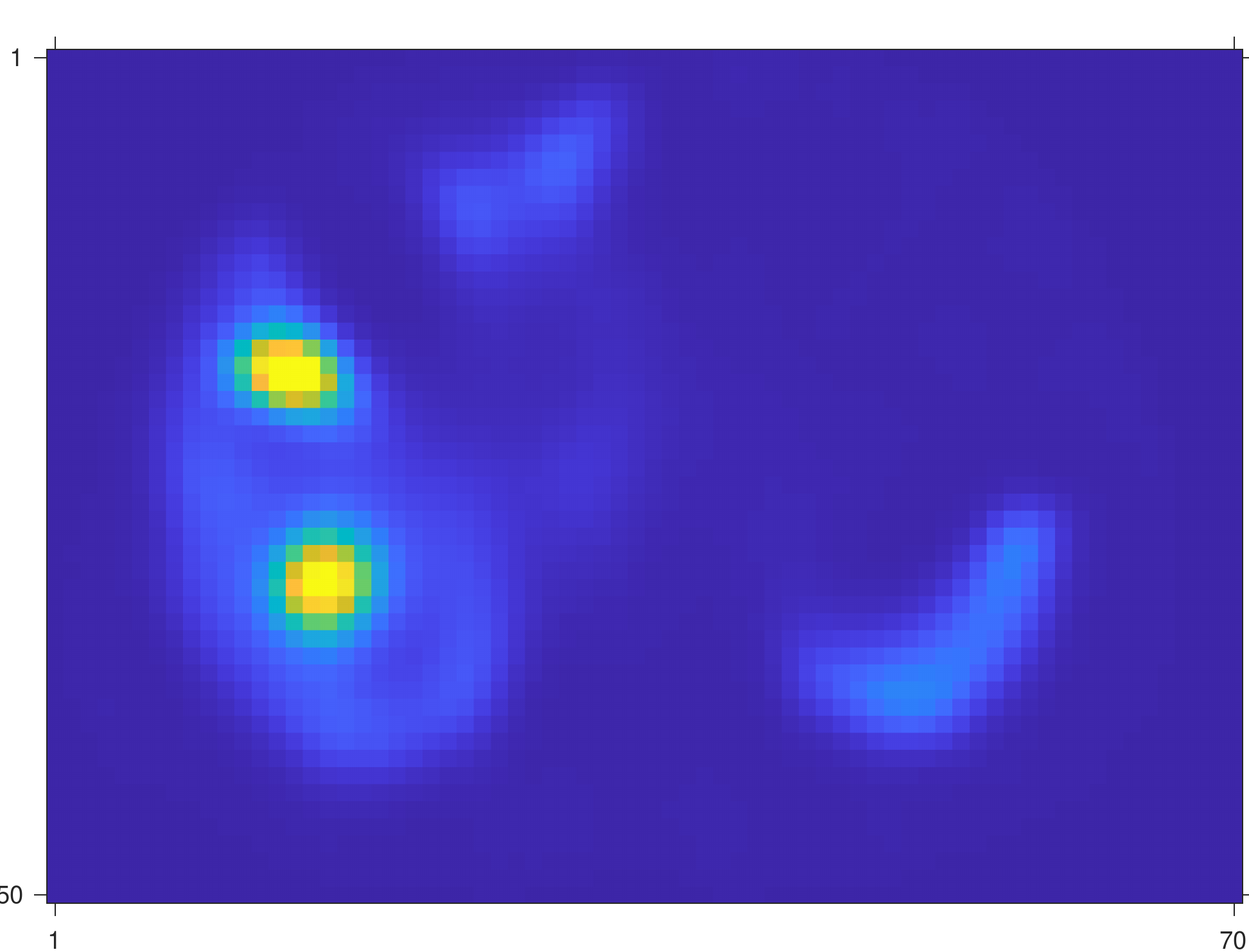}}
\subfloat[Seq-iter2]{\includegraphics[width=0.33\linewidth]{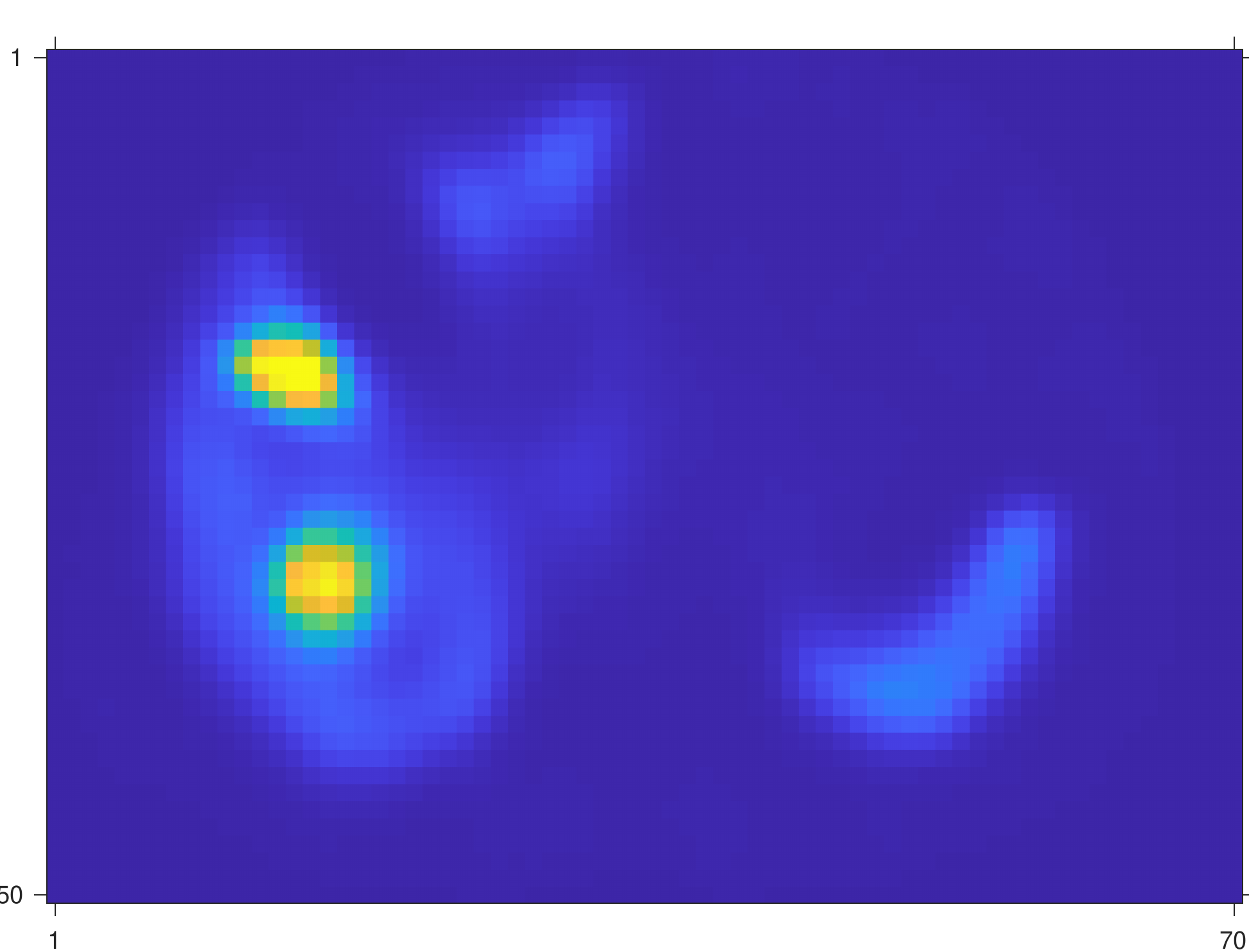}}
\subfloat[Seq-iter3]{\includegraphics[width=0.33\linewidth]{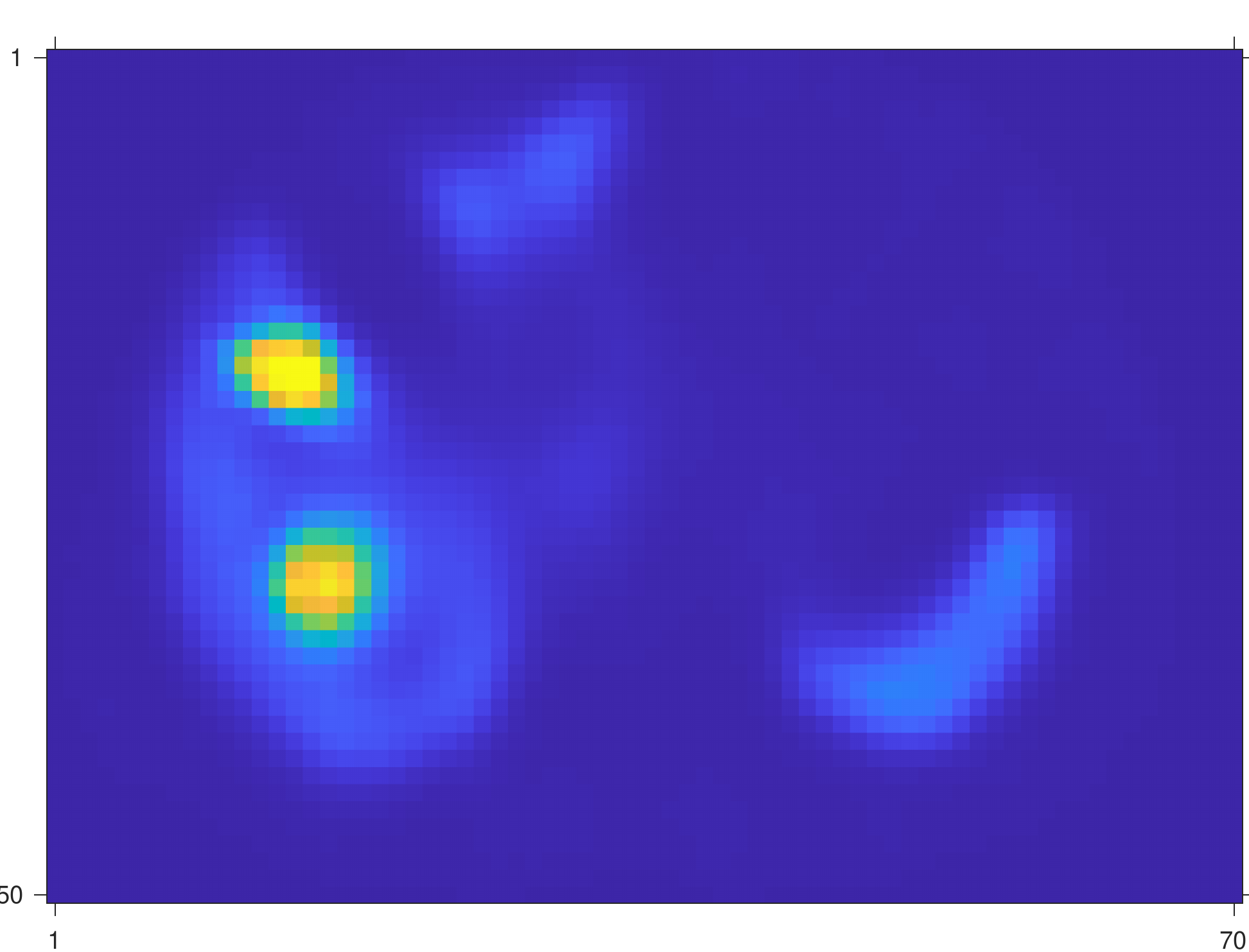}}
\\
\subfloat[Trun-iter1]{\includegraphics[width=0.33\linewidth]{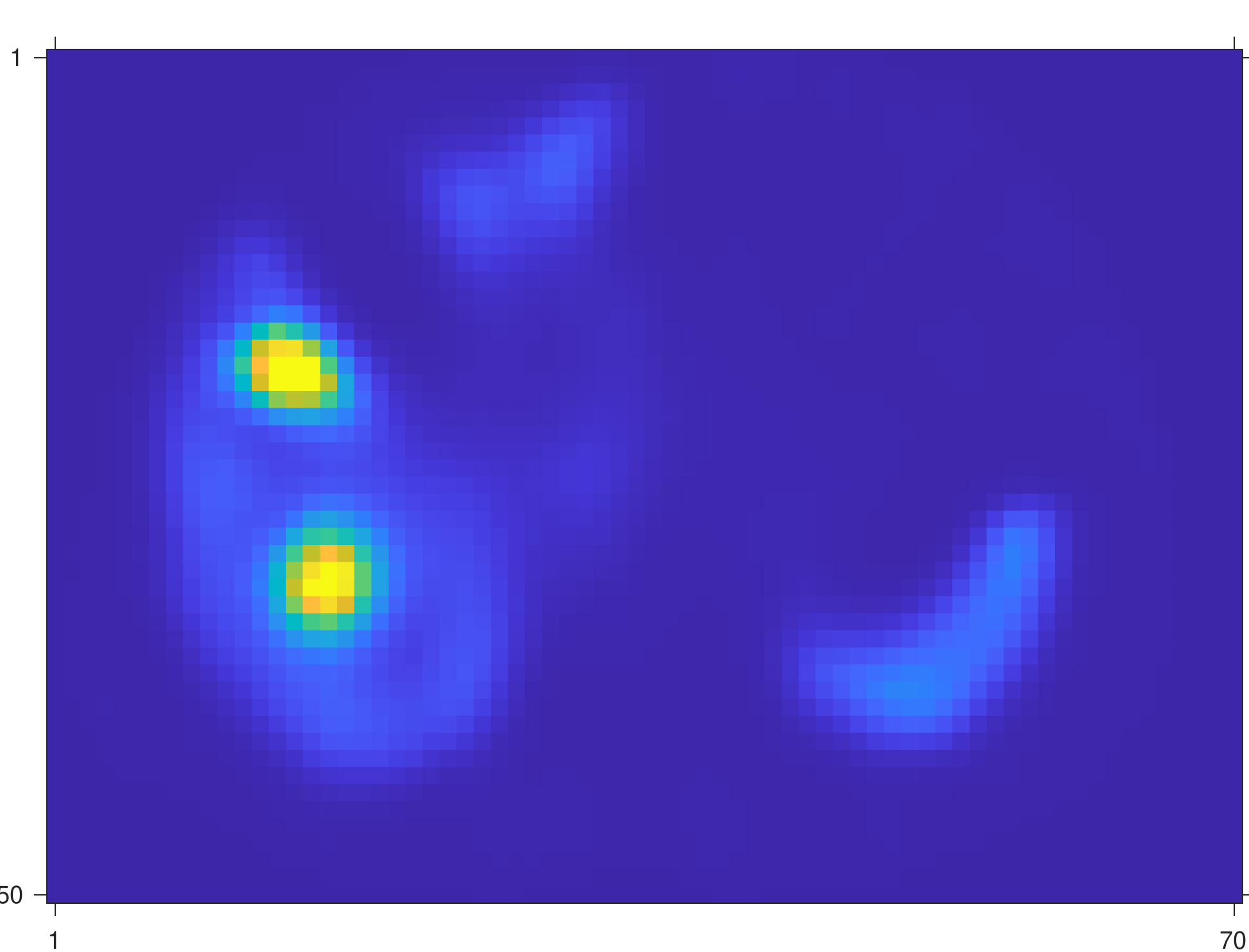}}
\subfloat[Trun-iter2]{\includegraphics[width=0.33\linewidth]{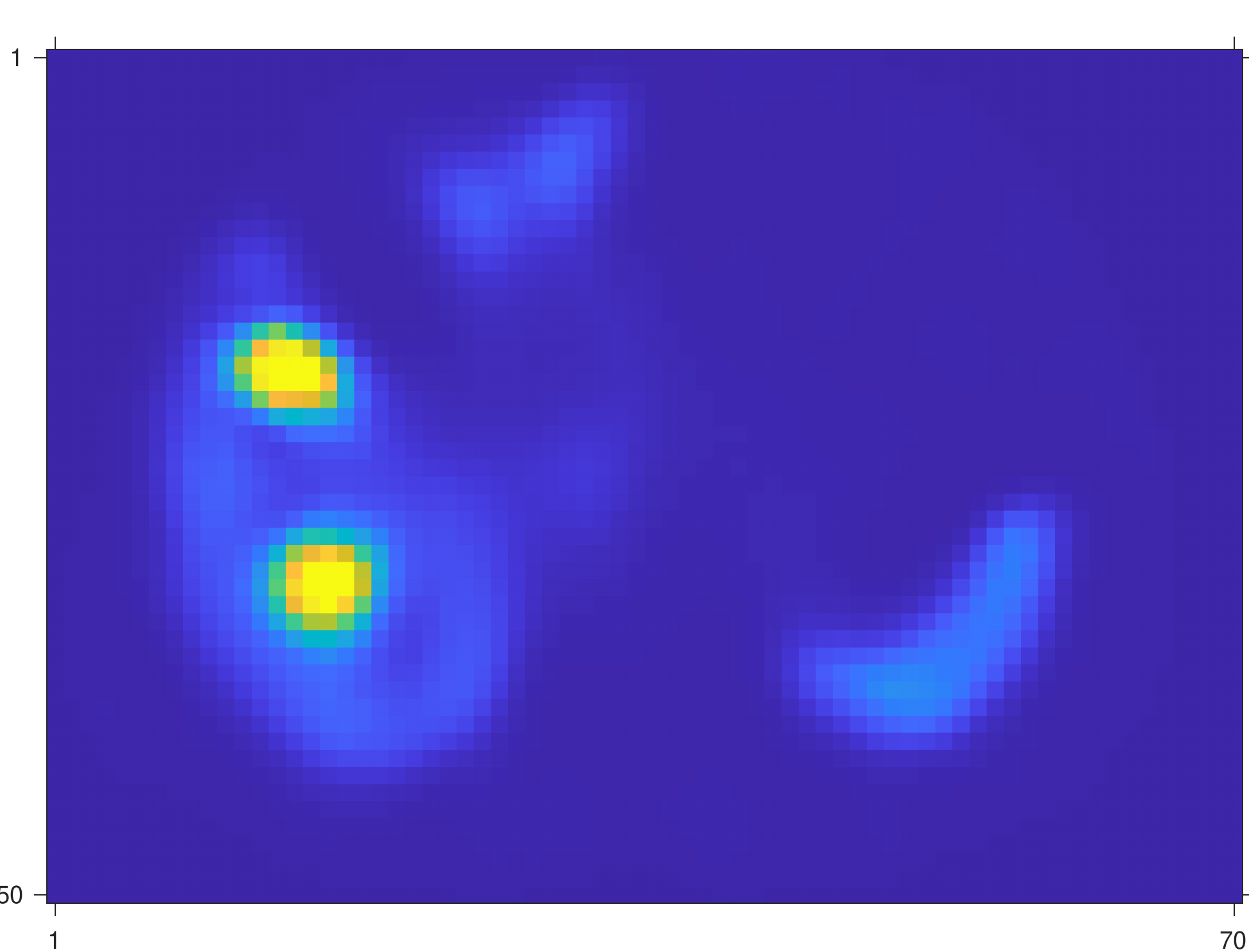}}
\subfloat[Trun-iter3]{\includegraphics[width=0.33\linewidth]{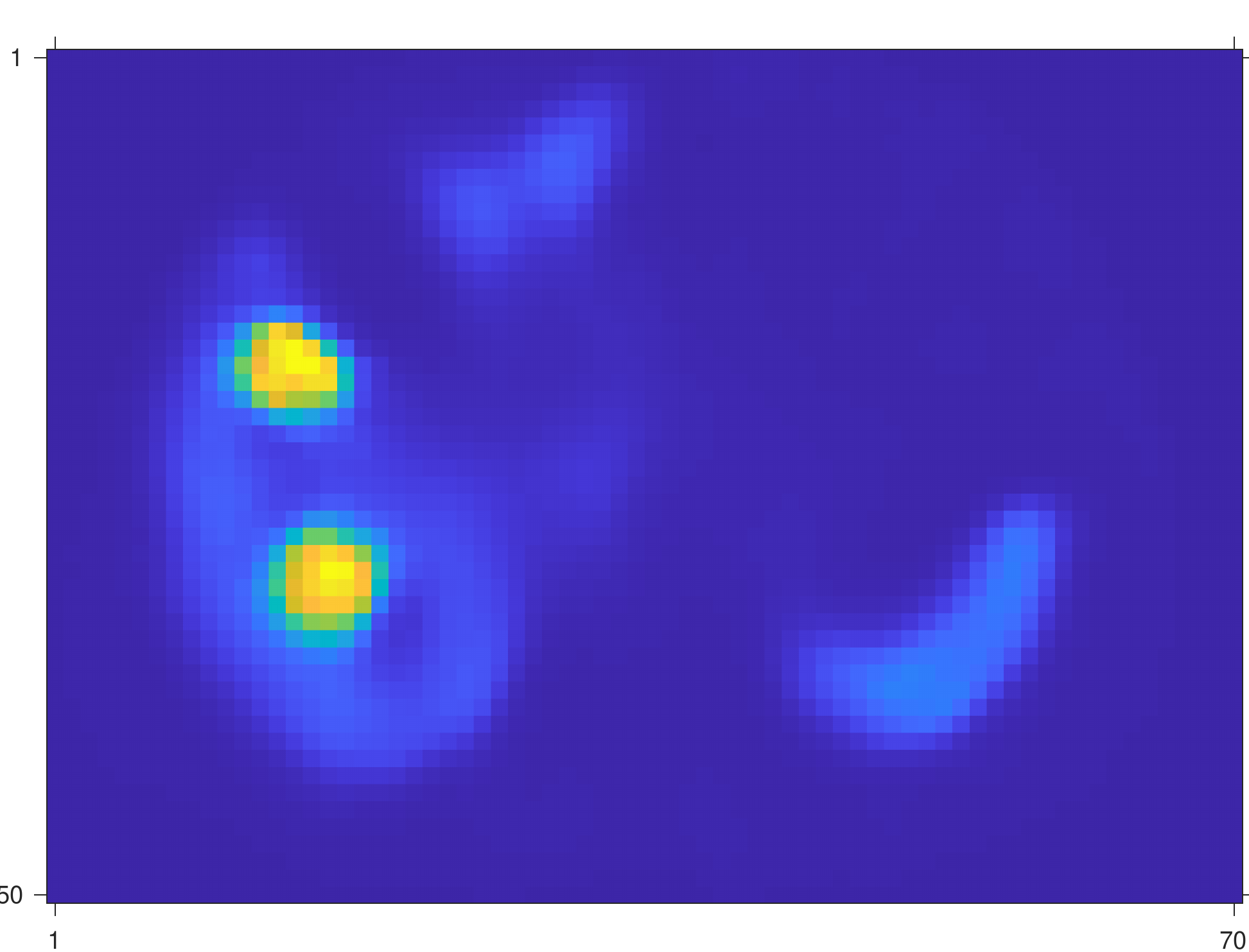}}
\\
\subfloat[End-iter1]{\includegraphics[width=0.33\linewidth]{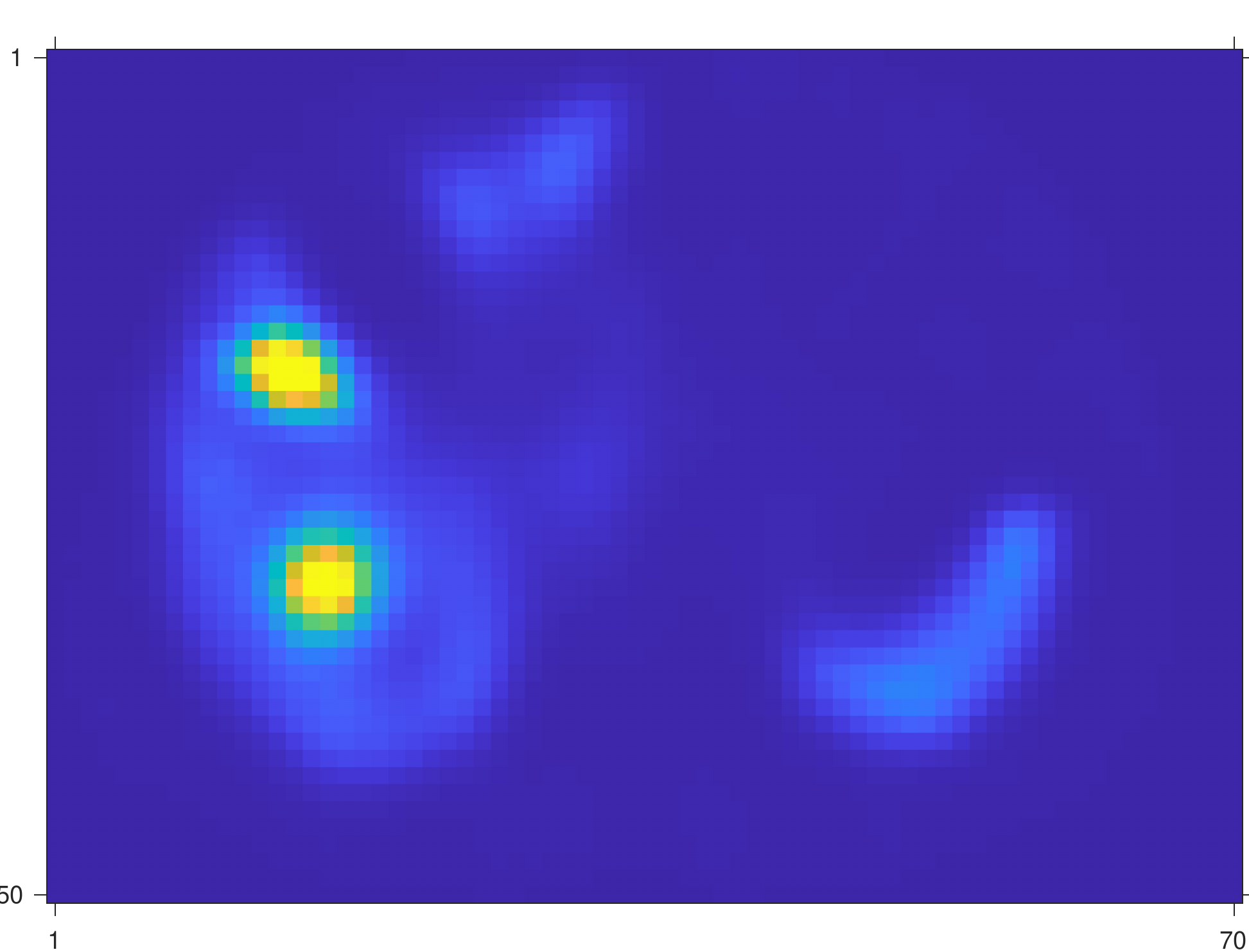}}
\subfloat[End-iter2]{\includegraphics[width=0.33\linewidth]{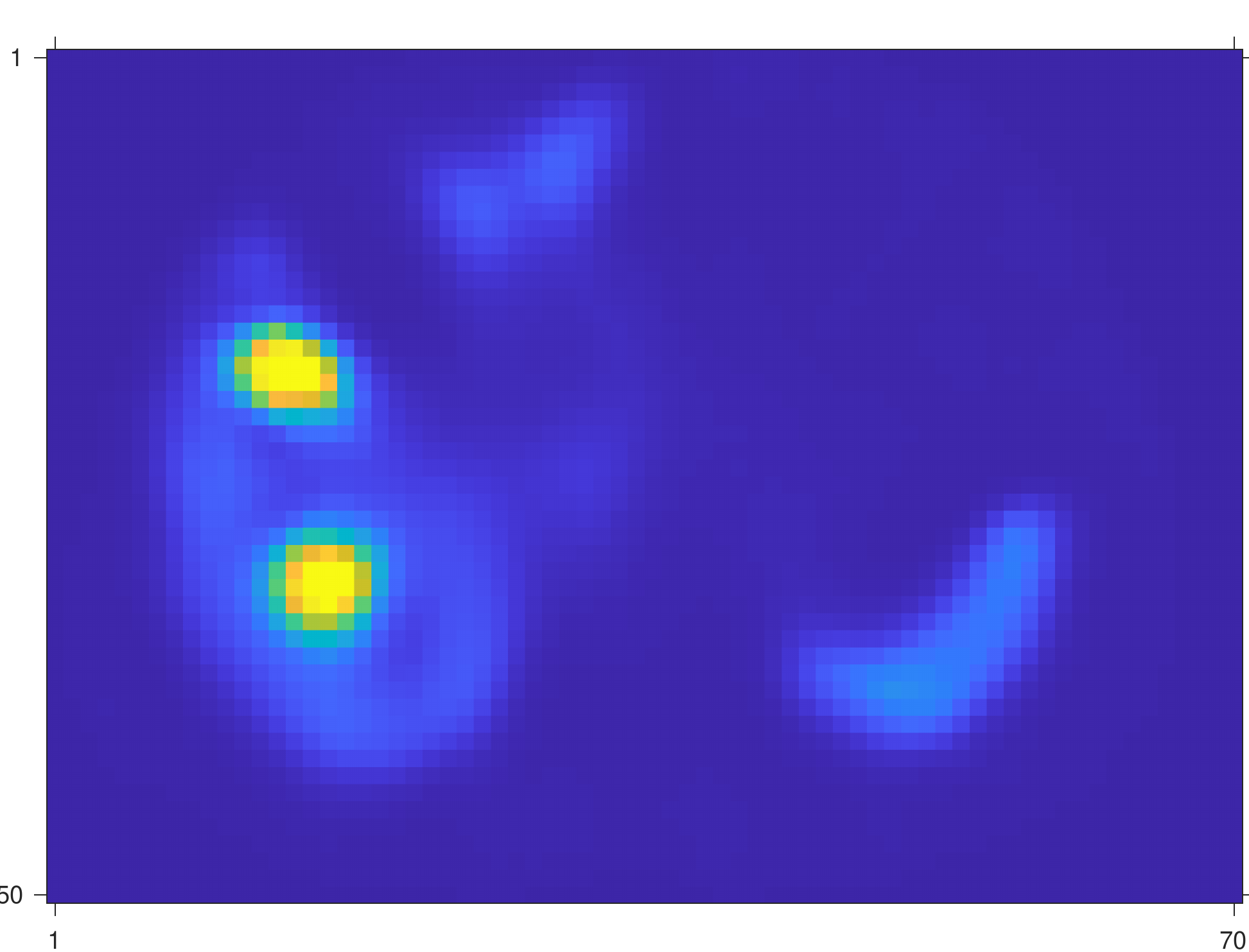}}
\subfloat[End-iter3]{\includegraphics[width=0.33\linewidth]{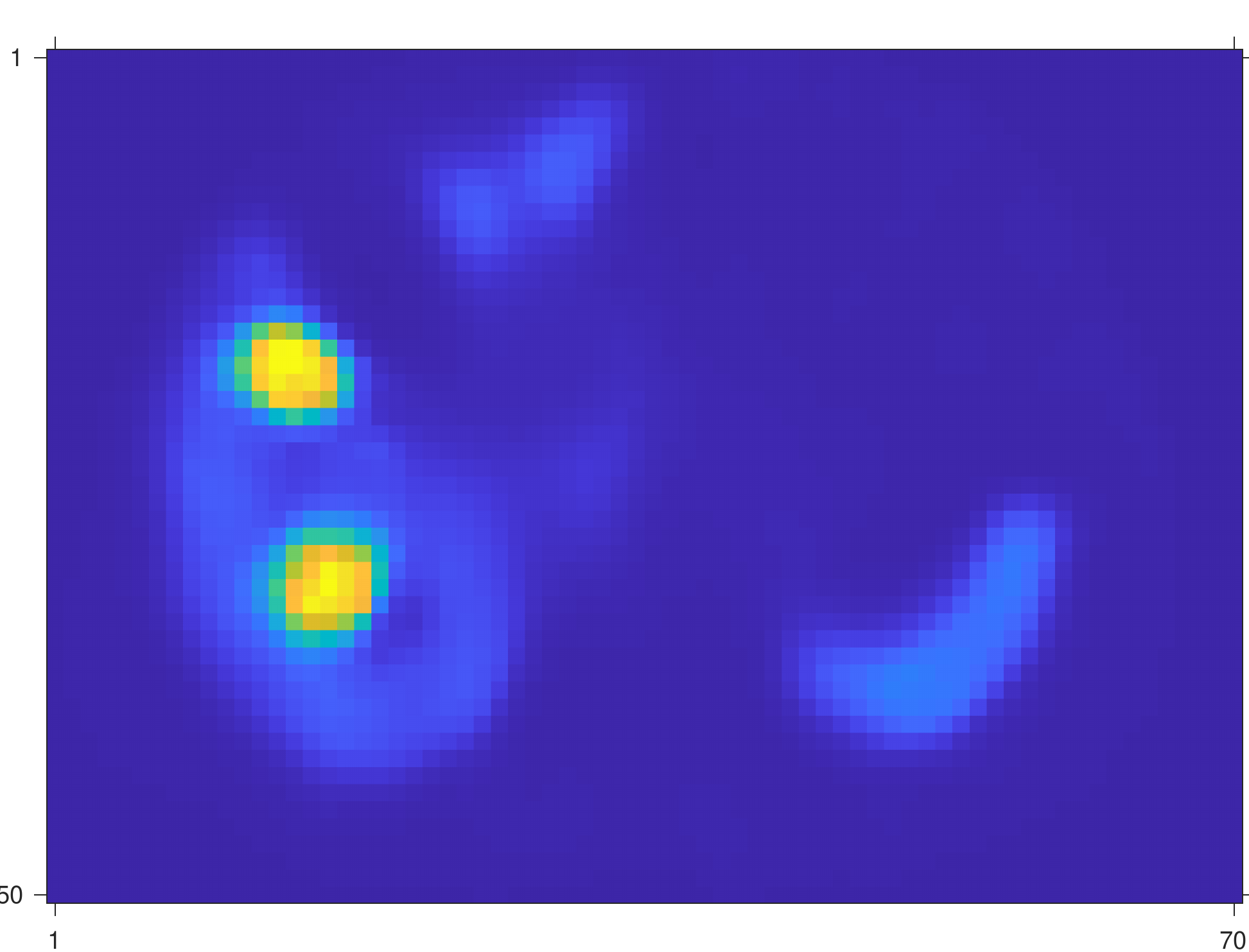}}
\caption{Visualization of intermediate iteration results of
different training methods. Subfigure (d)-(f): sequential training; (g)-(i): gradient truncation; (j)-(l): end-to-end training.}
\label{fig:iter}
\end{figure}

One potential problem 
associated with end-to-end training 
(and gradient truncation)
is that the results at intermediate iterations 
could be unfavorable,
because they are not directly trained by the targets
\cite{knoll:19:rot}. 
Here,
we examined the images 
at intermediate iterations
and did not observe such problems
as illustrated in \fref{fig:iter},
where images at each iteration 
gave a fairly accurate estimate to the true activity.
Perhaps under the shallow-network setting 
(\eg, 3 layers used here,
with only 3 outer iterations),
the network for each iteration was less likely  
to overfit the training data.
Another reason could be due to 
the non-shared weights setting
so that the network 
could learn suitable weights
for each iteration.

\section{Discussion}
\label{sec:discussion}

Training end-to-end CNN-based iterative algorithms
for SPECT image reconstruction
requires memory efficient 
forward-backward projectors
so that backpropagation can be less computationally expensive.
This work implemented a new SPECT projector
using Julia that is an open-source, high performance 
and cross-platform language.
With comparisons between Monte Carlo (MC)
and a Matlab-based projector,
we verified the accuracy, speed and memory-efficiency
of our Julia projector.
These favorable properties support efficient backpropagation
when training end-to-end unrolled iterative 
reconstruction algorithms.
Most modern DL algorithms 
process multiple data batches in parallel, 
so memory efficiency is of great importance
for efficient training and testing neural networks.
To that extent, our Julia projector
is much more suitable than the Matlab-based projector.

We used the CNN-regularized EM algorithm as an example
to test end-to-end training and other training methods
on different datasets including \lu XCAT phantoms,
\lu and \ytt virtual patient phantoms.
Simulation results demonstrated that
end-to-end training improved reconstruction 
quality on these datasets.
For example, 
end-to-end training improved the MAE of lesion/liver 
in \ytt phantoms
by 8.7\%/16.6\% and 7.2\%/12.4\%
compared to sequential training
and gradient truncation.
This improvement could be attributed to 
the correct gradient was used in backpropagation.
Although the end-to-end training yielded
the lowest reconstruction error on both \lu XCAT phantoms
and \ytt VP phantoms,
the reconstruction errors
on \lu VP phantoms were comparable with the gradient truncation.
This could be due to the choice of loss functions
and CNN architectures
in the EM algorithm, 
which we will explore in the future.
Also we noticed that
the recovery of the nonuniform activity in VP phantoms 
was generally higher than activity
for the XCAT phantom 
(MAE reported in \tref{tab:xcat} and \tref{tab:mae,lu177})
because the assigned ``true" activities at the boundaries of organs did not drop sharply, 
and instead, were blurred out. 
And therefore the OSEM 
algorithm was fairly competitive 
as reported in \tref{tab:mae,lu177};
in \ytt VP results, 
the OSEM performed worse than learning-based methods,
which could be attributed to the high 
downscatter associated with \ytt SPECT 
due to the continuous bremsstrahlung energy spectrum.
We found all learning methods
did not work very well for small tumors
(\eg, 5mL),
potentially due to the worse PV effect.
Reducing PV effects in SPECT images has been studied
extensively \cite{pretorius:09:dti,grings:22:tmo}.
Recently,
Xie \etal \cite{xie:22:sfp} trained a 
deep neural network
to learn the mapping between PV-corrected 
and non-corrected images.
Incorporating their network into our reconstruction model 
using transfer learning
is an interesting future direction.

Although promising results were shown in previous sections,
this work has several limitations.
First, we did not test numerous
hyperparameters and CNN architectures,
nor with a wide variety of phantoms and patients 
for different radionuclides therapies.
Secondly, our experiments used OSEM images as warm start
to the CNN-regularized EM algorithm,
where the OSEM itself was initialized with a uniform image.
We did not investigate using other images such as uniform images
as the start of the EM algorithm.
Using a uniform image to initialize the network
would likely require far more network iterations
which would be very expensive computationally
and therefore impractical.
Additionally, this paper used fixed regularization parameter 
($\beta$ in \eqref{e,reg}) rather than declaring $\beta$ 
as a trainable parameter.
We compared different methods for backpropagation,
which requires using the same cost function \eqref{e,cost,poisson}
for a fair comparison.
If one set $\beta$ as a trainable parameter,
then different methods could learn
different $\beta$ values,
leading to different cost functions.
However, the investigation of trainable $\beta$ values
is an interesting future work.
Another limitation is that we did not investigate
more advanced parallel computing methods
such as distributed computing using multiple computers
to further accelerate our Julia implementation
of SPECT forward-backward projector.
Such acceleration is feasible
using existing Julia packages
if needed.
The compute times reported in
\fref{fig:ap,vs,bfbp}
show that the method needs a few seconds
per 128 projection views
using 8 threads,
which is already feasible
for scientific investigation.

We also found there exists a trade-off 
between computational cost and reconstruction accuracy
for different training methods.
End-to-end training yielded reconstruction results
with the lowest MAE and NRMSE 
because the correct gradient was used 
during backpropagation.
Sequential training yielded worse results,
but it was significantly faster and more memory efficient
than the end-to-end training method.
It is notably faster because
it splits the whole training process and
trains each of neural networks separately,
and its backpropagation does not involve terms 
associated with the MLEM algorithm,
so sequential training is actually equivalent 
to training that neural network alone
without considering MLEM. 
Sequential training also used much less memory 
because the training was performed 
iteration by iteration, one network by one network,
and hence the memory limitation did not 
depend on the number of unrolled iterations
in the MLEM algorithm.
\section{Conclusion}
\label{sec:conclusion}
This paper presents a Julia implementation of 
backpropagatable SPECT forward-backward projector
that is accurate, fast and memory-efficient
compared to Monte Carlo (MC) and a previously developed 
analytical Matlab-based projector.
\comment{
Furthermore, our backpropagatable projector
enables end-to-end training of deep learning regularized 
unrolled iterative SPECT reconstruction algorithms 
\blue{as demonstrated}.
Such algorithms have to be trained by either
sequential training or gradient truncation previously.
}
Simulation results based on \lu XCAT phantoms,
\ytt and \lu virtual patient (VP) phantoms
demonstrate that:
1) End-to-end training yielded reconstruction images with
the lowest MAE
and NRMSE when tested on XCAT phantoms 
and \ytt VP phantoms, 
compared to other training methods
(such as sequential training and gradient truncation)
and OSEM.
2) For \lu VP phantoms, 
end-to-end training method yielded 
better results than sequential training and OSEM;
but was rather comparable with gradient truncation.
We also found there exists a trade-off between
computational cost and reconstruction accuracy 
in different training methods 
(\eg, end-to-end training and sequential training).
These results indicate that end-to-end training, 
which is feasible with our developed Julia projector,  
is worth investigating for SPECT reconstruction.

\section*{Acknowledgement}
All authors declare that they have no known conflicts of interest in terms of competing financial interests or personal relationships that could have an influence or are relevant to the work reported in this paper.
\section*{References}
\printbibliography[heading=none]
\end{document}